\documentclass[twocolumn,eqsecnum, twocolappendix]{aastex63}

\usepackage{amsmath}
\usepackage{fourier}
\usepackage{appendix}
\usepackage{comment}
\usepackage{cleveref}

\usepackage{hyperref}

\def\equationautorefname~#1\null{Equation (#1)\null}

\newcommand{\beq}{\begin{equation}}
\newcommand{\eeq}{\end{equation}}

\graphicspath{{./}{figures/}}

\shorttitle{Linear Analysis of a Magnetic Rotating Disk}
\shortauthors{Das \& Basu}

\let\oldhat\hat
\renewcommand{\vec}[1]{\boldsymbol{#1}} 
\renewcommand{\hat}[1]{\oldhat{\boldsymbol{#1}}}
\newcommand{\etaOD}{\eta_{\rm OD,0}}
\newcommand{\etaODt}{\tilde{\eta}_{\rm OD,0}}
\newcommand{\etaAD}{\eta_{\rm AD,0}}
\newcommand{\etaADt}{\tilde{\eta}_{\rm AD,0}}
\newcommand{\tauni}{\tau_{ni,0}}
\newcommand{\taunit}{\tilde{\tau}_{ni,0}}
\newcommand{\ceff}{C_{\rm eff,0}}
\newcommand{\cefft}{\tilde{C}_{\rm eff,0}}
\newcommand{\z}{Z_0}
\newcommand{\zt}{\tilde{Z}_0}

\newcommand{\vAt}{\tilde{V}_{A,0}}

\begin{document}

\title{Linear Stability Analysis of a Magnetic Rotating Disk with Ohmic Dissipation and Ambipolar Diffusion}

\author[0000-0002-7424-4193]{Indrani Das $^{\star}$}
\affiliation{Department of Applied Mathematics, University of Western Ontario, London, Ontario N6A 5B7, Canada}
\affiliation{Department of Physics and Astronomy, University of Western Ontario, London, Ontario N6A 3K7, Canada}
\email{$^{\star}$ idas2@uwo.ca}

\author[0000-0003-0855-350X]{Shantanu Basu $^{\dagger}$}
\affiliation{Department of Physics and Astronomy, University of Western Ontario, London, Ontario N6A 3K7, Canada}
\affiliation{Institute for Earth \& Space Exploration, University of Western Ontario, London, Ontario N6A 5B7, Canada}
\email{$^{\dagger}$ basu@uwo.ca}


\begin{abstract}
We perform a linear analysis of the stability of isothermal, rotating, magnetic, self-gravitating sheets that are weakly ionized. 
The magnetic field and rotation axis are perpendicular to the sheet. 
We include a self-consistent treatment of thermal pressure, gravitational, rotational, and magnetic (pressure and tension) forces together with two nonideal magnetohydrodynamic (MHD) effects (Ohmic dissipation and ambipolar diffusion) that are treated together for their influence on the properties of gravitational instability for a rotating sheet-like cloud or disk.
Our results show that there is always a preferred length scale and associated minimum timescale for gravitational instability. 
We investigate their dependence on important dimensionless free parameters of the problem: the initial normalized mass-to-flux ratio $\mu_0$, the rotational Toomre parameter $Q$, the dimensionless Ohmic diffusivity $\etaODt$, and the dimensionless neutral-ion collision time $\tilde{\tau}_{\rm{ni,0}}$ that is a measure of the ambipolar diffusivity. 
One consequence of $\etaODt$ is that there is a maximum preferred length scale of instability that occurs in the transcritical ($\mu_0 \gtrsim 1$) regime, qualitatively similar to the effect of $\tilde{\tau}_{\rm{ni,0}}$, but with quantitative differences. 
The addition of rotation leads to a generalized Toomre criterion (that includes a magnetic dependence) and modified length scales and timescales for collapse. 
When nonideal MHD effects are also included, the Toomre criterion reverts back to the hydrodynamic value. 
We apply our results to protostellar disk properties in the early embedded phase and find that the preferred scale of instability can significantly exceed the thermal (Jeans) scale and the peak preferred fragmentation mass is likely to be $\sim 10- 90 \ M_{\rm Jup}$.

\end{abstract}



\section{Introduction} \label{sec:intro}

For decades, theoretical studies have suggested that magnetic fields play an indispensable role in the formation and evolution of interstellar clouds, cloud cores, and protostellar disks \citep{mestel56,mouschovias78,shu87,shu99,mou1999,wurster18}.
Recent observations by the {\it Planck} satellite \citep{planckXXI2015, planck2016} have convincingly emphasized the importance of the magnetic field to the density structures on physical scales ranging from tens of parsecs to approximately one parsec in the nearby ($d < 450$ pc) well-known molecular clouds. 
They statistically evaluated the relative orientation between the magnetic field projected on the plane of sky obtained from the polarized thermal emission ($353 \> \rm{Hz}$) of magnetically-aligned dust grains with the maps of gas column density $N_{\rm{H}}$ and found that the magnetic field became oriented more nearly perpendicular to the elongations in column density maps when $N_{\rm{H}} \gtrsim 10^{22}$ cm$^{-2}$. This is consistent with self-gravity becoming important at these column densities but being not so important at lower column densities. By using the Davis-Chandrasekhar-Fermi (DCF) method \citep{davis1951,CF1953} to estimate the magnetic field strength, they also found that the large-scale (low density) magnetic field is quite strong relative to turbulence and self-gravity, with estimations that the turbulence is sub-Alfv\'enic (or close to Alfv\'enic) and the mass-to-flux ratio is subcritical \citep[see Table D.1 in][]{planck2016}. 
\citet{Pattle2017} used polarimetry to estimate a subcritical mass-to-flux ratio ($\sim 0.4$) on the large scale in the Orion A filament. 


\cite{fiedler1993} carried out a two-dimensional ($r,z$ in cylindrical coordinates) simulation of core formation and prestellar collapse in a molecular cloud with an initial subcritical mass-to-flux ratio. In this situation, ambipolar diffusion, the drift of neutrals through the plasma and magnetic field lines because of the imperfect coupling between the neutrals and charged species, can lead to core formation. The cloud has time to settle into a flattened structure with minor axis parallel to the background magnetic field. Based on this result, \cite{CM93,CM94} and \cite{BM94,BM95a,BM95b} studied ambipolar-diffusion-driven protostellar core formation and collapse using the ``thin-sheet'' approximation, with axially symmetric disks threaded by a vertical magnetic field, with hydrostatic equilibrium maintained along field lines at all times. 


The thin-sheet approximation was subsequently used by \citet[][see also \cite{ind00}]{basu04} for models of nonaxisymmetric, gravitationally collapsing cores in subcritical and supercritical clouds. 
\citet[][see also \cite{morton91}]{ciolek06} presented a linear stability analysis of isothermal, partially ionized, magnetic, self-gravitating sheets using the thin-sheet approximation. The preferred fragmentation scale typically has the largest super-Jeans value at transcritical (but mildly supercritical) values of the mass-to-flux ratio.
The predicted preferred fragmentation length scales obtained from this linear analysis were verified to agree with the average fragmentation scales of a large suite of nonlinear evolution calculations in the thin sheet approximation \citep{basu09b}.  Three-dimensional simulations of fragmentation including ambipolar diffusion \citep{KudohBasu3D2007,kudohbasu2011} showed that the general trends are robust.

Gravitational instability (hereafter GI) is also thought to be important in protostellar disks, as a pathway for the formation of stellar companions, brown dwarfs, or giant planets \citep[see review by][]{kratter2016}. Global numerical simulations of disks show that it can produce clumps of the appropriate masses \citep[e.g.,][]{stamatellos2009,vor10planet,basu12,vor16}. Simulations of the self-consistent formation of disks from the collapse of a prestellar core generally show that the disk mass is comparable to the central protostar mass in the early evolution of disks, making them susceptible to GI \citep{Vorobyov06,vor10,vor15}. 


Interest in the early (possibly GI dominated or influenced) evolution of disks has increased due to recent ALMA observations showing that they exist in the early class 0 stage of star formation \citep{sakai2014, ohashi2014, lefloch2015, plunkett2015, ching2016, tokuda2016, aso2017, lee2017, lee2018}. ALMA has also clarified the properties of disk structure in the later class I and II stages \citep{aso2015, bjerkeli2016, perez2016, alves2017}, including a vast array of substructure like gaps, rings, and spiral arms revealed by the DSHARP project \citep[see][]{andrews18, huang18}. These observations show that the process of planet formation is well underway soon after protostar and disk formation. The required rapid planet formation implies a possible important role for GI during the early embedded phase of disks.

Despite extensive work to date on hydrodynamic modeling and observations of disks, the complex role of magnetic fields is just beginning to be explored. Observationally,
magnetic fields are very difficult to detect in disks. An indirect detection through polarization of dust emission due to elongated magnetically-aligned grains is complicated by the polarization due to scattering that can dominate the signal at mm wavelengths \citep{kataoka15,kataoka16,yang16,yang16b}.


In order to understand disk formation and subsequent evolution, the nonideal MHD effects (Ohmic dissipation, ambipolar diffusion, Hall effect) are substantially key features. 
A nascent disk forms in a magnetically subdominant region where the magnetic field is primarily weakened by Ohmic dissipation (hereafter OD) and ambipolar diffusion (hereafter AD) \citep[e.g.,][]{dapp10, dapp12, tomida2015, masson2016, tsukamoto2018, tsukamoto2016, wurster2018b, Hirano2019, Hirano2020}. Without the OD that becomes the dominant form of magnetic dissipation at number densities above $10^{12}$ cm$^{-3}$, a disk may not even form \citep[e.g.,][]{allen2003, galli2006, mellon2008, li2014}; the so-called ``magnetic braking catastrophe''. 


In general then, a rotationally-supported circumstellar disk is formed around a newly born star in a relatively high density region where OD becomes important. 
In the Ohmic regime, the inductive effect of the plasma is restricted by the collisions that the charge carriers encounter (i.e., the resistivity), primarily with neutral particles in the partially ionized environment. The OD must then regulate the different (stable and unstable) modes that occur in the high density environment of protostellar disks. Thus it is important to study the effect of nonideal MHD in a rotating self-gravitating environment that is most applicable to the early evolution of disks. The Toomre criterion is modified due to magnetic fields \citep{lizano2010b}. OD and AD will modify it further still. The effect of GI in inducing giant planet formation 
\citep[e.g.,][]{bodenheimer1995, saigo2006, lizano2010b,machida2016} will be modified by these effects. Furthermore, the marginal state of instability described by the Toomre criterion is known to introduce low-amplitude fluctuations in global disk models \citep{vorobyov07} that can drive the angular momentum transport. Therefore, the effect of nonideal MHD on marginally stable modes is also important to clarify. Three-dimensional MHD simulations of disk formation starting from a prestellar core tend to show that the early evolution of disks is characterized by a significant magnetic field strength such that the mass-to-flux ratio is only mildly supercritical \citep{Hirano2020}.


In this paper, we present a linear stability analysis to explore the gravitational instability in protostellar disks. We adopt a sheetlike model that is isothermal, self-gravitating, weakly ionized, magnetic, and rotating. 
We investigate two nonideal MHD effects: Ohmic dissipation (OD) and ambipolar diffusion (AD). The interplay of OD and rotation are particularly interesting extensions of the analysis presented by \cite{ciolek06}.

Our paper is structured as follows. 
In \autoref{sec:formulation} and \ref{sec:Equations}, we describe the fundamental assumptions and derive the necessary system of governing equations for a model cloud. 
From \autoref{sec:LA} to \autoref{sec:statBlimit}, we describe the stability of the model cloud by linearizing and Fourier analyzing the governing equations, and present some results including the generalized (magnetically dependent) Toomre criterion and the stationary magnetic field limit. 
In \autoref{sec:results} we present numerical results of our model, and in
\autoref{sec:discussion} we relate our results to observations and models of protostellar disks. 
Finally, in \autoref{sec:summary}, we summarize and conclude the consequences of our findings. In the interest of completeness, this paper contains many equations and derivations. Note that every parameter written with a ‘prime' or ‘tilde' denotes their dimensionless form. 


\section{Analytic Considerations} \label{sec:model} 
\subsection{\rm{Physical Formulation}}  \label{sec:formulation}
We formulate model clouds as rotating, self-gravitating, partially ionized, isothermal, magnetic, planar thin sheets with infinite extent in the $x$- and $y$- directions and a local vertical half-thickness $Z(x,y,t)$. Our model follows a similar kind of formulation as adopted and described by  \cite{ciolek06} and \cite{basu09b, basu09a}, but with further interesting physics.

\begin{figure}
\epsscale{1}
\plotone{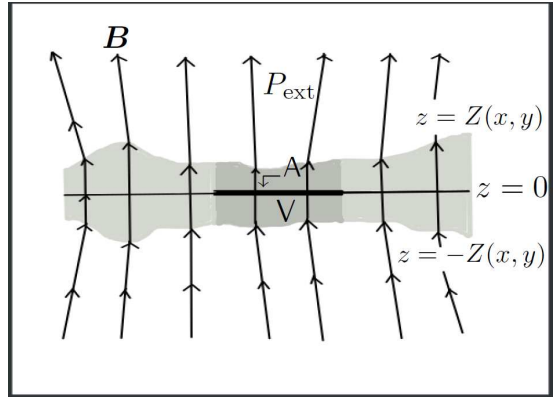}
\caption{Schematic diagram of the thin-disk model. An area $A$ is indicated by the thick dark line and can be seen edge-on in the $z=0$ plane. The associated volume $V$ is shown by dark shaded region, and adjacent lightly shaded regions are bounded by the curves $z = Z$ and $z =-Z$ and a hot, tenuous medium with external pressure $P_{\rm ext}$. The lines with arrows represent the magnetic field.}
\label{fig:thindisk}
\end{figure}

The configuration of the magnetic field threading such a cloud is
\begin{equation}
\begin{aligned}
    \vec{B}(x, y, z, t) {} & = B_{z,{\rm eq}} (x, y, t) \hat{z} \;\;\;\; {\rm for}\;\; |z|\; \leq \; Z(x, y, t), \\
                     & = \bigg[B_z (x, y, z, t)\hat{z}
                     + B_x (x, y, z, t)\hat{x} \\
                     & + B_y (x, y, z, t)\hat{y} \bigg] \;\;\;\; {\rm for} \;\; |z|\; >  Z(x, y, t),
\end{aligned}
\end{equation}
where $B_{z,\rm{eq}}$ is the magnetic field strength in the equatorial plane ($z=0$) of the cloud (see \autoref{fig:thindisk}). In the limit $|z| \rightarrow \infty$, $\vec{B} \rightarrow B_{\rm{ref}} \hat{z}$, where $B_{\rm{ref}}$ is a uniform reference magnetic field very far away from the sheet. 
From now on, all physical quantities are understood to be a function of time $t$.

The unit normal vectors to the upper and lower surfaces of the sheet are given by
\begin{equation}
    \hat{n} = \frac{\pm \hat{z} \; \mp \left[\left(\partial Z/ \partial x \right) \hat{x} + \left(\partial Z/ \partial y \right) \hat{y} \right]}{\left[1 + \left(\partial Z/ \partial x \right)^2 + \left(\partial Z/ \partial y \right)^2 \right]^{1/2}},
\end{equation}
where the upper sign refers to the upper surface and the lower sign to the lower surface.

Using the integral form of Gauss's law yields that the normal components of the magnetic field across the upper and lower surfaces of the sheet are continuous. This leads to
\begin{equation}
\begin{aligned}
    B_z\left(x,y, \pm Z \right) & - \; B_x \left(x,y,\pm Z \right) \frac{\partial Z}{\partial x}\\
                    & - \; B_y \left(x,y, \pm Z \right) \frac{\partial Z}{\partial y} = B_{z,{\rm{eq}}}(x,y).
\end{aligned}
\end{equation}

In our model, we adopt a velocity unit of $c_s$, the isothermal sound speed, and a column density unit of $\sigma_{n,0}$, the initial uniform column density. The length unit is $L_0 = c_s ^2/(2\pi G \sigma_{n,0})$, leading to a time unit $t_0= c_s/(2\pi G \sigma_{n,0})$, where $G$ is the universal gravitational constant.  
The mass unit is $M_0 = c_s^4/(4\pi^2 G^2 \sigma_{n,0})$ and
the unit of acceleration is $2\pi G\sigma_{n,0}$, which is the magnitude of the vertical gravitational acceleration above the planar sheet. The magnetic field strength unit is $B_0 = 2\pi G^{1/2}\sigma_{n,0}$. 
See \autoref{sec:app_params} for the numeric values of all these free parameters.

Vertical pressure equilibrium arises from a balance between thermal pressure and the combined contribution from self-gravitational pressure, magnetic pressure and any other external pressure, which yields
\begin{equation}
\begin{aligned}
    \rho_{n} & c_s ^2  = \frac{\pi}{2}G \sigma_{n} ^2 + \> P_{\rm{ext}} \\
             & + \frac{1}{8\pi} \left[B_{x,Z}^2 +B_{y,Z}^2 + \left\{B_{x,Z} \left(\partial Z/\partial x \right) + B_{y,Z} \left(\partial Z/\partial y \right) \right\}^2\right],
\end{aligned}
\label{eq:pbal}
\end{equation}
where $B_{x,Z} \equiv B_x(x,y, +Z)$, $B_{y,Z} \equiv B_y(x,y, +Z)$, and $\rho_{n}$ and $\sigma_{n}$ are the volume and column mass density of neutrals, respectively. The calculation of $B_{x,Z}$ and $B_{y,Z}$ is discussed in \cite{ciolek06}. Note that in molecular clouds, $\rho_n \gg \rho_i$, where $\rho_i$ is the ion density. 
Furthermore, $c_s = (k_B T /m_n )^{1/2}$ is the isothermal sound speed, $k_B$ is the Boltzmann constant, $T$ is the temperature and $m_n$ is the mean mass of a neutral particle ($m_n = 2.33 \> $amu). The evolution equations of our model include the effect of AD and OD.
Because of AD, neutrals can stay at least partially coupled to the magnetic field via neutral-ion collisions.
This is quantified by the time scale for collisions between neutrals and ions \citep[e.g.,][and references within]{BM94}:

\begin{equation}
\tau_{ni} \equiv 1.4 \frac{m_i + m_n}{m_i} \frac{1}{n_i \langle \sigma w \rangle_{i {\rm H}_2}},
\end{equation}
where $\langle\sigma w\rangle_{i {\rm H}_2}$ is the average collision rate between ions of mass $m_i$ (singly ionized Na, Mg, and HCO, for which we adopt a typical mass of 25 amu) and neutrals of mass $m_n$. We adopt a neutral-ion collision rate between ${\rm H}_2$ and $\rm{HCO}^{+}$ as $1.69 \times 10^{-9}\, \rm{cm}^3\, \rm{s}^{-1}$ \citep{mcdaniel1973}. 
These collisions transport knowledge of the magnetic field to the neutral particles via ions that are tied to the field lines. 
The factor $1.4$ arises because the inertia of helium is neglected in calculating the slowing-down time of the neutrals by collisions with ions \citep{ciolek06, mou1999}.

We adopt a constant power-law approximation for calculating the ion number density ($n_i$) in terms of the neutral number density ($n_n$):
\begin{equation}
n_i = \kappa \;  \Bigg(\frac{n_n}{10^5 \; {\rm cm}^{-3}} \Bigg)^k,
\end{equation}
where $\kappa\, (= 3 \times 10^{-3} \rm{cm^{-3}})$ and $k\, (= 1/2)$ are constants \citep[see][]{ciolek06, CM98powerk} . So, the ionization fraction ($\chi_i$) can be written as 
\begin{equation}
    \chi_i = \frac{n_i}{n_{n}} \approx 10^{-5} \> n_{n}^{-1/2},
\end{equation}
and is typically a very small number, $\approx 10^{-7}$
when $n_n = 10^4$ cm$^{-3}$. Molecular clouds are weakly ionized yet retain a relatively good (though imperfect) coupling between plasma and neutrals due to the enhanced Langevin cross section for ion-neutral collisions \citep[see][\S\ 27]{shu1992gas}.

In our formulation we include the additional nonideal MHD effect of OD. It is a measure of the decoupling of the charged species from the magnetic field, due to resistivity arising from collisions of the charge carriers with neutrals. Collisions of the charged species with each other is neglected as we are studying a weakly ionized plasma. 
The conductivity for each charged species $s=e,i$ can be written as
\begin{equation}
    \sigma_s = \frac{n_s q_s^2  \tau_{sn}}{m_s} ,
\end{equation}
where $n_s$ is the number density of each charged species (we can assume $n_e \approx n_i$ due to charge neutrality), $q_s$ is the charge of each species, $m_s$ is the mass of each species, and $\tau_{sn}$ is the mean collision time of each charged species with neutrals (see \autoref{sec:coll_time_app}). 
We define conductivity $\sigma_c = \sum_{s= e, i} \sigma_s$, and the electron contribution is expected to dominate. 
So, finally, the expression of Ohmic diffusivity ($\eta_{\rm{OD}}$) can be written as
\begin{equation}
    \eta_{\rm{OD}} = \frac{c^2}{4\pi \sigma_c},
\end{equation} 
where $c$ is the speed of the light. Note that $(4\pi \eta_{\rm{OD}})/{c^2}$ is the Ohmic resistivity, i.e., the inverse of the conductivity, and leads to the 
well-known form of Ohm's law:
\begin{equation}
   \vec{E}_{n} = \frac{4\pi \eta_{\rm{OD}}}{c^2} \vec{j} ,
\end{equation}
where $\vec{j}$ is the electric current density and $\vec{E}_n$ is the electric field in the reference frame of the neutrals \citep[for more details, see][]{dapp12}.
 


\subsection{{{Fundamental Equations}}} \label{sec:Equations}
The system of equations for the model cloud are derived \citep[see][]{ciolek06} by integrating the fundamental MHD equations over the vertical direction (i.e., from $z_{\rm{lower}} = -Z(x,y)$ to $z_{\rm{upper}}= Z(x,y)$). Doing the same for the equation of mass continuity yields

\beq
    \frac{\partial \sigma_n}{\partial t} + {\vec{\nabla}}_p \cdot (\sigma_n \> \vec{v}_n) = 0,
\label{eq:masscontequn}    
\eeq
where $\sigma_n (x,y) = \int_{-Z} ^{Z} \rho_n (x,y) dz $.
Next, we consider the equation of force using the total stress tensor (thermal plus Maxwell) 
\beq
\vec{T} = -\left[\rho_n c_s^2 + \frac{B^2}{8\pi} \right]\vec{1} + \frac{\vec{B B}}{4\pi} ,
\label{eq:stresstensor}
\eeq
here, $\vec{1}$ is identity tensor. The force equation (per unit area) in the rotating frame of reference for the neutrals is given by,

\beq
\begin{aligned}
    \frac{\partial}{\partial t} (\sigma_n \> \vec{v}_n) + {\boldsymbol{\nabla}}_p \cdot \> (\sigma_n \vec{v}_n & {} \> \vec{v}_n) = \vec{F}_T + \vec{F}_{\rm{Mag}}\\
                                                                                                           & + \sigma_n \left[\vec{g}_p - 2(\vec{\Omega} \times \vec{v}_n) + \vec{\Omega} \times (\vec{\Omega} \times \vec{r})\right],
\end{aligned}
\label{eq:forceequn}
\eeq
where

\begin{equation}
    \vec{F}_T = -C^2 _{\rm{eff}} \vec{\nabla}_p \sigma_n,
\end{equation}

\begin{equation}
    \vec{F}_{\rm{Mag}} = \frac{B_{z,\rm{eq}}}{2\pi} \left(\vec{B}_p-Z\nabla_p B_{z,\rm{eq}} \right) +\mathcal{O} \left(\vec{\nabla}_p Z \right),
\label{eq:Fmagequn}    
\end{equation}

\begin{equation}
    Z = \frac{\sigma_n}{2 \rho_n},
\label{eq:equn_Z_sigma}    
\end{equation}

\begin{equation}
    \vec{g}_p = -\nabla_p \psi,
\end{equation}

\begin{equation}
    \psi = \mathcal{F}^{-1} \left[-2\pi G \frac{\mathcal{F}(\sigma_n)}{k} \right],
\end{equation}

\begin{equation}
    \vec{B}_p = -\nabla_p \Psi,
\end{equation}

\begin{equation}
    \Psi = \mathcal{F}^{-1} \left[\frac{\mathcal{F} \left(B_{z,\rm{eq}}-B_{\rm{ref}} \right)}{k} \right] ,
\end{equation}

\begin{equation}
    C^2_{\rm{eff}} = \frac{\pi}{2} G \sigma^2_{n} \frac{\left[3P_{\rm{ext}}+ \frac{\pi}{2} G \sigma^2_{n}\right]}{\left[P_{\rm{ext}}+\frac{\pi}{2} G \sigma^2_{n}\right]^2}  c_s ^2.
\label{eq:ceff}    
\end{equation}
In the above equations, $\vec{r} = x\hat{x} + y\hat{y}$, $\vec{\nabla}_p \equiv \hat{x}\partial/\partial x + \hat{y}\partial/\partial y$  is the planar gradient operator and $\psi$ and $\Psi$ are the gravitational and magnetic potential, respectively. Here, $\mathcal{F}(f)$ and $\mathcal{F}^{-1}(f)$ represent the forward and backward Fourier transform of a function $f$, respectively. The
$C_{\rm{eff}}$ is the local effective sound speed which includes the effect of an external pressure. In the absence of $P_{\rm{ext}}$, $C_{\rm{eff}}$ is reduced to the isothermal sound speed $c_s$.
The $\vec{v}_n (x,y) = v_{n,x} (x,y) \hat{x} + v_{n,y} (x,y) \hat{y}$ is the velocity of neutrals in the plane of the sheet. The planar sheet is rotating with an angular velocity $\Omega$ about the $z$-axis, so that $\vec{\Omega} = \Omega \hat{z}$. The magnetic field and rotation axis are perpendicular to the sheet. Here, $2(\vec{\Omega} \times \vec{v}_n)$ and $\vec{\Omega} \times (\vec{\Omega} \times \vec{r})$ are the Coriolis and centrifugal acceleration terms, respectively.
A more complete expression of $\vec{F}_{\rm{Mag}}$ can be written showing the $\mathcal{O}(\nabla_p Z)$ terms explicitly (see \autoref{eq:fmagx}, \ref{eq:fmagy}; also \cite{ciolek06}). 
The vertical $z$-wavenumber $k$ ($>0$) is presented as a function of $k_x$ and $k_y$, which are the $x$-, and $y$-wavenumbers in the plane of the sheet such that $k \equiv k_z = (k_x^2 +k_y^2)^{1/2}$. 
By a sheet being thin we mean that for any physical quantity $f (x, y, z, t)$, the criterion $f/ \nabla_p f \gg Z$ is satisfied.

The advection of magnetic flux for our model is described by the magnetic induction equation,
\beq
    \frac{\partial \vec{B}_{z,\rm{eq}}}{\partial t} = {\vec{\nabla}_p} \times \left(\vec{v}_i \times \vec{B}_{z,\rm{eq}} \right) - \> \vec{\nabla}_p \times \left(\eta_{\rm{OD}} \> \vec{\nabla}_p \times \vec{B}_{z,\rm{eq}} \right),
\label{eq:inductionequn}    
\eeq
where 
\begin{equation}
    \vec{v}_{i} = \vec{v}_{n} + \frac{\tau_{ni}}{\sigma_n} \>  {\vec{F}_{\rm{Mag}}}.
\label{eq:velocityequn_in}    
\end{equation}
In the above equations, $\eta_{\rm OD}$ and $\tau_{ni}$ are the Ohmic diffusivity, neutral-ion collision time, respectively.
The $\vec{v}_i (x,y)$ is the ion velocity such that  $\vec{v}_i (x,y) = v_{i,x} (x,y) \hat{x} + v_{i,y} (x,y) \hat{y}$. 
Finally, we obtain a simplified form of these equations by separating the $x$- and $y$- components. Doing that for \autoref{eq:masscontequn}, \autoref{eq:forceequn}, \autoref{eq:Fmagequn}, and \autoref{eq:inductionequn}, yields

\begin{equation}
     \frac{\partial \sigma_n}{\partial t} + \frac{\partial}{\partial x}(\sigma_n v_{n,x}) + \frac{\partial}{\partial y} (\sigma_n v_{n,y}) = 0 ,
\label{eq:masscontequn_xy}
\end{equation}


\begin{equation}
\begin{aligned}
    \frac{\partial}{\partial t} (\sigma_n v_{n,x}) {} & + \frac{\partial}{\partial x} (\sigma_n v_{n,x}^2)+ \frac{\partial}{\partial x}  (\sigma_n v_{n,x} v_{n,y})\\
                                                     &  = \sigma_n \> g_x - C_{\rm{eff}} ^2 \frac{\partial \sigma_n}{\partial x} + F_{{\rm Mag}, x} + 2 \sigma_n \Omega v_{n,y},
\end{aligned}   
\label{eq:forceequn_x}
\end{equation}

\begin{equation}
\begin{aligned}
    \frac{\partial}{\partial t} (\sigma_n v_{n,y}) {} & + \frac{\partial}{\partial y} (\sigma_n  v_{n,x} v_{n,y})+ \frac{\partial}{\partial y}  (\sigma_n v_{n,y}^2) \\
                                                     & = \sigma_n \> g_y - C_{\rm{eff}} ^2 \frac{\partial \sigma_n}{\partial y} + F_{{\rm Mag}, y} - 2 \sigma_n \Omega v_{n,x},
\end{aligned}
\label{eq:forceequn_y}
\end{equation}

\begin{equation}
\begin{aligned}
    F_{{\rm Mag},x} {} & = \frac{B_{z, \rm{eq}}}{2\pi} \> \left(B_{x,Z} - Z\> \frac{\partial B_{z, \rm{eq}}}{\partial x} \right) \\
                      & + \frac{1}{4\pi} \frac{\partial Z}{\partial x} \Biggr[B^2_{x,Z} +B^2_{y,Z} +  2 B_{z,\rm{eq}} \left(B_{x,Z} \frac{\partial Z}{\partial x}+ B_{y,Z} \frac{\partial Z}{\partial y} \right)\\
                      & \hspace{3 cm}+ \left(B_{x,Z} \frac{\partial Z}{\partial x}+ B_{y,Z} \frac{\partial Z}{\partial y} \right) ^2 \Biggr],
\end{aligned}          
\label{eq:fmagx}
\end{equation}

\begin{equation}
\begin{aligned}
    F_{{\rm Mag},y} {} & = \frac{B_{z, \rm{eq}}}{2\pi} \> \left(B_{y,Z} - Z\> \frac{\partial B_{z, \rm{eq}}}{\partial y}\right) \\
                      & + \frac{1}{4\pi} \frac{\partial Z}{\partial y} \Bigg[B^2_{x,Z} +B^2_{y,Z} +  2 B_{z,\rm{eq}} \left(B_{x,Z} \frac{\partial Z}{\partial x}+ B_{y,Z} \frac{\partial Z}{\partial y}\right)\\
                      & \hspace{3 cm} + \left(B_{x,Z} \frac{\partial Z}{\partial x}+ B_{y,Z} \frac{\partial Z}{\partial y} \right)^2 \Bigg],
\end{aligned} 
\label{eq:fmagy}
\end{equation}

\begin{equation}
    \begin{aligned}
    \frac{\partial B_{z,\rm{eq}}}{\partial t} = -\frac{\partial}{\partial x} {} & \left(B_{z,\rm{eq}} v_{i,x} \right)  -\frac{\partial}{\partial y} \left(B_{z,\rm{eq}} v_{i,y} \right)\\
                                                                                & + \left[\frac{\partial}{\partial x}\left(\eta_{\rm{OD}} \frac{\partial B_{z,\rm{eq}}}{\partial x}\right)+ \frac{\partial}{\partial y} \left(\eta_{\rm{OD}} \frac{\partial B_{z,\rm{eq}}}{\partial y}\right)\right].
    \end{aligned}
\label{eq:inductionequn_xy}    
\end{equation}
Note that in the force equations we no longer consider the centrifugal term [$\vec{\Omega} \times (\vec{\Omega} \times \vec{r}) = -\Omega^2(x\hat{x} + y\hat{y})$]. This is because we assume that the centrifugal force is balanced in the background state by a gravitational force produced by an unspecified mass distribution. This is a form of the ``Jeans swindle'', to rely on a force balance in the uniform background state \citep[see][\S\ 5.6.1]{binneytremaine2008}.



\subsection{Stability of the model: Linearization and Analysis} \label{sec:LA}

Starting with a static uniform background, any physical quantity of the thin-sheet equations can be expanded by
writing it via 
\beq
f(x,y,t) = f_0 + \delta f_a e^{i\left( k_x x + k_y y -\omega t \right)},
\eeq
where $f_0$ is the unperturbed background state, $\delta f_a$ is the amplitude of the perturbation. 
$k_x$, $k_y$, and $k$ are the $x$-, $y$-, and $z$- wavenumbers, respectively, and $\omega$ is the complex angular frequency. 
With this Fourier analysis, $\partial/\partial t \rightarrow -i\omega$, $\partial/\partial x \xrightarrow{} i k_x$ , and $\partial/\partial y \rightarrow i k_y$. 
For assumed small-amplitude perturbations such that $\left|\delta \>f_a \right| \ll f_0 $, and retaining the linearized form of the perturbed quantities from Eqs. \ref{eq:masscontequn_xy}, 
\ref{eq:forceequn_x}, \ref{eq:forceequn_y} and \ref{eq:inductionequn_xy}, the following equations are obtained 
\beq
\omega \> \delta \sigma'_{n} =  k_x \> c_s \> \delta v'_{n,x} + k_y \>  c_s \> \delta v'_{n,y} \>\>,
\label{eq:lin_mass_con}
\eeq

\beq
\begin{aligned}
\omega \> c_s \> \delta v'_{n,x} ={} & \frac{k_x}{k} \> \left[C_{\rm{eff,0}} ^2 \> k - 2 \pi G \sigma_{n,0} \right] \> \delta \sigma'_{n} + i \> 2\Omega c_s \delta v'_{n,y} \\
                                    & + \frac{k_x}{k} \> \left[\> 2 \pi G \sigma_{n,0} \> \mu_0 ^{-1} + k\>  V_{A,0} ^2 \> \mu_0 \>\right] \>\delta B'_{z,\rm{eq}} \>\>,
\end{aligned}
\eeq

\beq
\begin{aligned}
\omega \> c_s \> \delta v'_{n,y} ={} &  \frac{k_y}{k} \> \left[C_{\rm{eff,0}} ^2 \> k - 2 \pi G \sigma_{n,0}\right] \> \delta \sigma'_{n} - i \> 2\Omega c_s \delta v'_{n,x} \\
                                    & + \frac{k_y}{k} \> \left[\> 2 \pi G \sigma_{n,0} \> \mu_0 ^{-1} + k\>  V_{A,0} ^2 \> \mu_0 \>\right] \>\delta B'_{z, \rm{eq}} \>\>,
\end{aligned}
\eeq

\beq
\begin{aligned}
\omega \> {} & \delta B'_{z, \rm{eq}}  = \frac{k_x}{\mu_0} \> c_s \> \delta v'_{n,x}  +  \frac{k_y}{\mu_0} \> c_s \> \delta v'_{n,y} \\
             & - i \left[\etaOD \> k^2 + \>\tau_{ni,0} \> \left(2\pi G \sigma_{n,0} \mu_0 ^{-2}k + k^2 \> V_{A,0}^2 \right)\right] \> \delta B'_{z,\rm{eq}} ,
\end{aligned}
\label{eq:lin_induction}
\eeq
where 
the perturbed eigenfunctions $\delta \sigma_{n}$, $\delta v_{n,x}$  (and $\delta v_{n,y}$), $\delta B_{z, \rm{eq}}$ are normalized by $\sigma_{n,0}$, $c_s$ and $B_0 \, (= 2\pi G^{1/2}\sigma_{n,0})$, respectively such that $\delta \sigma'_{n} = \delta \sigma_{n}/\sigma_{n,0}$, $\>\delta v'_{n,x} = \delta v_{n,x}/c_s$, $\delta v'_{n,y} = \delta v_{n,y}/c_s \> $, and $\> \delta B'_{z, \rm{eq}} = \delta B_{z, \rm{eq}}/B_0$. Here, $\tau_{ni,0}$, $\etaOD$, $C_{\rm{eff,0}}$, and $\sigma_{n,0}$, $\rho_{n,0}$  represent the initial uniform component of neutral-ion collision time, the Ohmic diffusivity, the local effective sound speed, the mass column density of the sheet, and the volume density, respectively. 
The quantities $\tau_{ni,0}$ and $\etaOD$ are regarded as measures of AD and OD, respectively. 
From \autoref{eq:velocityequn_in} one obtains
\begin{equation}
    v_{i,x} = v_{n,x} + \frac{\tau_{ni,0}}{\sigma_n} \> \left(\frac{\rho_{n,0}}{\rho_n} \right)^{1/2} \> {F_{{\rm Mag},x}} ,
\label{eq:velocityequn_in_xcomp} 
\end{equation}
\begin{equation}
    v_{i,y} = v_{n,y} + \frac{\tau_{ni,0}}{\sigma_n} \> \left(\frac{\rho_{n,0}}{\rho_n} \right)^{1/2} \> {F_{{\rm Mag},y}} ,
\label{eq:velocityequn_in_ycomp}    
\end{equation}
where $v_{i,x}$, $v_{i,y}$, $v_{n,x}$, $v_{n,y}$ have been discussed earlier. 
The above equations introduce the normalized initial mass-to-flux ratio of the background reference state,
\beq
\mu_0 \equiv 2\pi G^{1/2} \frac{\sigma_{n,0}}{B_{\rm{ref}}} = \frac{1}{\tilde{B}_{\rm{ref}}} ,
\label{eq:mu}
\eeq
where 
$\tilde{B}_{\rm{ref}} = B_{\rm{ref}}/B_0$, and $(2\pi G^{1/2})^{-1}$ is the critical mass-to-flux ratio for gravitational collapse in the adopted model \citep{nakano1978, ciolek06}, and $B_{\rm{ref}}$ is the magnetic field strength of the background reference state that is equal to the initial uniform component of the magnetic field strength in the equatorial plane of the cloud ($B_{z,\rm{eq},0}$). Regions with $\mu_0 < 1$ are defined as subcritical, regions with $\mu_0 > 1$ are defined to be supercritical, and regions with $\mu_0 \approx 1$ are transcritical.
Furthermore, $V_{A,0}$ is the initial uniform Alfv\'en speed,
\begin{equation}
    V^2_{A,0} \equiv \frac{B^2_{\rm{ref}}}{4 \pi \rho_{n,0}} = 2\pi G \sigma_{n,0} \mu_0 ^{-2} Z_0 \>\>.
\end{equation}
The initial uniform component of the ambipolar diffusivity can be expressed as 
\begin{equation}
    \etaAD = V^2_{A,0} \tau_{ni,0} = 2\pi G \sigma_{n,0} \mu_0 ^{-2} Z_0  \tauni.
    \label{eq:etaAD}
\end{equation}
The initial vertical half-thickness is
\begin{equation}
    Z_0 = \frac{\sigma_{n,0} c^2 _s}{\pi G \sigma_{n,0}^2 +  2 P_{\rm{ext}}}.
\label{eq:Z0}    
\end{equation}
From now on, we use the following form of the pressure balance equation 
\beq
\rho_{n,0} c_s ^2 = \frac{\pi}{2}G \sigma_{n,0} ^2 + \> P_{\rm{ext}},
\label{eq:pbal_lin}
\eeq
obtained by linearizing \autoref{eq:pbal}.


\subsection{Dispersion Relation }\label{sec:DR}
A gravitationally unstable mode occurs if one of the imaginary parts of the complex angular frequency ($\omega_{\rm{IM}}$) leads to a growing solution, i.e., $\omega_{\rm{IM}} > 0$. 
The growth time of such an instability is obtained from the relation $\tau_g = 1/\omega_{\rm{IM}}$. The dispersion relation is found from the following system of equations:
\begin{equation}
\begin{bmatrix}
-\omega & k_x c_s & k_y c_s & 0\\
\frac{k_x}{k}A_1 & -\omega \> c_s & 2i\Omega c_s & \frac{k_x}{k} A_2 \\
\frac{k_y}{k} A_1 & -2i\Omega c_s & -\omega \> c_s & \frac{k_y}{k} A_2 \\
0 &  \frac{k_x}{\mu_0} c_s &  \frac{k_y}{\mu_0} c_s & -[\omega + i(\theta + \gamma)]
\end{bmatrix}
\begin{bmatrix}
\delta \sigma'_{n} \\
\delta v'_{n,x} \\
\delta v'_{n,y} \\
\delta B'_{z,\rm{eq}} \\
\end{bmatrix}
= 0    \;,
\label{eq:matrix}
\end{equation}
where 
\begin{equation}
    A_1 = \left(C_{\rm{eff,0}} ^2 k - 2 \pi G \sigma_{n,0} \right),
\end{equation}

\begin{equation}
    A_2 = \left(2 \pi G \sigma_{n,0} \> \mu_0 ^{-1} + k\>  V_{A,0} ^2 \> \mu_0 \right),
\end{equation}

\begin{equation}
    \gamma = \etaOD \> k^2,
\label{eq:gamma}
\end{equation}

\begin{equation}
    \theta = \tau_{ni,0} \left(2\pi G \sigma_{n,0}  \mu_0 ^{-2} k + k^2 V_{A,0} ^2 \right) = \etaAD \frac{(k+ Z_0 k^2)}{Z_0},
\label{eq:theta}    
\end{equation} 
(see \autoref{sec:etak_dis} for more discussion on $\etaAD$ and $\etaOD$).
Now, solving the determinant of the above matrix, the dispersion relation is
\begin{equation}
\begin{aligned}
     \left(\omega + i \> [\theta + \gamma] \right)  \big(\omega^2 {} & - C^2_{\rm{eff,0}} \> k^2  + 2\pi G \sigma_{n,0} k - 4\Omega^2 \big) \\
                                                     & = \omega \> \left[2\pi G \sigma_{n,0} k \mu_0 ^{-2} + k^2 \> V^2 _{A,0} \right] \;.
\end{aligned}
\label{eq:DR}
\end{equation}
In the limit of flux-freezing (${\tau}_{ni,0} \xrightarrow \> 0$, \ $\etaOD \xrightarrow \> 0$),
\begin{equation}
\begin{aligned}
    \omega^2 - C^2_{\rm{eff,0}} \> k^2  {} & + 2\pi G \sigma_{n,0} k - 4\Omega^2\\
                                           & = 2\pi G \sigma_{n,0} k \mu_0 ^{-2} + k^2 \> V^2 _{A,0} \;.
\end{aligned}
\label{eq:ffDR}
\end{equation}
In the limit of OD only (${\tau}_{ni,0} \xrightarrow \> 0$),
\begin{equation}
\begin{aligned}
    \left(\omega + i \> \gamma \right) \big(\omega^2 {} & - C^2_{\rm{eff,0}} \> k^2  + 2\pi G \sigma_{n,0} k - 4\Omega^2 \big)\\
                                        & = \omega \> \left(2\pi G \sigma_{n,0} k \mu_0 ^{-2} + k^2 \> V^2 _{A,0} \right) \;.
\end{aligned}
\end{equation}
In the limit of AD only ($\etaOD \xrightarrow \> 0$),
\begin{equation}
\begin{aligned}
    \left(\omega + i \> \theta \right) \big(\omega^2 {} & - C^2_{\rm{eff,0}} \> k^2  + 2\pi G \sigma_{n,0} k - 4\Omega^2 \big)\\
                                         & = \omega \> \left(2\pi G \sigma_{n,0} k \mu_0 ^{-2} + k^2 \> V^2 _{A,0} \right) \;.
\end{aligned}
\end{equation}
In the limit of flux-freezing, the gravitationally unstable mode corresponds to one of the roots of $\omega^2 < 0$ and occurs for $\mu_0 >1$. The growth time for this mode becomes a function of $\Omega$ and $\mu_0$ and can be written as
\begin{equation}
    \tau_g = \frac{\lambda}{2\pi \left[G \sigma_{n,0} (1-\mu_0 ^{-2})(\lambda - \lambda_{MS}) -\> \frac{\Omega^2 \lambda^2}{\pi^2} \right]^{1/2}} \;,
\end{equation}
for $\lambda \geq \lambda_{MS}$, where
\begin{equation}
    \lambda_{\rm{MS}} = \frac{\left(C^2_{\rm{eff,0}} + V^2 _{A,0} \right)}{ G \sigma_{n,0} \left(1-\mu_0 ^{-2} \right) } .
\end{equation}
The minimum growth time for the unstable mode occurs at the preferred magnetosonic length scale $\lambda_{\rm{MS,m}} = 2\lambda_{\rm{MS}}$. 
As $\mu_0 \rightarrow \infty$, this implies negligible magnetic support ($\tilde{B}_{\rm{ref}} \rightarrow 0$). In this regime, the growth time $\tau_{g,T}$ is dependent on the critical thermal length scale $(\lambda_{\rm T} \equiv C^2_{\rm{eff,0}} /(G\sigma_{n,0}))$ as follows:
\begin{equation}
    \tau_{g,T} = \frac{\lambda} {2\pi \left[G \sigma_{n,0} \left(\lambda- \lambda_{\rm T} \right) -\> \frac{\Omega^2 \lambda^2}{\pi^2} \right]^{1/2}} \;.
\label{eq:fftaug}    
\end{equation}
In this regime ($\mu_0\gg1$), the minimum growth time for the unstable mode occurs at the preferred thermal length scale $\lambda_{T,m} = 2\lambda_{\rm T}$. 

After including the nonideal MHD effects, i.e., OD and AD, the gravitationally unstable mode still corresponds to one of the roots of the full dispersion relation (\autoref{eq:DR}) and all of them are obtained numerically. Because it is a cubic equation, an analytic expression of the growth time cannot be written down as simply as \autoref{eq:fftaug}.



\subsection{Normalization} \label{sec:norm}

The model we use can be characterized by several dimensionless free parameters in addition to $\mu_0$ (see \autoref{eq:mu}). 
We normalize all length scales by $L_0$ and timescales by $t_0$, mass by $M_0$, column densities by $\sigma_{n,0}$, magnetic field strength by $B_0$. 
We define a dimensionless form of the initial neutral-ion collision time $\Tilde{\tau}_{ni, 0} = {\tau}_{ni, 0}/t_0 = \left(2\pi G \sigma_{n,0} \tau_{ni, 0} \right)/c_s$ and a dimensionless external pressure $\Tilde{P}_{\rm{ext}} \equiv (2 P_{\rm{ext}})/ (\pi G \sigma^2 _{n,0} )$. The dimensionless local effective sound speed is 
\begin{equation}
    \tilde{C}_{\rm{eff},0} = \frac{C_{\rm{eff},0}}{c_s} = \frac{\left(1+3\tilde{P}_{\rm{ext}}\right)^{\frac{1}{2}}}{\left(1+\tilde{P}_{\rm{ext}}\right)}.
\label{eq:ceff_tilde}    
\end{equation}
The dimensionless Alfv\'en wave speed is
 \begin{equation}
     \tilde{V}_{A,0} = \frac{{V}_{A,0}}{c_s} =  \tilde{Z}_0^{1/2} \tilde{B}_{\rm{ref}} = \tilde{Z}_0^{1/2}  \mu_0 ^{-1},
 \end{equation}
We define the normalized ambipolar diffusivity 
\begin{equation}
    \etaADt = \etaAD \left(t_0 /L_0^2 \right)= \vAt^2 \tilde{\tau}_{ni,0} = \tilde{Z}_0 \mu_0^{-2}\tilde{\tau}_{ni,0},
    \label{eq:etaADt}
\end{equation}
and a normalized Ohmic diffusivity $\etaODt = \etaOD \> \left(t_0 / L_0 ^2 \right)$. 
Here, $\tilde{Z}_0$ is the normalized local vertical half-thickness and is written as 
\begin{equation}
    \tilde{Z}_0 = \frac{Z_0}{L_0} = \frac{2}{\left(1 + \Tilde{P}_{\rm{ext}} \right)},
\label{eq:equn_Z0_Pext}    
\end{equation}
(note that $\tilde{Z}_0 = 2$ with no external pressure). 
See \autoref{sec:figA} for more discussion on $\cefft$ and $\zt$.
The normalized isothermal magnetosonic speed in our units is written as
\begin{equation}
     \tilde{V}_{\rm{MS},0} = \left(\tilde{C}^2_{\rm{eff},0} + {\tilde{V}}^2_{A,0} \right)^\frac{1}{2} = \tilde{C}_{\rm{eff},0} \left[1 + \frac{2}{\mu_0 ^2} \frac{(1+\tilde{P}_{\rm{ext}})}{(1+3\tilde{P}_{\rm{ext}})} \right]^\frac{1}{2} .
\label{eq:v_MS}     
\end{equation}
 
The normalized form of governing equations (see \Crefrange{eq:lin_mass_con}{eq:lin_induction}) are following
\beq
\omega' \> \delta \sigma'_{n} =  k'_x \> \delta v'_{n,x} + k'_y \> \delta v'_{n,y} \>\>,
\eeq

\beq
\begin{aligned}
\omega' \> \delta v'_{n,x}  = {} & \frac{k'_x}{k'} \> \left[\tilde{C}_{\rm{eff,0}} ^2 \> k' - 1\right] \> \delta \sigma'_{n} + \> i \> Q \delta v'_{n,y}\\
& + \> \frac{k'_x}{k'} \> \mu_0 ^{-1} \left[1  + \tilde{Z}_0 k' \right] \>\delta B'_{z,\rm{eq}} \>\>,
\end{aligned}
\eeq

\beq
\begin{aligned}
\omega' \> \delta v'_{n,y} = {} & \frac{k'_y}{k'} \> \left[\tilde{C}_{\rm{eff,0}} ^2 \> k' - 1 \right] \> \delta \sigma'_{n} - \> i \> Q \delta v'_{n,x} \\
& + \> \frac{k'_y}{k'} \>  \mu_0 ^{-1} \left[1 \> + \tilde{Z}_0  k' \right] \>\delta B'_{z, \rm{eq}} \>\>,
\end{aligned}
\eeq

\beq
\begin{aligned}
\omega' \> \delta B'_{z, \rm{eq}}  = \frac{k'_x}{\mu_0} \> \delta v'_{n,x}  +  \frac{k'_y}{\mu_0} \> \delta v'_{n,y} \> - i \left[ \tilde{\gamma} + \tilde{\theta} \right] \> \delta B'_{z,\rm{eq}} \>\> .
\end{aligned}
\eeq
So, the normalized form of the dispersion relation is
\begin{equation}
\begin{aligned}
    \left(\omega' + i[\Tilde{\theta}+\Tilde{\gamma}] \right) \big(\omega'^2  {} & - \tilde{C}^2 _{\rm{eff,0}} k'^2 + k' -Q^2 \big)\\
                                                                                &= \omega' \> \left(k' \mu_0 ^{-2} + \tilde{Z}_0 k'^2 \mu_0 ^{-2}\right) \;,
\end{aligned}
\label{eq:normdr}
\end{equation}
where 
$k' = k L_0$, $\omega' = \omega t_0$, and
\begin{equation}
    \Tilde{\gamma} = \gamma t_0 = \etaODt \> k'^2,
\end{equation}
\begin{equation}
    \Tilde{\theta} = \theta t_0 = \Tilde{\tau}_{ni, 0} \> \mu_0 ^{-2} \> \left(k' + \tilde{Z}_0 k'^2 \right) = \etaADt \frac{\left(k' + \tilde{Z}_0 k'^2 \right)}{\tilde{Z}_0}.
\end{equation}
In the limit of flux-freezing ($\tilde{\tau}_{ni,0} \xrightarrow \> 0$, $\etaODt \xrightarrow \> 0$), 
\begin{equation}
    \omega'^2 - \tilde{C}^2 _{\rm{eff,0}} k'^2  + k' -Q^2 =  k' \mu_0 ^{-2}  + \tilde{Z}_0 k'^2 \mu_0 ^{-2} .
\label{eq:ffnorm}
\end{equation}
In the limit of OD only ($\tilde{\tau}_{ni,0} \xrightarrow \> 0$),
\begin{equation}
\begin{aligned}
    \left(\omega' + i\Tilde{\gamma} \right) \> \big(\omega'^2 {} & - \tilde{C}^2 _{\rm{eff,0}} k'^2 + k'-Q^2 \big)\\
                                                 & = \omega' \> \left(k' \mu_0 ^{-2} + \tilde{Z}_0 k'^2 \mu_0 ^{-2} \right) \;.
\end{aligned}
\end{equation}
In the limit of AD only ($\etaODt \xrightarrow \> 0$),
\begin{equation}
\begin{aligned}
    \left(\omega' + i\Tilde{\theta}\right) \> \big(\omega'^2 {} & - \tilde{C}^2 _{\rm{eff,0}} k'^2 + k' -Q^2 \big)\\
                                                 & = \omega' \> \left(k' \mu_0 ^{-2} + \tilde{Z}_0 k'^2 \mu_0 ^{-2}\right) \;.
\end{aligned}
\end{equation}
Here, we represent the effect of rotation in terms of the Toomre parameter
\begin{equation}
    Q \equiv \frac{c_s \Omega}{\pi G \sigma_{n,0}} 
\end{equation}
\citep{toomre1964}. 

Similarly, under flux-freezing the normalized form of the growth time of the gravitationally mode can be written as
\begin{equation}
    \tau'_{g} = \frac{\lambda'}{\left[2\pi \left(1-\mu_0 ^{-2} \right) \left(\lambda'- \lambda'_{\rm{MS}}\right) - \>\> Q^2 \lambda'^2 \right]^{1/2}} \;,
\label{eq:normtaug}
\end{equation}
for $\lambda' \geq \lambda'_{MS}$ and for $\mu_0 >1$, 
\begin{equation}
    \lambda'_{\rm{MS}} = 2\pi \frac{\left(\tilde{C}^2 _{\rm{eff,0}} + \tilde{Z}_0 \mu_0 ^{-2}\right)}{\left(1-\mu_0 ^{-2} \right)} \;.
\label{eq:normlambdag}    
\end{equation}
The minimum growth time for the unstable mode occurs at $\lambda'_{\rm{MS,m}} = 2\lambda'_{\rm{MS}}$. Note that, $k'_{\rm{MS,m}} = k'_{\rm{MS}}/2$.
The dimensionless thermal growth time ($\tau'_{g,T}$) is
\begin{equation}
    \tau'_{g,T} = \frac{\lambda'}{\left[2\pi (\lambda'- \lambda'_{\rm T}) - \>\> Q^2 \lambda'^2 \right]^{\frac{1}{2}}} \;,
\label{eq:normtauthermal}
\end{equation}
where the dimensionless critical thermal length scale is defined as
\begin{equation}
\lambda'_{\rm T} = 2\pi \tilde{C}^2 _{\rm{eff,0}} = \pi\frac{\left(1+3\tilde{P}_{\rm{ext}}\right)}{\left(1+\tilde{P}_{\rm{ext}}\right)} \tilde{Z}_0 \> .
\label{eq:normlambdathermal}  
\end{equation}
Interestingly, we notice that in the flux-frozen case the normalized shortest growth timescale is different for each different normalized rotation rate $Q$, as obtained in \autoref{eq:normtaug} and \ref{eq:normtauthermal}. However, we find that the corresponding normalized preferred length scale remains the same irrespective of any rotation as shown in \autoref{eq:normlambdag} and \ref{eq:normlambdathermal} (see also \autoref{fig:ODandQ} for relevant discussion).  
When $\mu_0 \rightarrow \infty$ and $Q=0$, the minimum growth time 
for the unstable mode occurs at $\lambda'_{\rm{T,m}} = 2\lambda'_{\rm{T}}$ 
and it yields a growth time
\begin{equation}
     \tau'_{g,T,m} = \left({\frac{2\lambda'_{\rm T}}{\pi}}\right)^{1/2} = 2\> \tilde{C} _{\rm{eff,0}}\, , 
\end{equation}
which is the same as the dimensionless dynamical (free-fall) timescale ($t'_{\rm d} = t_{\rm d}/t_0 = \zt$, or $t_{\rm d} = Z_0/c_s $) when $\tilde{P}_{\rm{ext}}=0$.
For the highly supercritical regime ($\mu_0 \gg 1$), under the asymptotic limit ($\lambda' \gg \lambda'_{\rm{T}}$), 
\begin{equation}
    \tau'_{g} \ \rightarrow \ \left(\frac{\lambda'}{2\pi}\right)^{1/2},
\label{eq:normtaug_asymplim}    
\end{equation}
as obtained from \autoref{eq:normtaug} for the case of no rotation ($Q=0$). 
This behavior is seen in \autoref{fig:fluxfrozen}(a) and \autoref{fig:fig1AD} for very large length scales and $\mu_0 > 1$.
These results show that in the limit $Q = 0$, the isothermal sheet has a thermal length scale ($\lambda'_{\rm{T}} = 2\pi \tilde{C}^2_{\rm eff,0}$, effectively the ``Jeans length'') and a preferred thermal length scale $\lambda'_{\rm{T,m}} = 4\pi \tilde{C}^2_{\rm eff,0}$. 
Similarly, in the limit $\tilde{P}_{\rm{ext}} \rightarrow  0$, it becomes  $\lambda_{\rm{T,m}} = 4 \pi \; L_0 = 2\pi Z_0 $) and thermal (Jeans) timescale $\tau'_{\rm{T,m}} = 2$ (similarly, $\tau_{\rm{T,m}} = 2L_0 / c_s = Z_0 /c_s)$. 
We use $\lambda'$ as an independent variable since the characteristic dispersion relation for our eigensystem is only a function of  $k' \equiv k'_z = (k_x^{'2} + k_y^{'2})^{1/2}$. Under this approximation, all the perturbations are independent of the planar angle of propagation $\alpha$ ($= \tan^{-1}[k'_y /k'_x]$).

\subsection{Generalized Toomre Criterion}\label{sec:toomreQeff}
We derive a generalized Toomre criterion in terms of a generalized rotation parameter ($Q_{\rm{eff}}$) that includes a magnetic dependence (see \autoref{sec:gen_Toomre_criterion_app}). In the limit of flux-freezing, the expression is 
\begin{equation}
    Q_{\rm{eff}}= \Omega \frac{\left({C}^2 _{\rm{eff,0}} +  V_{A,0} ^{2}\right)^{1/2}}{\pi G \sigma_{n,0} \left(1-\mu_0 ^{-2}\right)} = Q \frac{\left(\tilde{C}^2 _{\rm{eff,0}} + \tilde{Z}_0\, \mu_0 ^{-2}\right)^{1/2}}{\left(1-\mu_0 ^{-2}\right)}\;
\label{eq:normQeff} 
\end{equation}
(see \cite{lizano2010b} for a similar expression). The above equation shows that $Q_{\rm{eff}}$ has a direct dependence on the mass-to-flux-ratio ($\mu_0$) as well as on the isothermal magnetosonic speed $\left({C}^2 _{\rm{eff,0}} +  V_{A,0} ^{2}\right)^{1/2}$ (see \autoref{eq:v_MS}).


\begin{figure}[ht!]
\epsscale{1}
\plotone{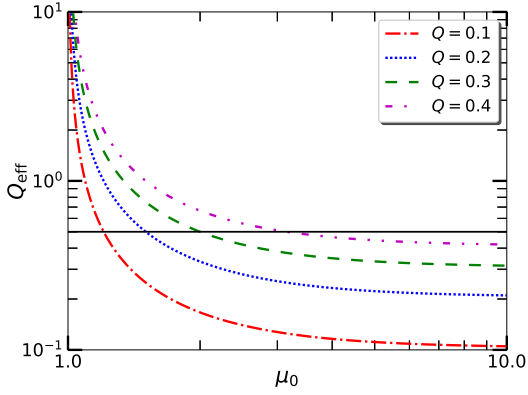}{}
\caption{Normalized generalized rotation parameter $(Q_{\rm eff})$ as a function of $\mu_0$ for different values of $Q =0.1$ (red), 0.2 (blue), 0.3 (green), 0.4 (magenta) with flux-freezing. The black solid line represents the instability cutoff and occurs at $Q_{\rm{eff}} = 1/2$ under flux-freezing.}
\label{fig:Qeff}
\end{figure}

\begin{figure}[hbt!]
\epsscale{1}
\plotone{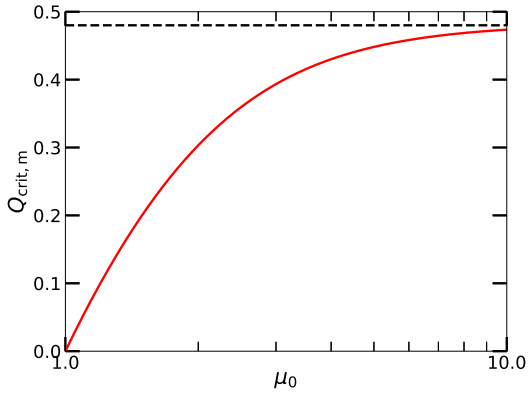}{}
\caption{Modified critical rotation parameter $Q_{\rm crit,m}$ (that has a magnetic dependence) as a function of $\mu_0$. The solid line shows the case for flux-freezing such that instability occurs for $Q < Q_{\rm crit,m}$. The dashed line shows critical value of instability in the hydrodynamic limit (i.e., $Q_{\rm crit,m} \rightarrow 1/(2 \tilde{C}_{\rm{eff}})$; see \autoref{eq:Q_crit_mag}.}
\label{fig:Qcrit}
\end{figure}

In the regime where the normalized mass-to-flux ratio approaches infinity, implying negligible magnetic support, and for no external pressure (i.e., $C_{\rm{eff,0}} = c_s$) one can show that
\begin{equation}
\begin{aligned}
    Q_{\rm{eff}} {} & \rightarrow \frac{c_s \Omega}{\pi G \sigma_{n,0}} =Q .
\end{aligned}    
\end{equation}
%
We evaluate the generalized Toomre instability criterion that yields 
\begin{equation}
    Q_{\rm{eff}} < \frac{1}{2}\, ,
\label{eq:Qlimit}     
\end{equation}
or, equivalently,
\begin{equation}
   Q <  Q_{\rm crit,m} = \frac{1}{2}  \frac{\left(1-\mu_0 ^{-2} \right)}{\left(\tilde{C}^2 _{\rm{eff,0}} + \tilde{Z}_0\, \mu_0 ^{-2} \right)^{1/2}} \, .
\label{eq:Q_crit_mag}    
\end{equation}
See the derivation in \autoref{sec:gen_Toomre_criterion_app}.

\autoref{fig:Qeff} shows the normalized magnetic Toomre $Q$ factor ($Q_{\rm{eff}}$) as a function of normalized mass-to-flux ratio ($\mu_0$) in the flux-freezing limit for four different values $Q=0.1, 0.2, 0.3, 0.4$. The solid line represents the cutoff value of $1/2$ that implies no unstable mode can occur for those values of $\mu_0$ for which $Q_{\rm{eff}} \geq 1/2$, as long as flux-freezing prevails.

One can show (see \autoref{fig:Qcrit} and also \autoref{tab:Qcritical}) that in the hydrodynamic limit with no external pressure ($\tilde{C}_{\rm{eff,0}} = 1$), $Q_{\rm crit,m}$ reduces to the critical Toomre instability limit. 
\autoref{fig:Qcrit} presents the magnetic critical limit of $Q$ (i.e., $Q_{\rm crit,m}$) obtained under the limit of flux-freezing as a function of $\mu_0$.
The dotted line represents the critical boundary in the hydrodynamic limit for a nonzero $\tilde{P}_{\rm ext}$, which is $1/(2 \tilde{C}_{\rm{eff}})$ (see \autoref{eq:Q_crit_mag}).
We also show that the magnetic dependent critical bound ($Q_{\rm crit,m}$) goes back to the hydrodynamic value in the regime $\mu_0 \gg 1$.
This above criteria can easily be acquired from the following dispersion relation 
\begin{equation}
\begin{aligned}
    \omega^2 {} & = C_{\rm eff,0}^2 k^2 - 2\pi G \sigma_{n,0} k + 4\Omega^2, \\
   \rm{or},\hspace{0.2cm}  \omega'^2  & = k'^2 - k' + Q^2 ,
\end{aligned}
\label{eq:toomreDR}    
\end{equation}
which is same as the dispersion relation (\autoref{eq:ffDR} or \autoref{eq:ffnorm}) for an isothermal planar sheet in the hydrodynamic limit.

\begin{figure}
\epsscale{1}
\plotone{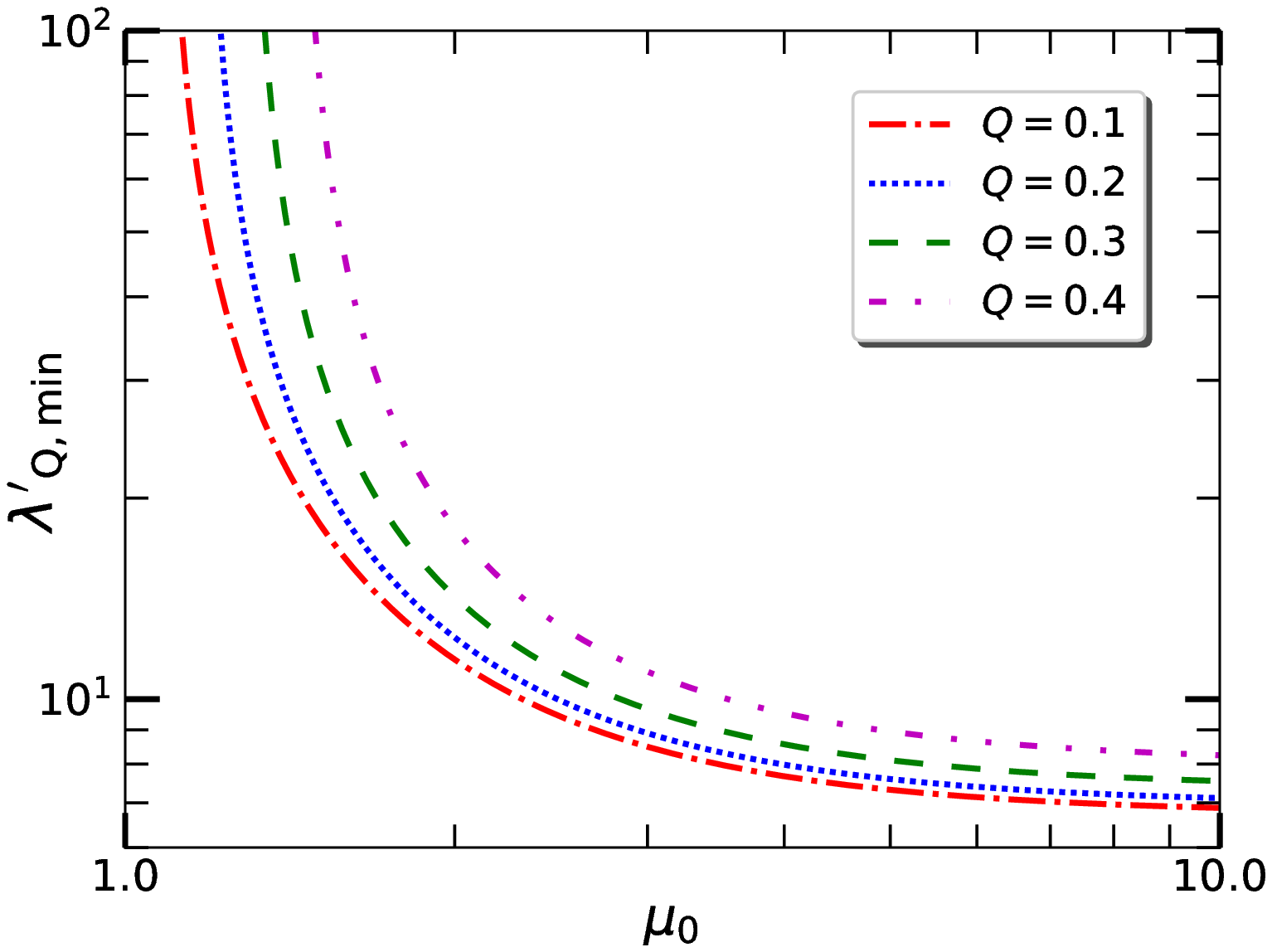}{(a)}
\plotone{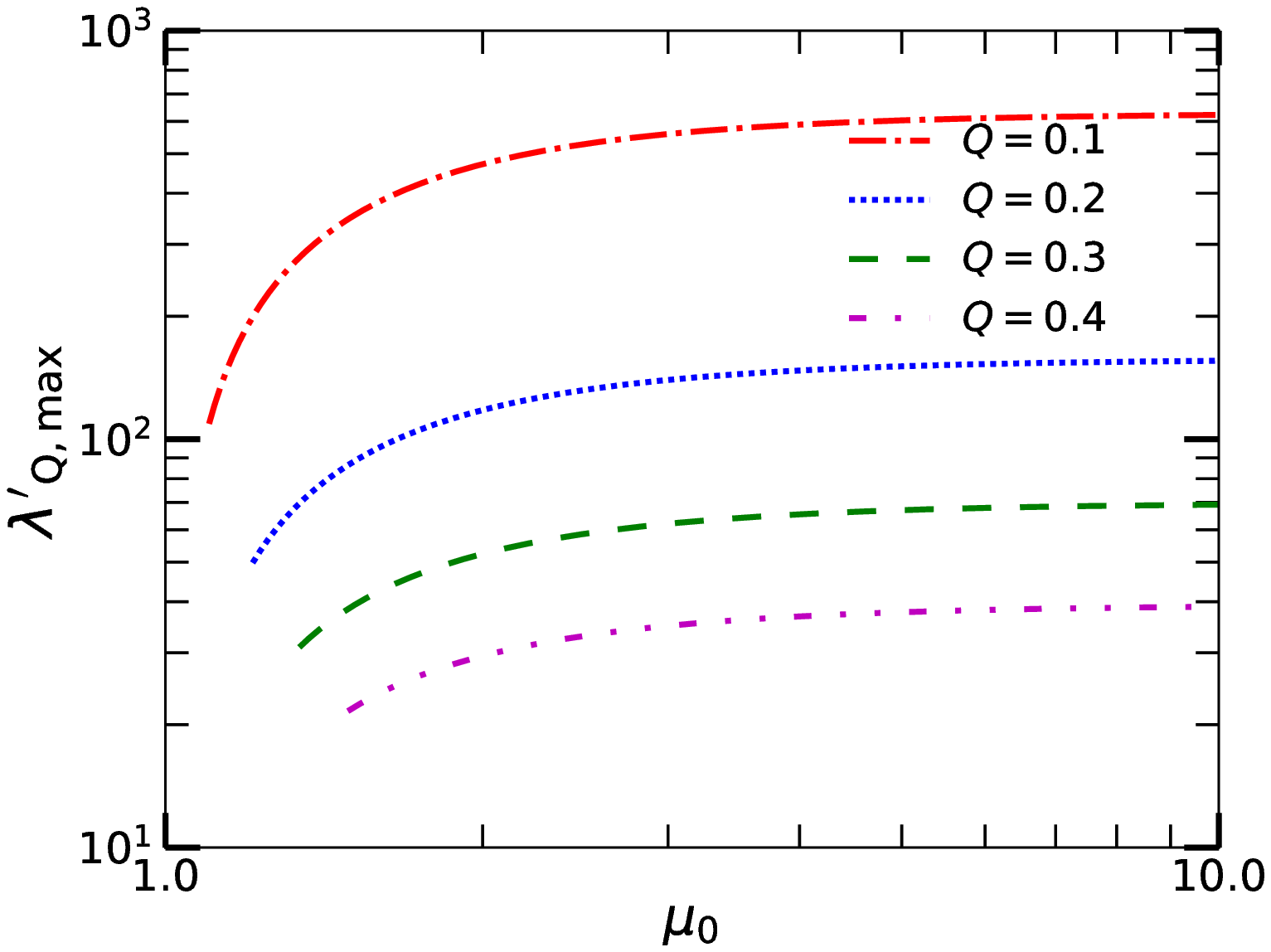}{(b)}
\caption{Normalized maximum and minimum wavelength for rotationally modulated instability as a function of $\mu_0$. 
(a) is for minimum wavelength, $\lambda'_{\rm Q, min}$ (see \autoref{eq:min_lambda}) and (b) is for maximum wavelength, $\lambda'_{\rm{Q, max}}$ (see \autoref{eq:max_lambda}).}
\label{fig:lambda_mag_mu}
\end{figure}

We discuss the effect of rotation on the lower and upper limits of the unstable range of wavelengths. From the dispersion relation under flux-freezing (\autoref{eq:ffnorm}), setting $\omega'^2 =0$ we obtain 
\begin{equation}
    k'_{\pm, {\rm Q}} = k'_{\rm{MS,m}} \bigg[1 \pm \sqrt{1 - 4 Q_{\rm{eff}} ^2} \bigg],
\label{eq:kmag}    
\end{equation}
where 
\begin{equation}
    k'_{\rm{MS,m}} = \frac{2\pi}{\lambda'_{\rm{MS,m}}} = \frac{(1-\mu_0^{-2})}{2 (\cefft^2 + \vAt^2)} \>,
\end{equation}
see \autoref{sec:norm} for a detailed discussion on $\lambda'_{\rm{MS,m}}$. Here, $'+'$ and $'-'$ signs belong to the minimum (maximum, i.e., $\lambda'_{\rm{Q, max}}$) and maximum (minimum, i.e., $\lambda'_{\rm{Q, min}}$) wavenumbers (wavelengths) for rotationally modulated instability, respectively. 
Under the approximation $4 Q_{\rm{eff}}^2 \ll 1$, 
$\lambda'_{\rm Q, min}$ and $\lambda'_{\rm Q, max}$ can be obtained from the above relation. It follows that

\begin{equation}
    \lambda'_{\rm{Q, min}} = \frac{2\pi}{\left(1 - Q_{\rm{eff}}^2 \right)} \; \frac{\left(\tilde{C}^2 _{\rm{eff,0}} + \tilde{Z}_0 \mu_0 ^{-2} \right)}{\left(1-\mu_0 ^{-2}\right)} = \frac{\lambda'_{\rm{MS}}}{\left(1 - Q_{\rm{eff}}^2 \right)} \, ,
\label{eq:min_lambda}
\end{equation}

\begin{equation}
    \lambda'_{\rm{Q, max}} = \frac{2\pi}{Q_{\rm{eff}}^2} \; \frac{\left(\tilde{C}^2 _{\rm{eff,0}} + \tilde{Z}_0 \mu_0 ^{-2}\right)}{\left(1-\mu_0 ^{-2}\right)} = \frac{\lambda'_{\rm{MS}}}{Q_{\rm{eff}}^2}\, .
\label{eq:max_lambda}
\end{equation}
We see that the lower and upper limits of unstable wavelengths gradually increase and decrease for higher rotation, as seen from \autoref{fig:lambda_mag_mu}(a) and (b), respectively. This suggests that rotation stabilizes not only the longer wavelengths but also the smaller wavelengths. Hence, when adding rotation, the total range of unstable length scales is reduced.

In the hydrodynamic limit ($\tilde{B}_{\rm{ref}} \rightarrow 0$ ; ${\mu}_0 \rightarrow \infty$), \autoref{eq:kmag} reduces to


\begin{equation}
    k'_{\pm, {\rm Q}} = \frac{1}{2 \tilde{C}^2 _{\rm{eff,0}}} \bigg[1 \pm \sqrt{1 - 4Q^2 \tilde{C}^2 _{\rm{eff,0}}} \> \bigg].
\label{eq:khydro_pext}    
\end{equation}
Similarly, under the approximation $4 Q^2 \tilde{C}^2 _{\rm{eff,0}} \ll 1$ in the hydrodynamic limit, we find

\begin{equation}
    \lambda'_{\rm{Q, min}} = \frac{2\pi}{\left(1 - Q^2 \tilde{C}^2 _{\rm{eff,0}} \right)} \tilde{C}^2 _{\rm{eff,0}},
\label{eq:min_lambda_hydro_pext}    
\end{equation}

\begin{equation}
    \lambda'_{\rm{Q, max}} = \frac{2\pi}{Q^2},
\label{eq:max_lambda_hydro_pext}    
\end{equation}
which are similar to \autoref{eq:min_lambda} and \autoref{eq:max_lambda} for $\mu_0 \rightarrow \infty$.
In the limit $\tilde{P}_{\rm{ext}} \rightarrow 0$, \autoref{eq:khydro_pext} becomes
\begin{equation}
    k'_{\pm, {\rm Q}}  = \frac{1}{2} \bigg[1 \pm \sqrt{1 - 4Q^2} \bigg],
\end{equation}
which can be directly obtained from \autoref{eq:toomreDR}.
Now, under the approximation $4Q^2 \ll 1$, we get 
\begin{equation}
    \lambda'_{\rm Q, min} = \frac{2\pi}{1 - Q^2},
\end{equation}  

\begin{equation}
    \lambda'_{\rm Q, max} = \frac{2\pi}{Q^2},
\end{equation}
which are counterparts to 
\autoref{eq:min_lambda_hydro_pext} and \autoref{eq:max_lambda_hydro_pext} in the limit $\tilde{P}_{\rm{ext}} \rightarrow 0$ (i.e., $\tilde{C}_{\rm{eff},0} =1$).

\subsection{{Stationary Magnetic Field Limit}} \label{sec:statBlimit}

In the limit of stationary magnetic field, $\omega \> \delta B'_{z,\rm{eq}} \, \rightarrow 0$, we discuss the respective cases of Ohmic dissipation (OD) and ambipolar diffusion (AD). 
Under the stationary magnetic field limit, we obtain the normalized dispersion relation for the case of only OD ($\taunit=0$, \ $Q=0$) is
\begin{equation}
    \omega'^2 + \omega' \> \frac{i}{\etaODt} \frac{\left(1+ \tilde{Z}_0 k' \right)}{k' \mu_0^2} - \left(\tilde{C}^2_{\rm{eff,0}} k'^2 - k' \right) = 0 \;\;,
\label{eq:OD_DR_statB}     
\end{equation}
which yields a growth timescale of OD
\begin{equation}
     \tau'_{g,\rm{OD}} = \frac{2 \etaODt \lambda'}{\left[ \left(\frac{\lambda' \left(\lambda'+2\pi \tilde{Z}_0 \right)}{2\pi \mu_0 ^2} \right)^2 + 8\pi  \etaODt ^2 \left(\lambda' - \lambda'_{\rm T} \right)\right]^{\frac{1}{2}} - \left( \frac{\lambda' \left(\lambda'+2\pi \tilde{Z}_0 \right)}{2\pi \mu_0 ^2} \right)}  \;.
\label{eq:OD_taug_statB}         
\end{equation}
See \autoref{sec:statBapp} for a derivation of \autoref{eq:OD_DR_statB}.
Minimizing $\tau'_{g,\rm{OD}}$ of \autoref{eq:OD_taug_statB} with respect to $\lambda'$ yields 
\begin{equation}
    \lambda'_{\rm preferred,OD} =  \lambda'_{\rm T}.
\label{eq:lambdagmpref_OD}
\end{equation}
Furthermore, one obtains
\begin{equation}
    \tau'_{g,\rm{OD}} \rightarrow \infty \hspace{0.4cm} {\rm at \>\> \lambda'_{\rm preferred,OD} = \lambda'_{\rm T}},
\label{eq:taugmpref_OD}    
\end{equation}
and this feature is illustrated later in \autoref{fig:ADonly_ODonly}(a) and (b). 
The remnant thermal pressure makes the timescale of the contraction driven by OD to be infinitely long in the regime $\mu_0 \ll 1$.

Under a similar approximation in the regime of only AD ($\etaODt=0$, \ $Q=0$), the resulting normalized dispersion relation is
\begin{equation}
    \omega'^2 + \omega' \; \frac{i}{\tilde{\tau}_{ni,0}} - \left(\tilde{C}^2_{\rm{eff,0}} k'^2 - k' \right) = 0, 
\label{eq:AD_DR_statB}    
\end{equation}
\citep[see also][]{ciolek06}. From the above relation of AD, one finds that an unstable mode exists for $\lambda' > \lambda'_{\rm T}$, and has a growth timescale of AD
\begin{equation}
    \tau'_{g,\rm{AD}} = \frac{2 \tilde{\tau}_{ni,0} \lambda'}{\left[\lambda'^2 + 8\pi  \tilde{\tau}_{ni,0} ^2 (\lambda' - \lambda'_{\rm T})\right]^{\frac{1}{2}} -\lambda'} \;.
\label{eq:AD_taug_statB}     
\end{equation}
See \autoref{sec:statBapp} for a derivation of \autoref{eq:AD_DR_statB}.
We further carried out the following calculation by minimizing $\tau'_{g,\rm{AD}}$ from \autoref{eq:AD_taug_statB} with respect to $\lambda'$ which yields 
\begin{equation}
    \lambda'_{\rm preferred,AD} = 2 \lambda'_{\rm T}\; .
\label{eq:lambdagmpref_AD}
\end{equation}
Furthermore, we obtain
\begin{equation}
    \tau'_{g,\rm{AD}} = \frac{4 \tilde{\tau}_{ni,0} \lambda'_{\rm T}}{\left[4\lambda^{'2}_{\rm T} + 8\pi  \tilde{\tau}_{ni,0} ^2 \lambda'_{\rm T} \right]^{\frac{1}{2}} -2 \lambda'_{\rm T}}  \hspace{0.4cm} {\rm at \>\> \lambda'_{\rm preferred,AD} = 2 \lambda'_{\rm T}},
\label{eq:taugmpref_AD}    
\end{equation}
and this value is independent of $\mu_0$, which is illustrated later in \autoref{fig:ADonly_ODonly}(c) and (d).

In the asymptotic limit ($\lambda' \gg \lambda'_{\rm T}$) from \autoref{eq:OD_taug_statB} and \autoref{eq:AD_taug_statB} we deduce  
\begin{equation}
\begin{aligned}
    \tau'_{g,\rm{OD}} & = \frac{2\etaODt}{\frac{\left(\lambda'+2\pi \tilde{Z}_0 \right)}{2\pi \mu_0 ^2} \left[ \left\{ 1 + \frac{8\pi  \etaODt ^2  \left(\lambda' - \lambda'_{\rm T} \right)}{\left(\frac{\lambda'(\lambda'+2\pi \tilde{Z}_0)}{2\pi \mu_0 ^2}\right)^2} \right\}^{\frac{1}{2}} - 1 \right]}  \rightarrow \frac{\lambda^{'2}}{4\pi^2 \mu_0^2 \etaODt},
\end{aligned}
\label{eq:OD_taug_statB_asymplim}
\end{equation}

\begin{equation}
    \tau'_{g,\rm{AD}} = \frac{2\taunit}{\left[ \left\{1 +  \frac{8\pi  \tilde{\tau}_{ni,0} ^2 (\lambda' - \lambda'_{\rm T})}{\lambda^{'2}} \right \}^{\frac{1}{2}} -1 \right]} \rightarrow \frac{\lambda'}{2\pi \taunit},
\label{eq:AD_taug_statB_asymplim}
\end{equation}
respectively, (using $(1+x)^{1/2} \approx 1+ x/2$ for $x\ll1$). This behavior of the growth timescales is seen in \autoref{fig:fig1AD} for very large length scales and $\mu_0<1$. In this figure we see the slope of the curves for OD are steeper than the case for AD.
Also we see that as $\etaODt \rightarrow \infty$, $\tau'_{g,\rm{OD}} \rightarrow \lambda'/\left[2\pi \left(\lambda'-\lambda'_{\rm T} \right)\right]^{1/2}$, 
and when $\tilde{\tau}_{ni,0} \rightarrow \infty$, $\tau'_{g,\rm{AD}} \rightarrow \lambda'/\left[2\pi \left(\lambda'-\lambda'_{\rm T} \right)\right]^{1/2}$.
This is identical to \autoref{eq:normtauthermal} when $\mu_0 \rightarrow \infty$, i.e., when $\tilde{B}_{\rm{ref}} \rightarrow 0$ and $Q=0$. 
In the regime $\mu_0 \ll 1$, the minimum growth time for OD and AD occur at the preferred wavelength $\lambda'_{\rm T}$ and $2\lambda'_{\rm T}$, respectively as defined above.
The limit $\etaODt \rightarrow \infty$ corresponds to an extremely high rate of collisions encountered by the charged particles such that they become decoupled from the magnetic field. 
On the other hand, $\tilde{\tau}_{ni,0} \rightarrow \infty$ corresponds to the case when there is no collisional coupling between the neutrals and the ions (and hence with the magnetic field). The ions are completely ‘‘invisible’’ to the neutrals in this situation, and there is no transmission of magnetic force to neutrals via neutral-ion collisions. 

\section{Numerical Results} \label{sec:results}
In subsequent sections we generate figures based on solutions of the normalized dispersion relation as described in (\autoref{sec:norm}). Henceforth the normalized wavelength $\lambda'$ is attributed as $\lambda'_g$ which means $ 2 \pi /k'_g$; ``g'' corresponds to the ``growth mode''.

\subsection{Flux-frozen Model} \label{sec:ffmodel}

\begin{figure*}[ht!]
\gridline{\fig{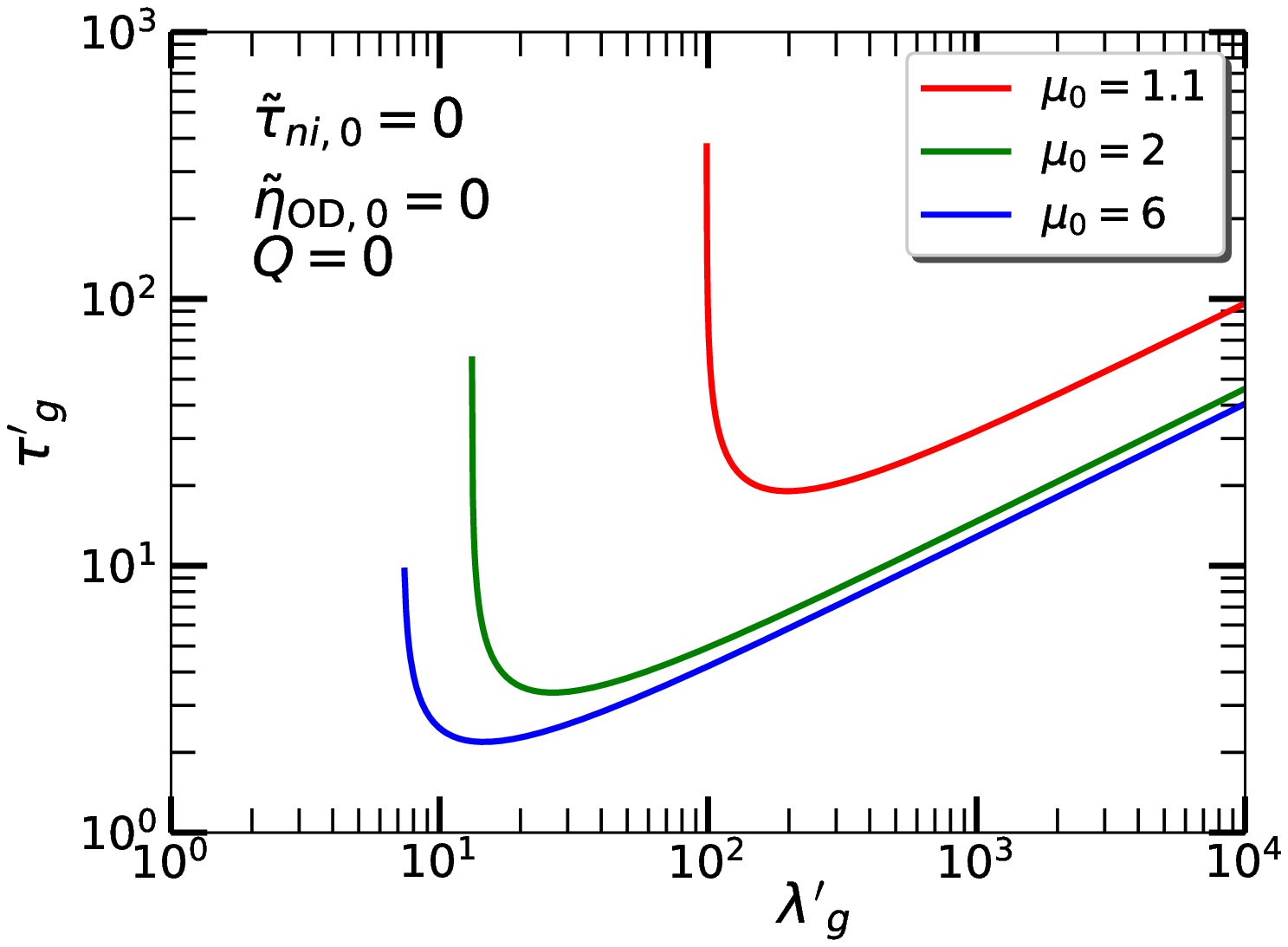}{0.45\textwidth}{(a)}
          \fig{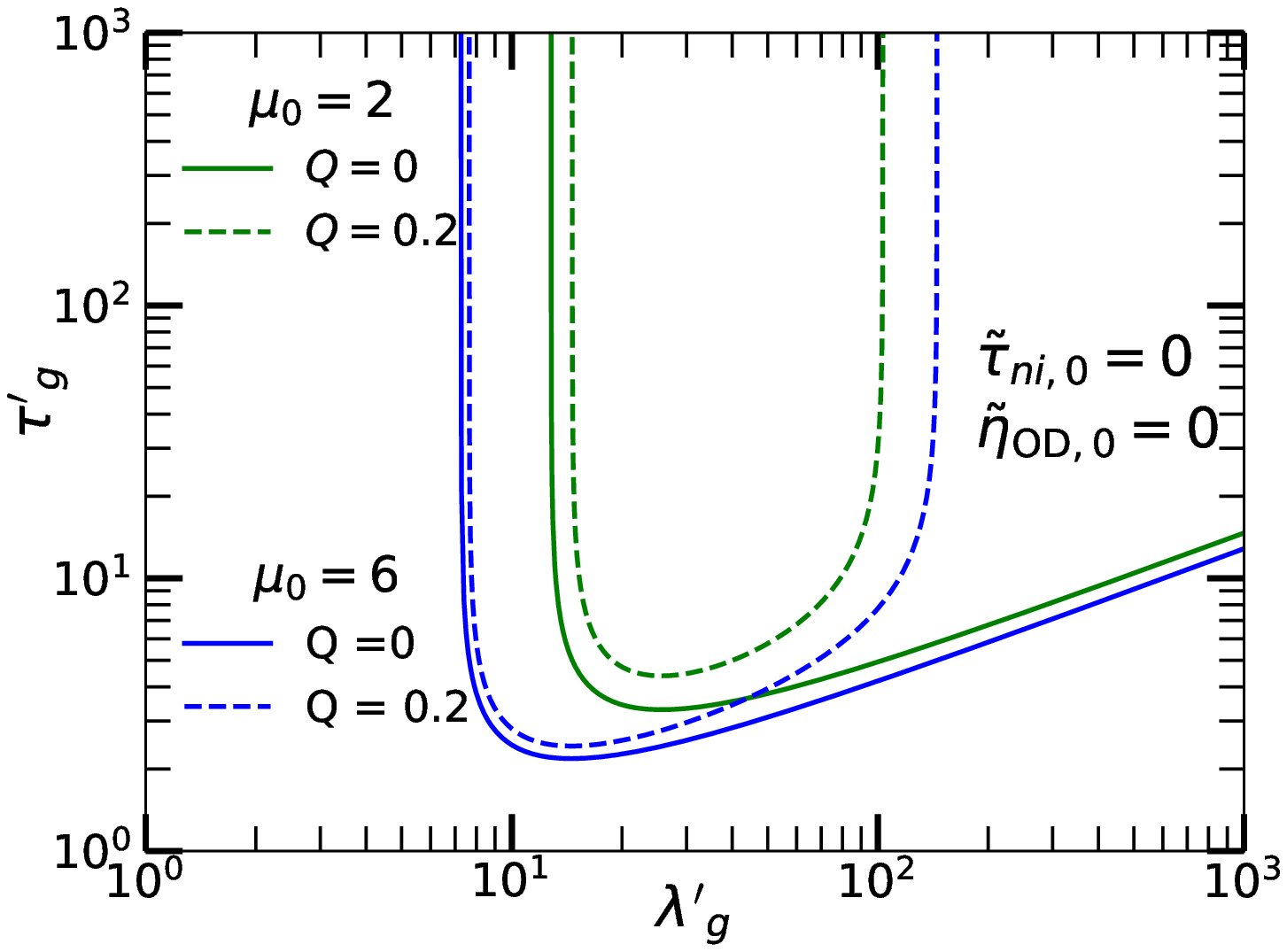}{0.45\textwidth}{(b)}
            }
\caption{Normalized growth time of the gravitationally unstable mode ($\tau'_{g} = \tau_{g}/t_0$) as a function of normalized wavelength ($\lambda'_{g}= \lambda_{g}/L_0$) for flux-frozen models ($\etaODt = 0$, $\tilde{\tau}_{ni,0}=0$) with different values of the normalized mass-to-flux-ratio.
The left panel (a) shows the case with $Q=0$ for fixed $\mu_0$ = 1.1 (red), 2 (green), and 6 (blue). The right panel (b) shows the cases with $Q=0$ and $Q=0.2$ for $\mu_0$ = 2 (green) and 6 (blue).}
\label{fig:fluxfrozen}
\end{figure*}

\autoref{fig:fluxfrozen} shows the instability growth time $\tau'_g$ ($=\tau_g/t_0$) as a function of the wavelength $\lambda'_g$ ($=\lambda_g/L_0$) for flux-frozen cases.  \autoref{fig:fluxfrozen}(a) shows the case without rotation (as obtained by \cite{ciolek06}), whereas \autoref{fig:fluxfrozen}(b) shows the growth time for supercritical clouds with $\mu_0 =2$ and $6$, 
including rotation (in terms of the $Q$ parameter) as obtained from our model (see \autoref{eq:ffnorm}). Here, $Q=0$ lines serve as a reference point. In the limit of large length scale, $\tau_g$ varies as $\lambda_g^{'1/2}$, which can be seen from \autoref{eq:normtaug_asymplim}. We notice that adding a small rotation ($Q=0.2$) causes the gravitational collapse timescale to be comparatively longer than the case without rotation. Instability occurs for those length scales that are not stabilized by the thermal, magnetic, and rotational support. We see that rotation plays a significant role to stabilize the longer wavelengths. Furthermore, along with thermal pressure, rotation also helps to stabilize the smaller length scales, as discussed earlier in \autoref{sec:toomreQeff} (see also \autoref{fig:lambda_mag_mu}). Hence, the range (or span) of unstable wavelengths has been reduced from both the left hand side (shorter end of the length scales) and the right hand side (longer side of the length scales). Later, in \autoref{sec:dis}, we discuss the effect of the magnetic field in creating a modified value of the critical rotation parameter.

\subsection{Theoretical Models with Nonideal MHD}{\label{sec:nonideal_theory_results}}

\begin{figure*}[ht!]
\gridline{\fig{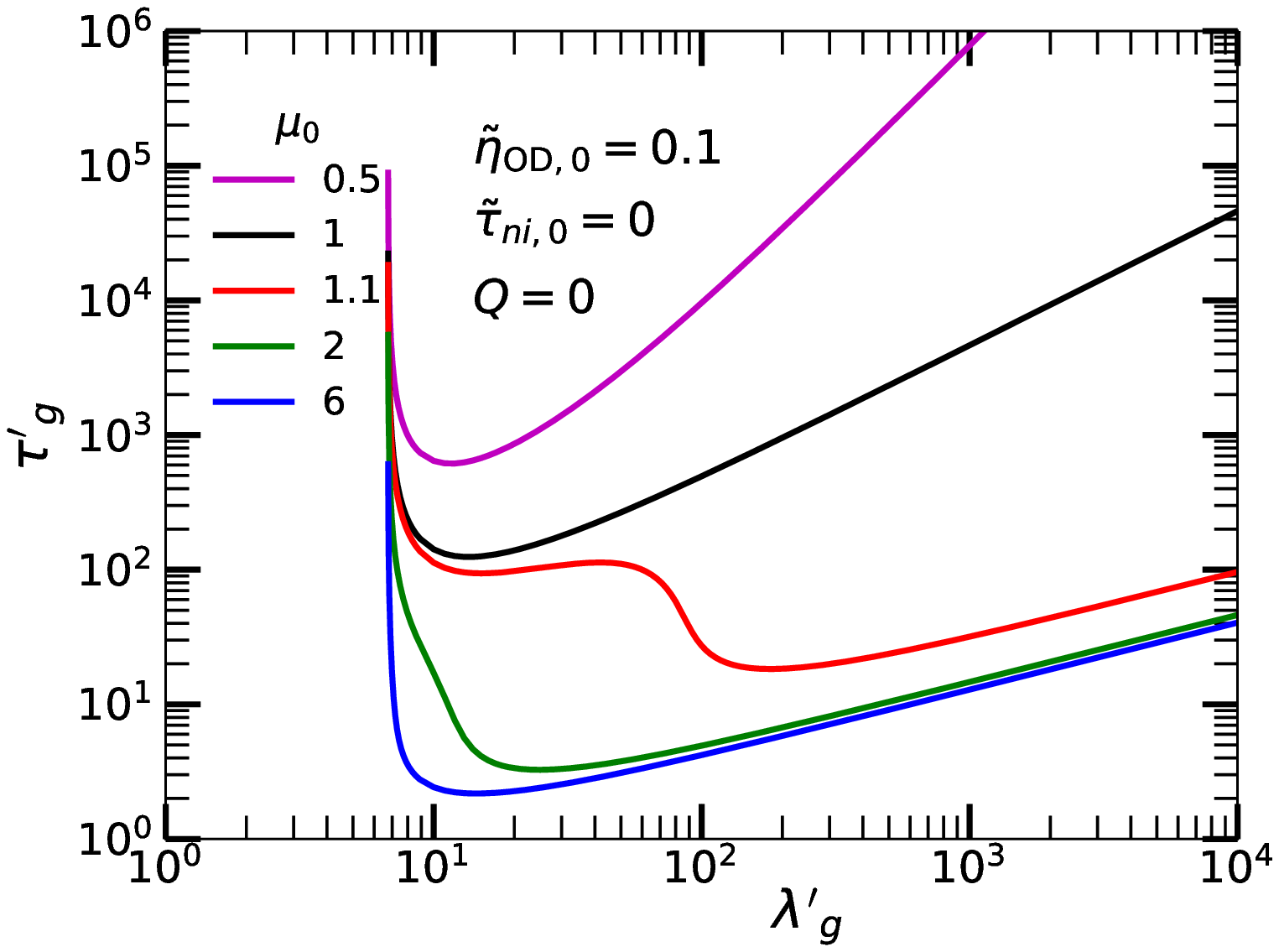}{0.33\textwidth}{(a)}
          \fig{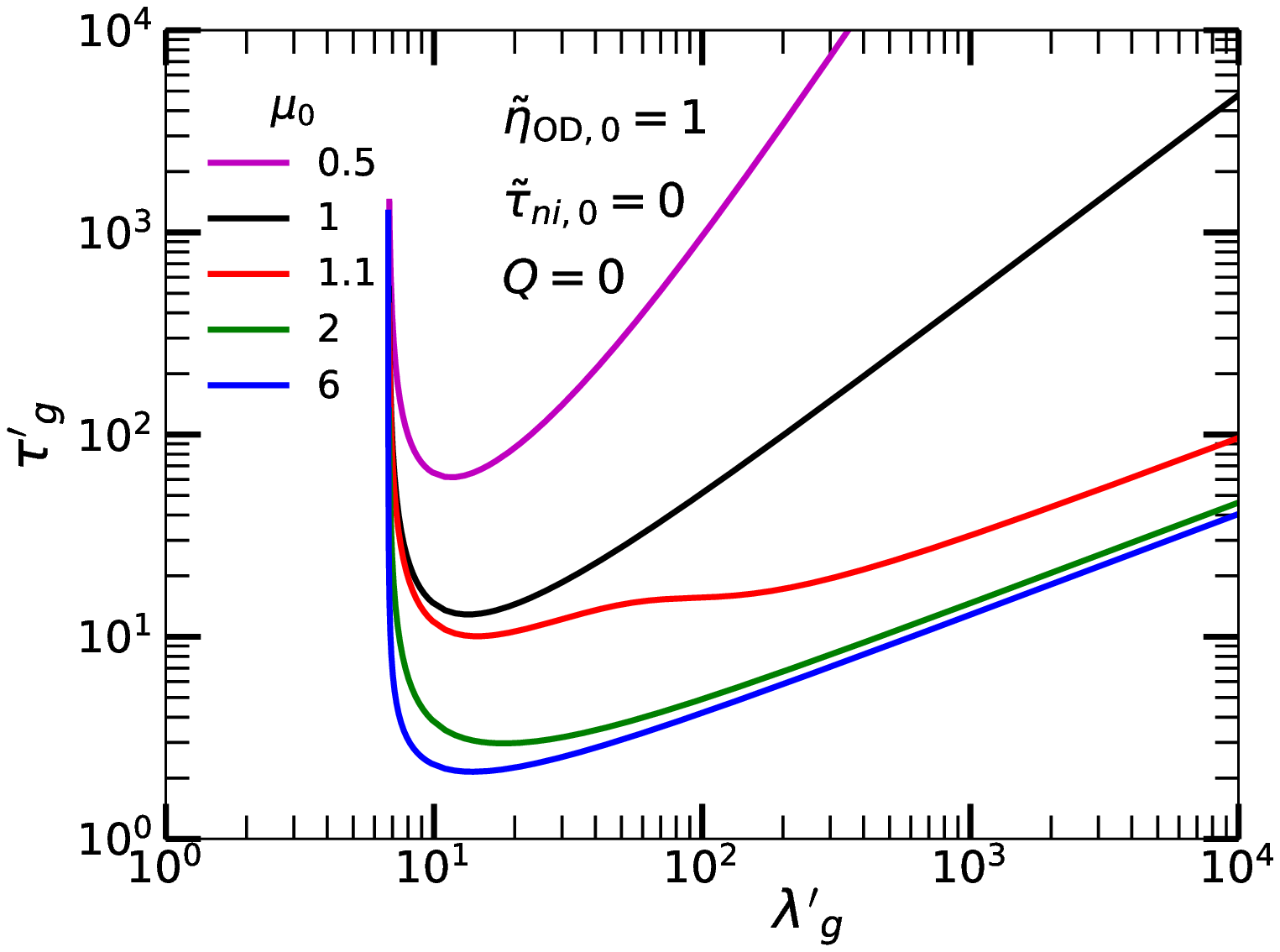}{0.33\textwidth}{(b)}
          \fig{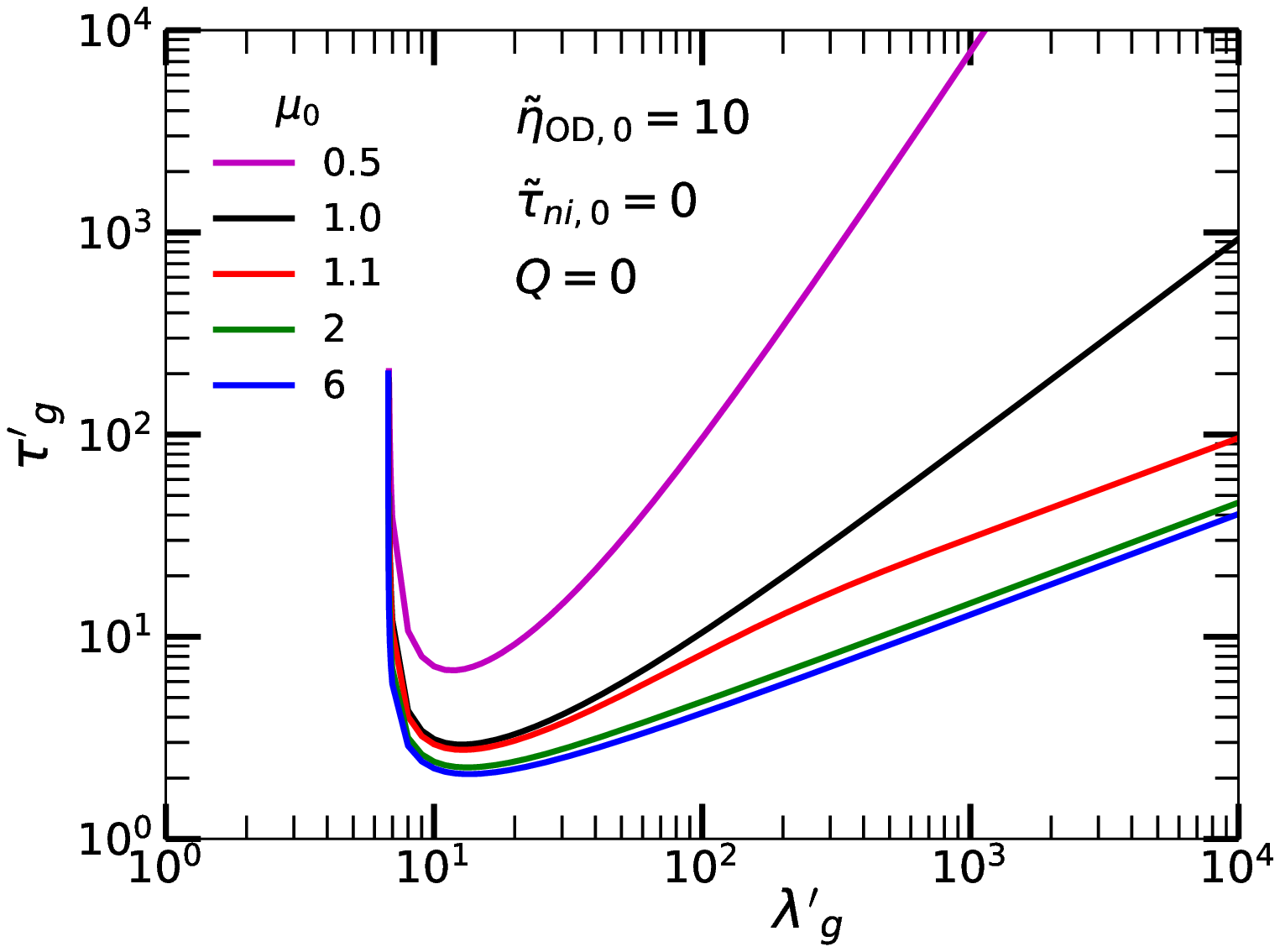}{0.33\textwidth}{(c)}
          }
\gridline{\fig{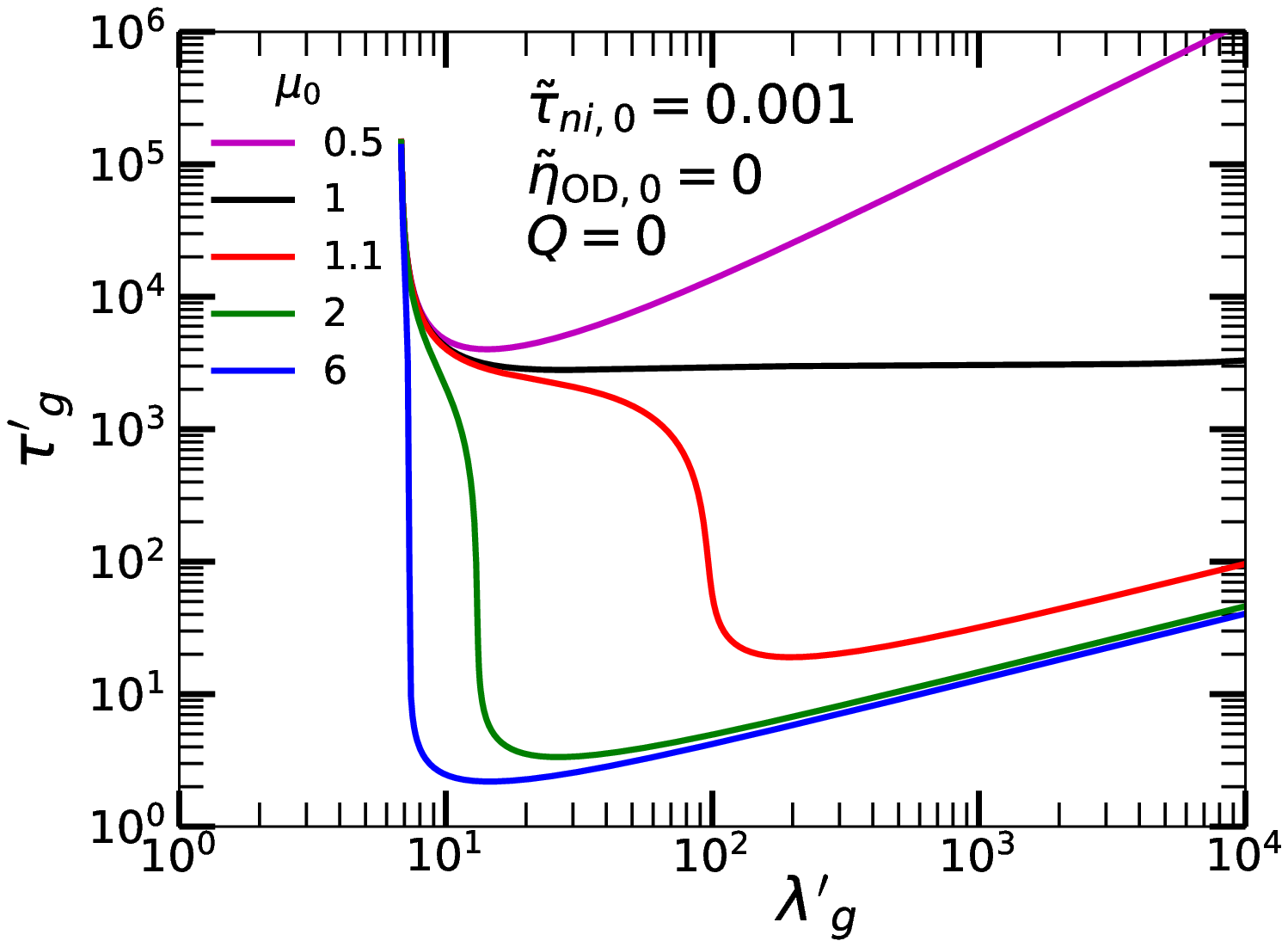}{0.33\textwidth}{(d)}
          \fig{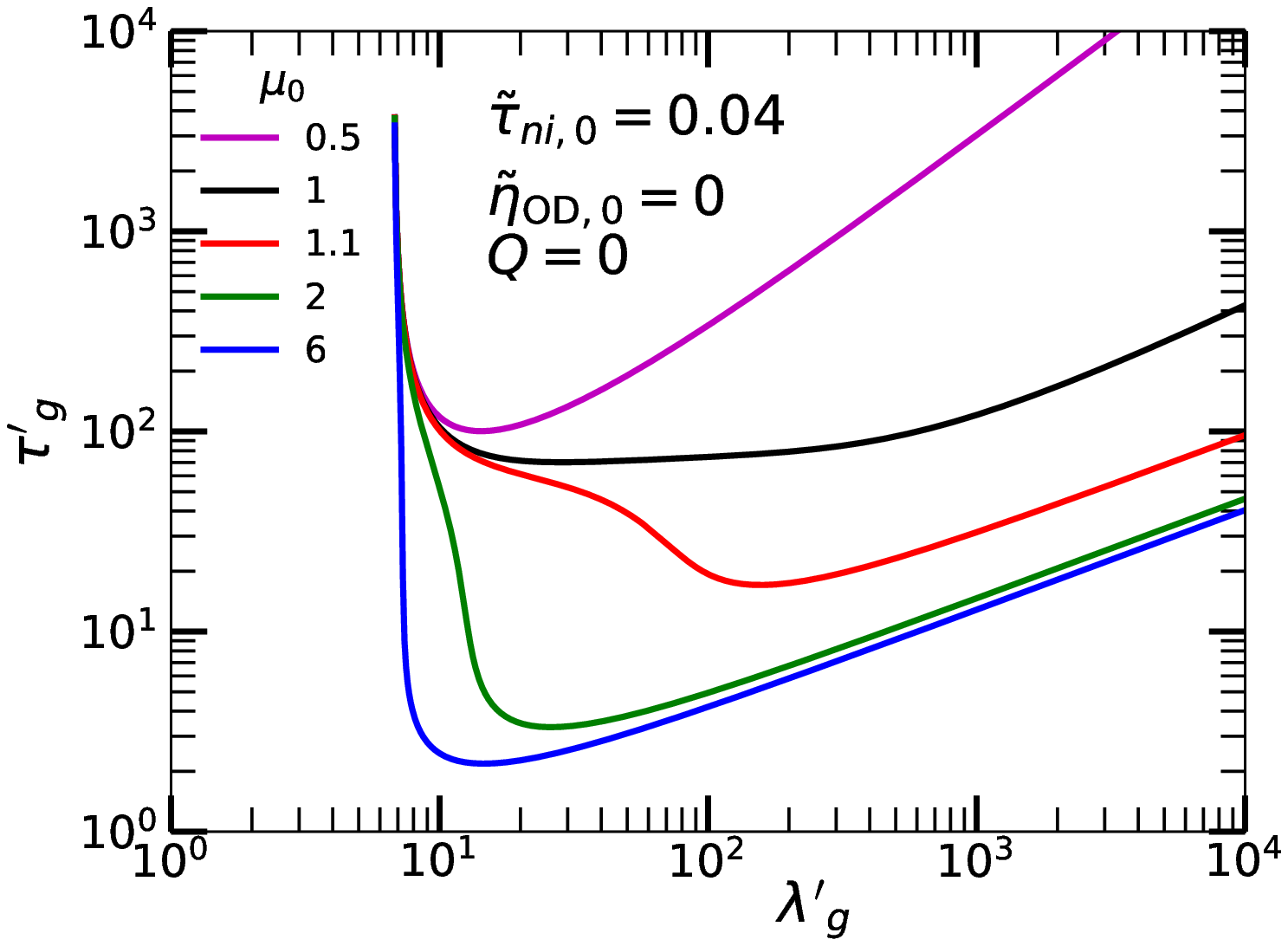}{0.33\textwidth}{(e)}
          \fig{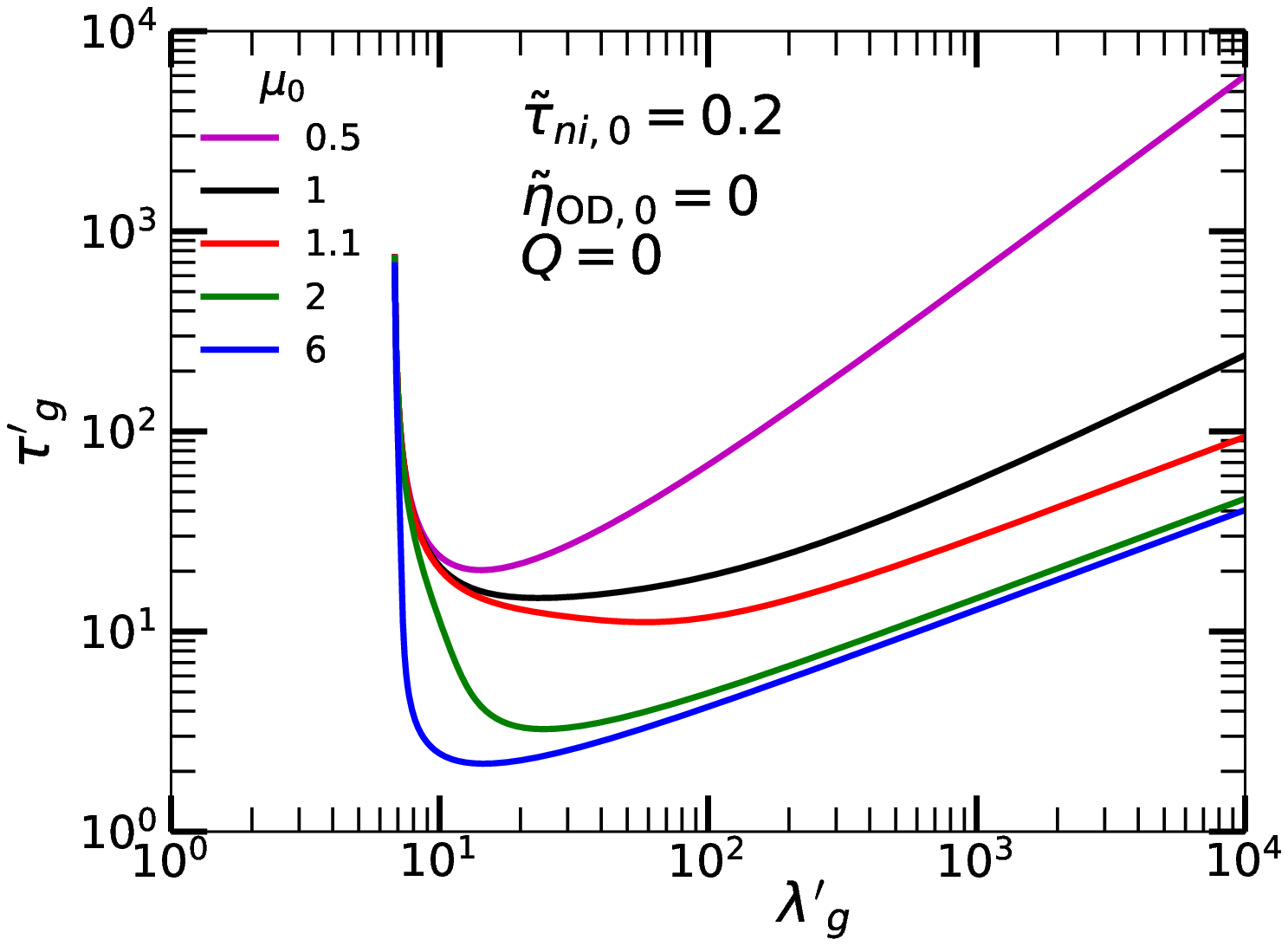}{0.33\textwidth}{(f)}
          }
\caption{Normalized growth time of the gravitationally unstable mode ($\tau'_{g} = \tau_{g}/t_0$) as a function of normalized wavelength ($\lambda'_{g}= \lambda_{g}/L_0$). Each panel shows a model with a fixed $\mu_0$ = 0.5 (magenta), 1 (black), 1.1 (red), 2 (green), 6 (blue). 
Figures in the upper panel (a),(b),(c) show the cases for normalized Ohmic diffusivity $\etaODt=0.1$, $1$, and $10$, respectively.
Figures in the lower panel (d),(e),(f) show the cases for neutral-ion collision time $\tilde{\tau}_{ni,0}=0.001$, $0.04$, and $0.2$, respectively. }
\label{fig:fig1AD}
\end{figure*}



\begin{figure*}[ht!]
\gridline{\fig{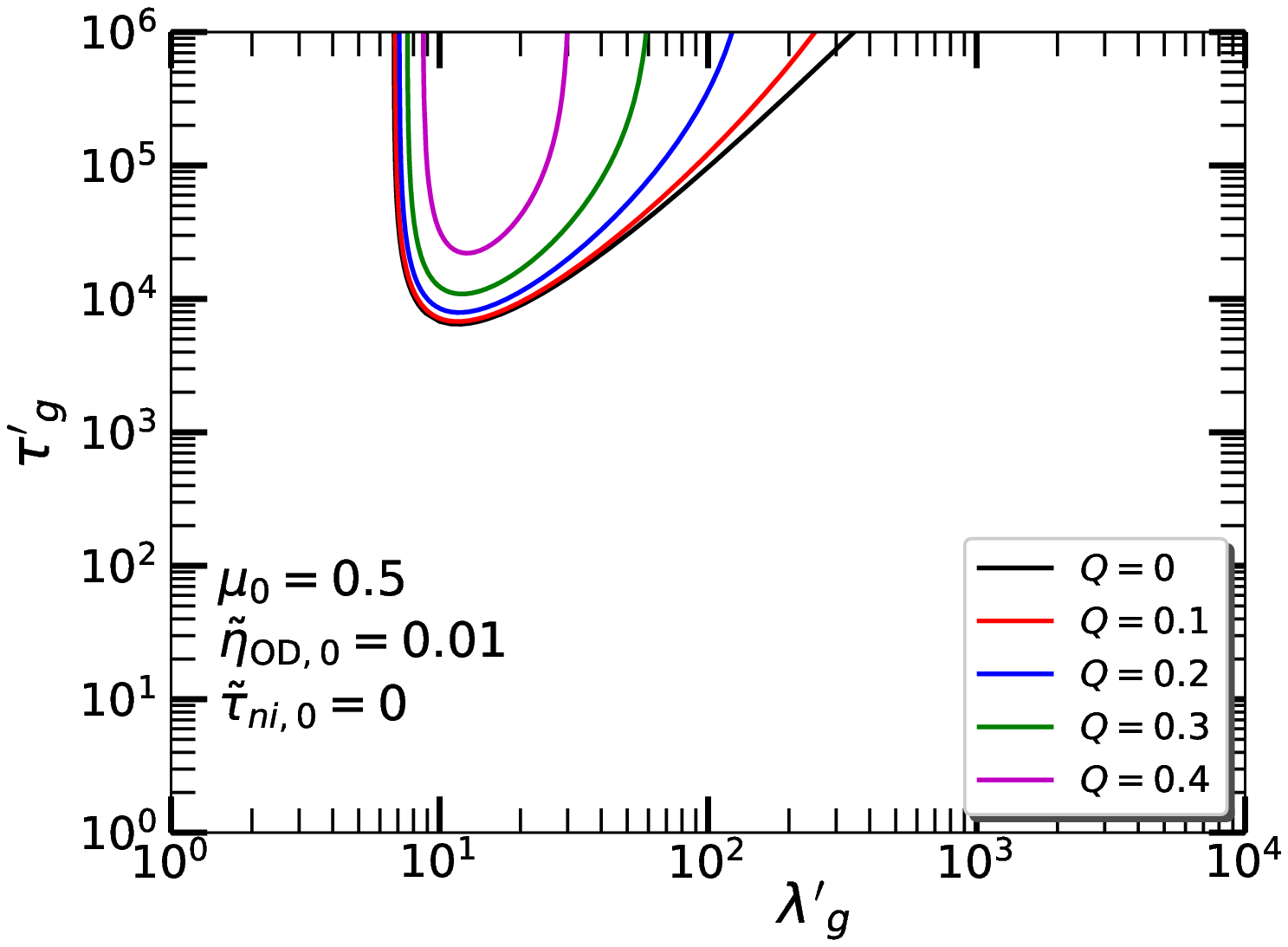}{0.245\textwidth}{(a)}
          \fig{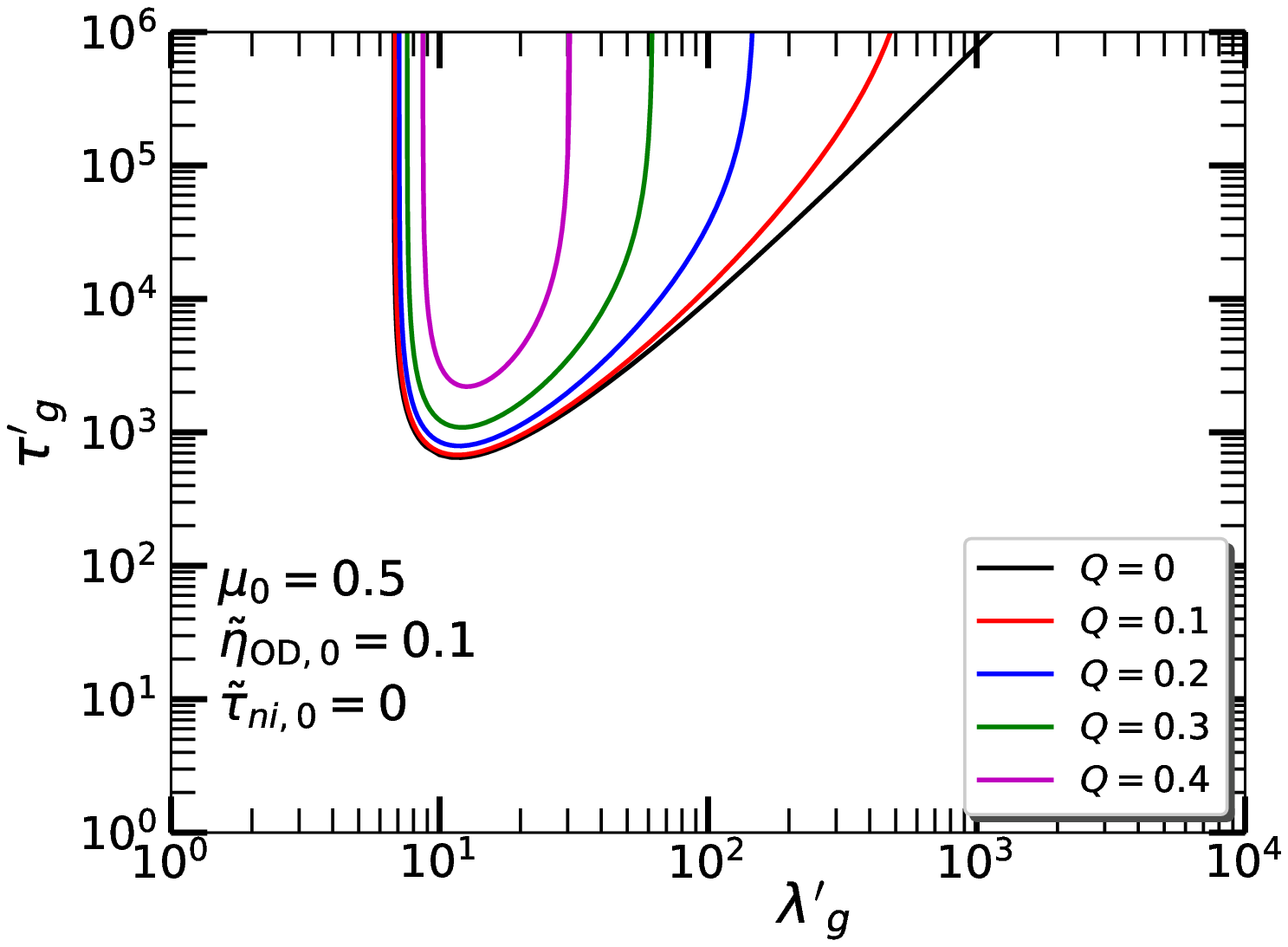}{0.245\textwidth}{(b)}
          \fig{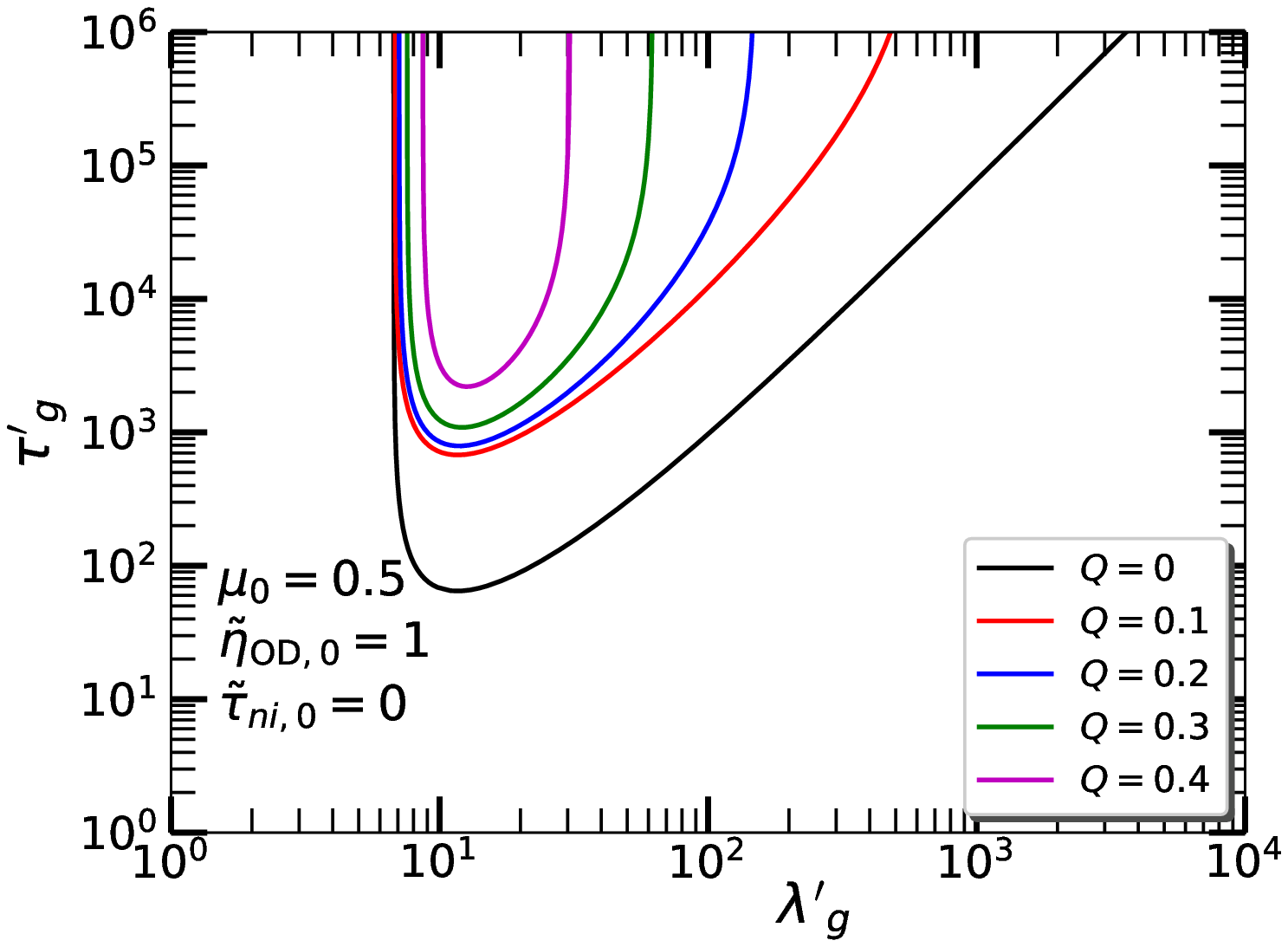}{0.245\textwidth}{(c)}
          \fig{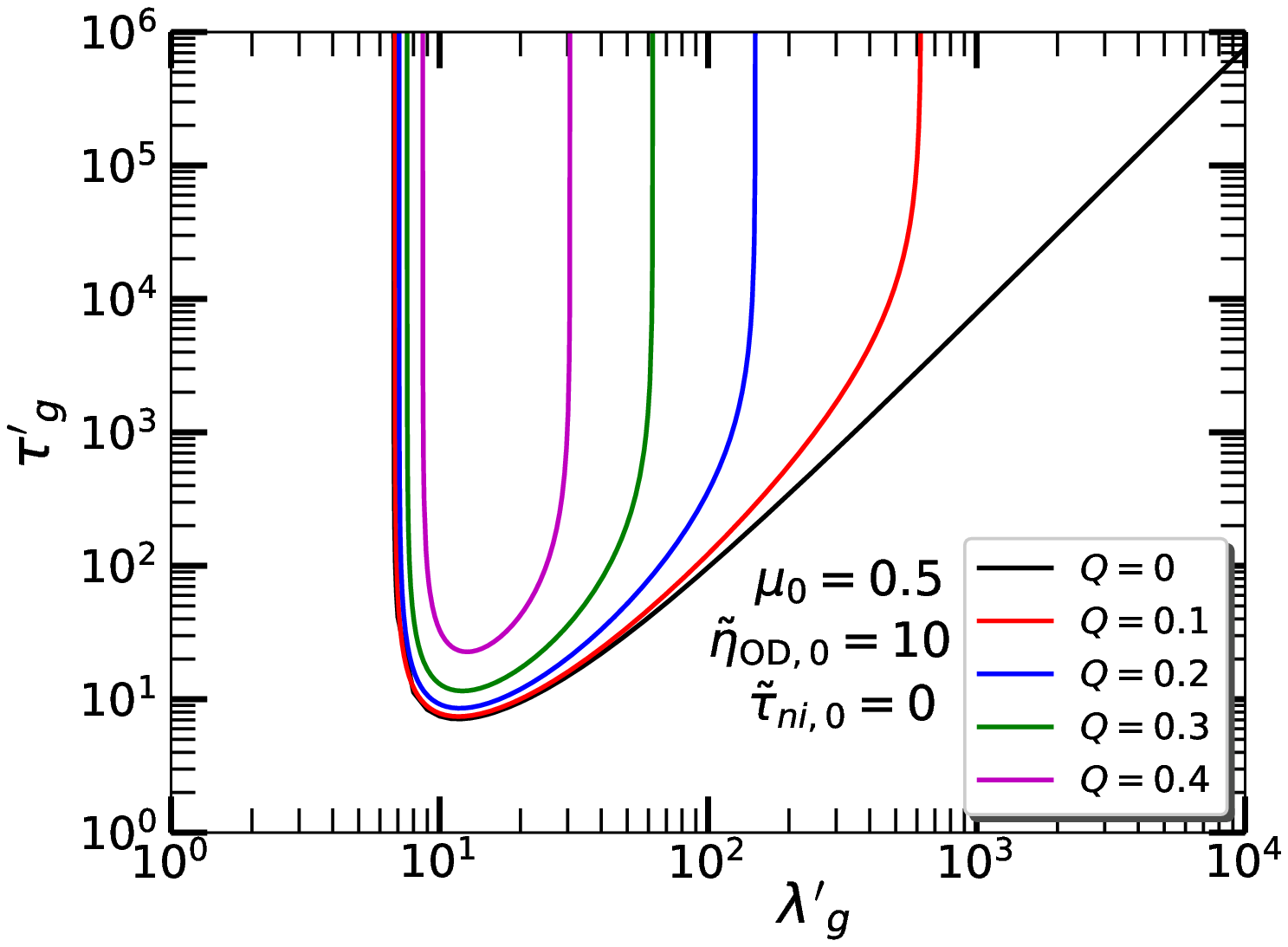}{0.245\textwidth}{(d)}
          }
\gridline{\fig{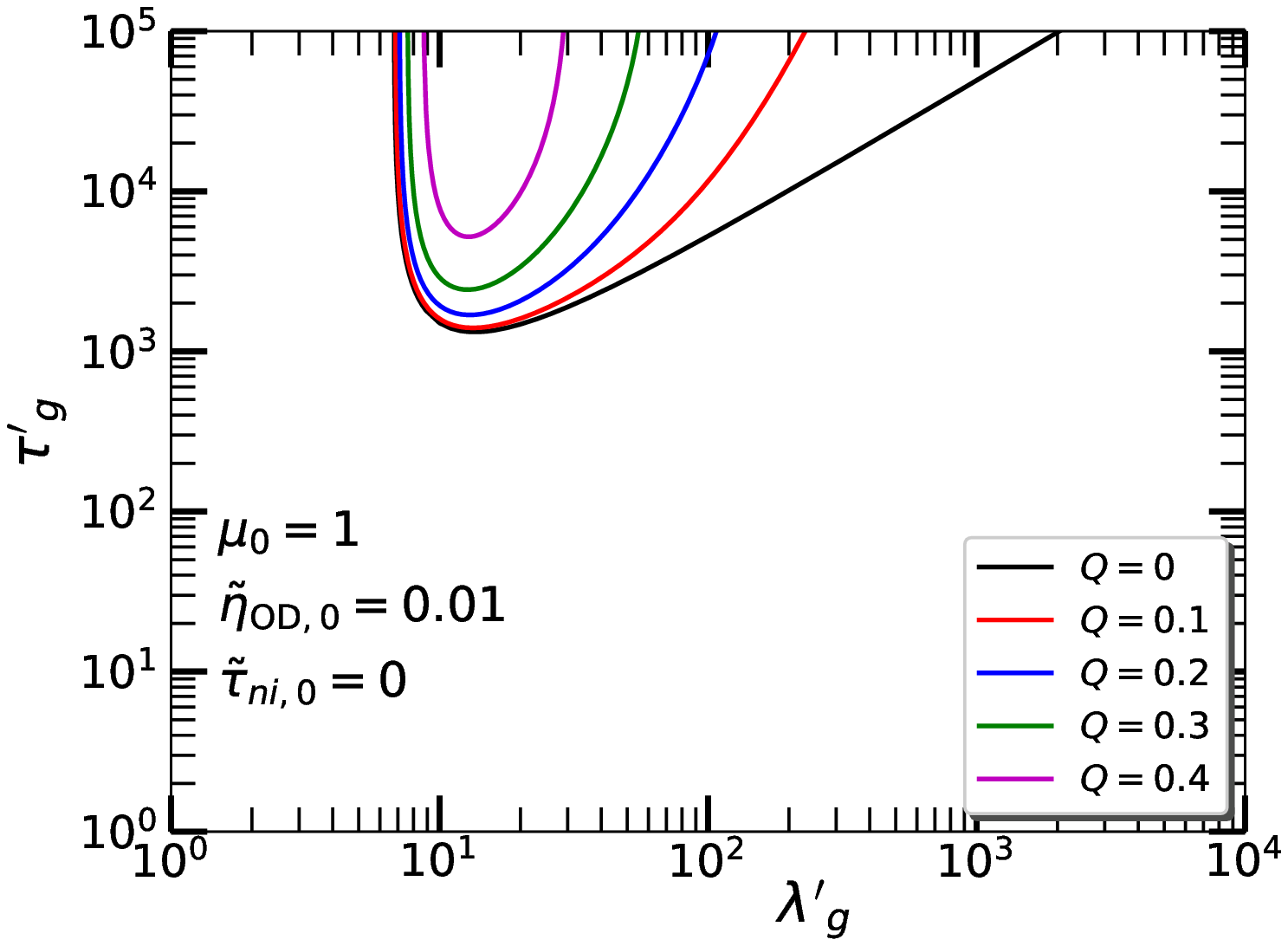}{0.245\textwidth}{(e)}
          \fig{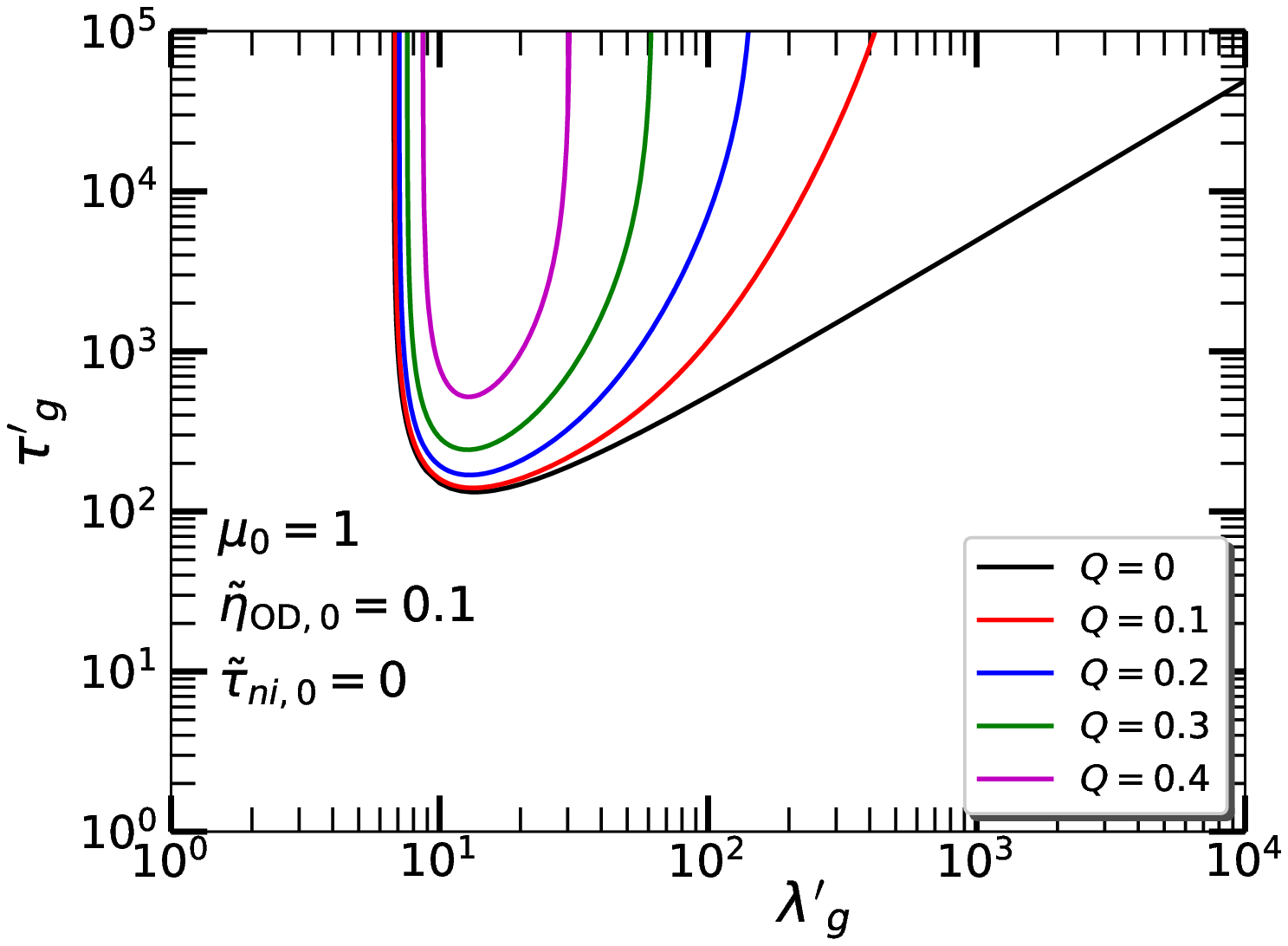}{0.245\textwidth}{(f)}
          \fig{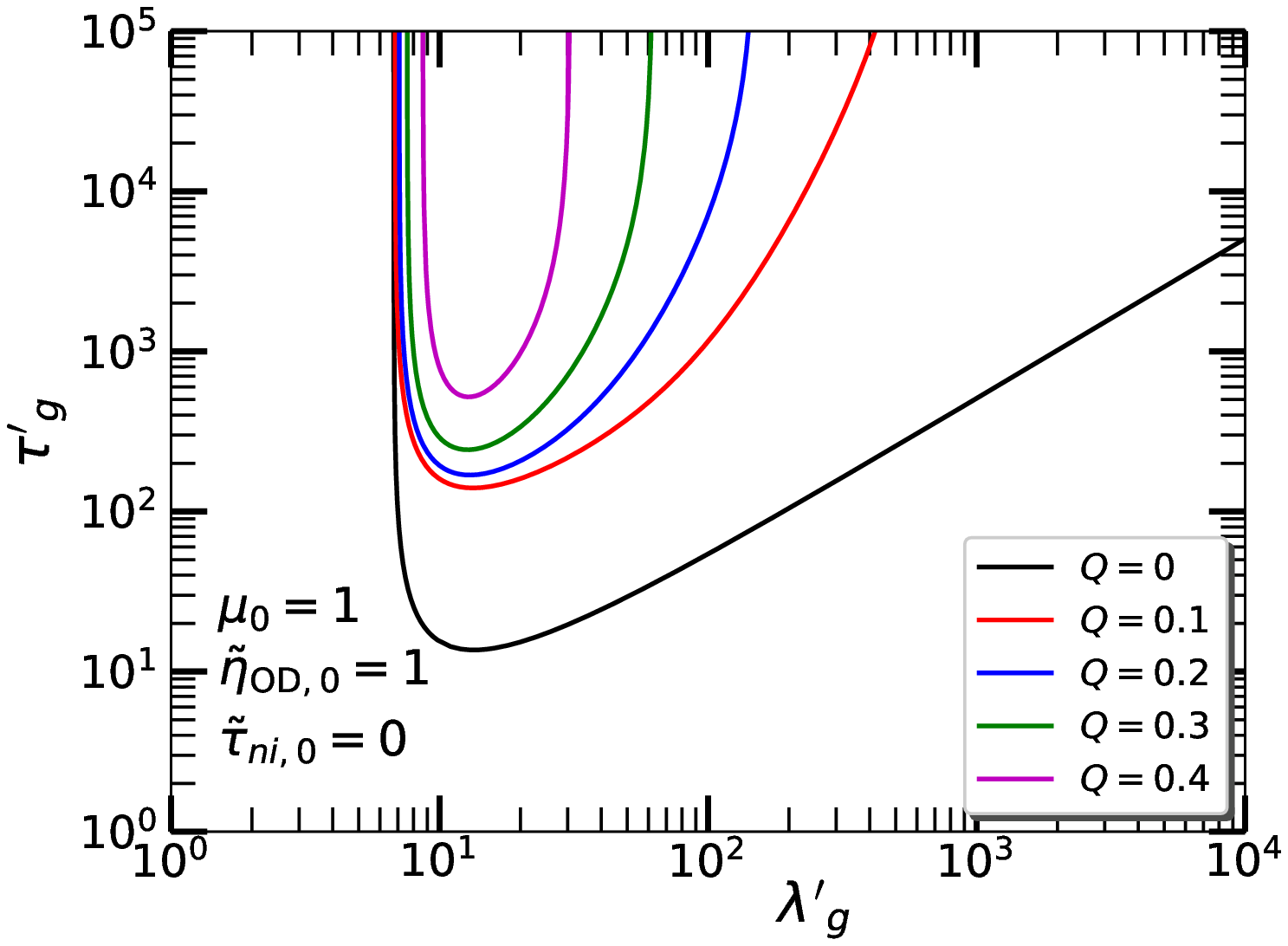}{0.245\textwidth}{(g)}
          \fig{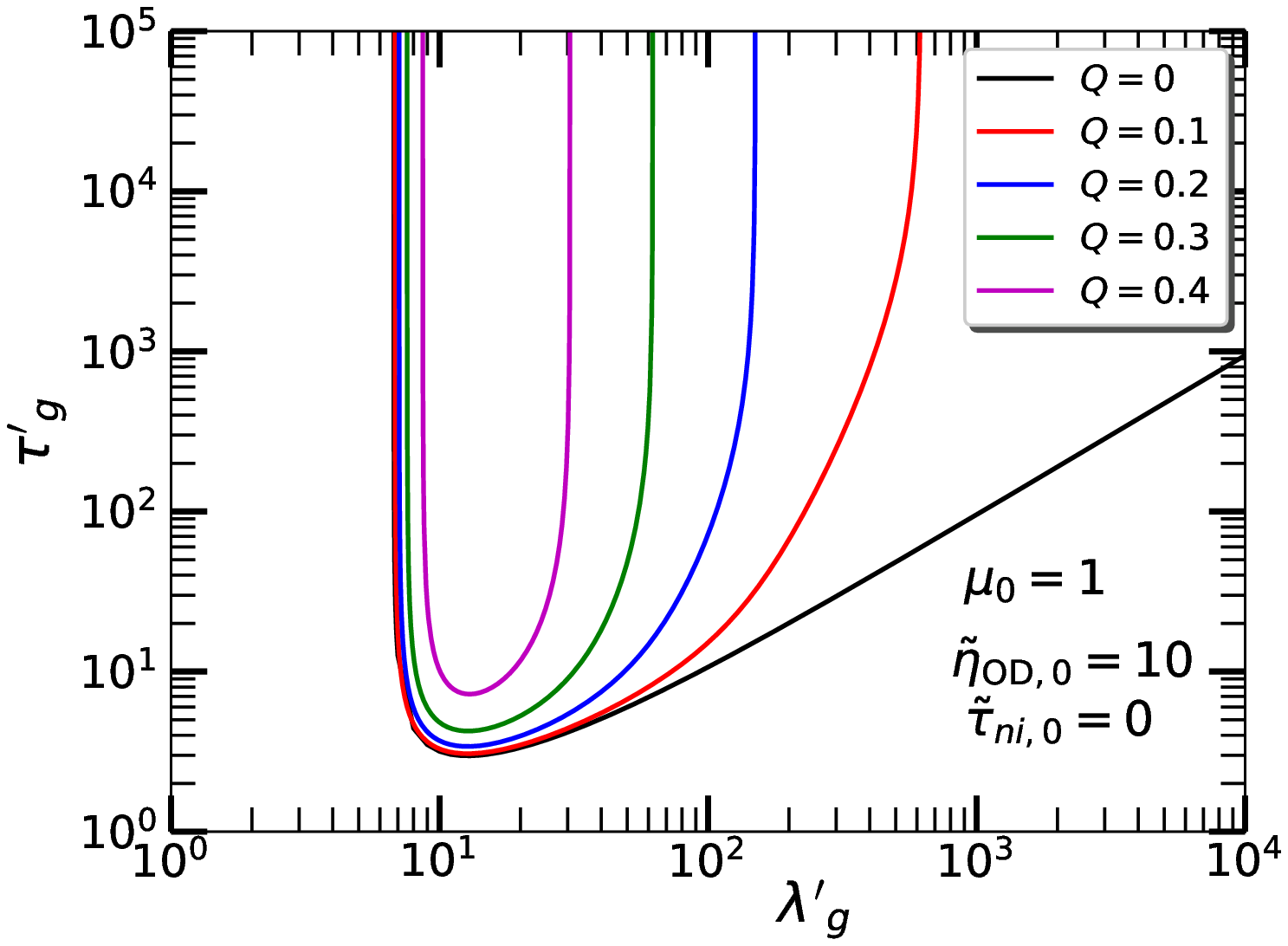}{0.245\textwidth}{(h)}
          }
\gridline{\fig{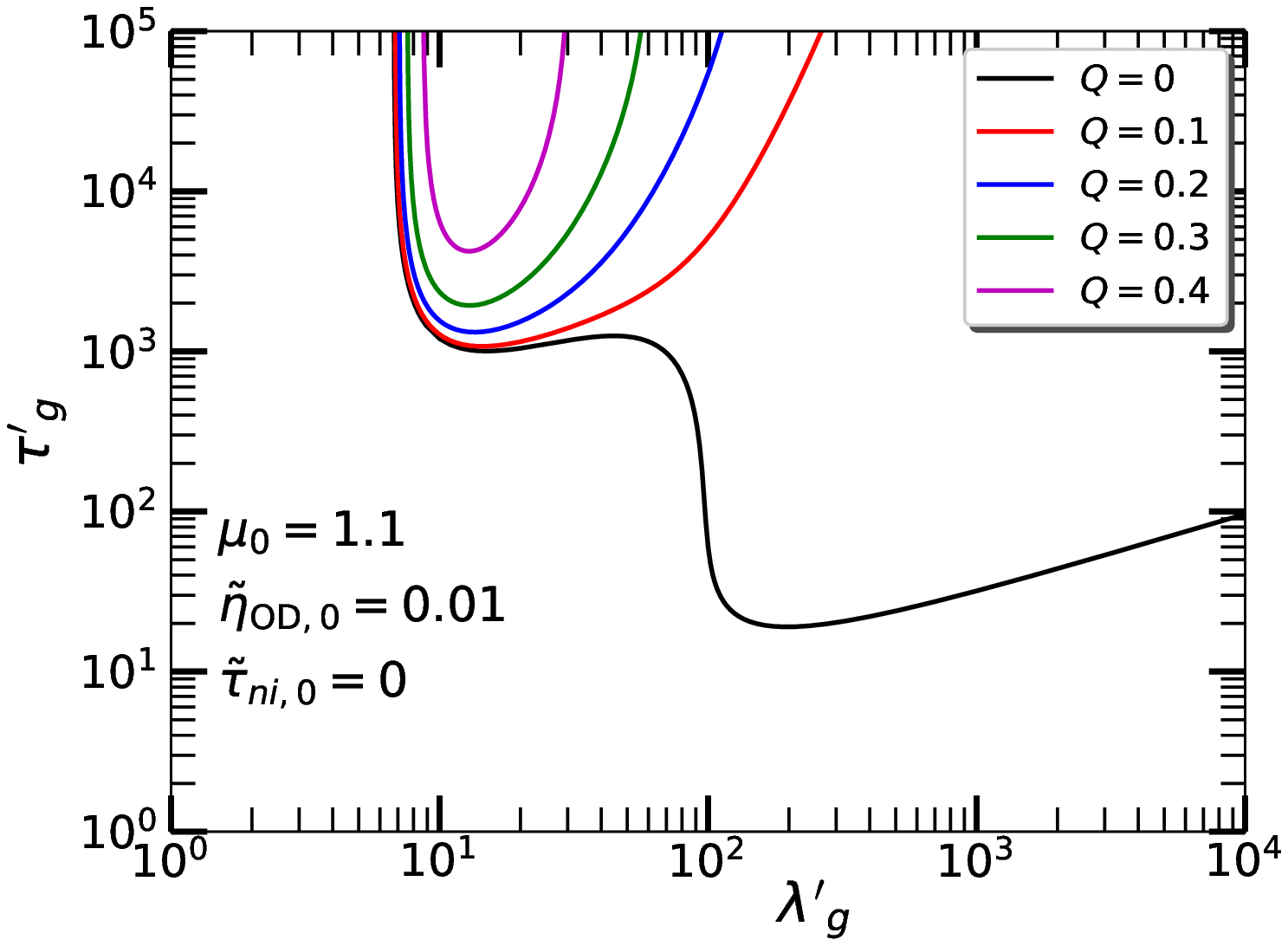}{0.245\textwidth}{(i)}
          \fig{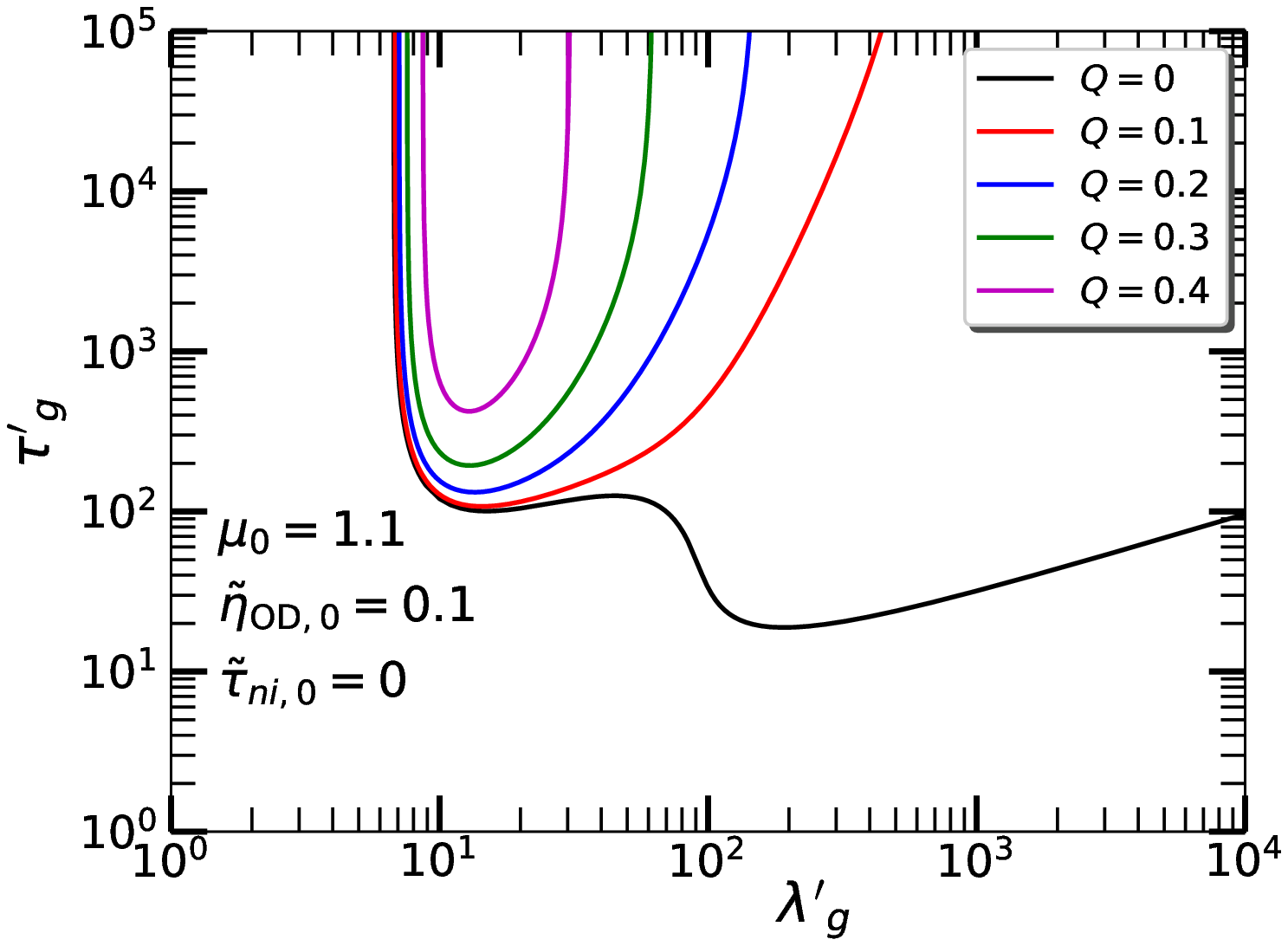}{0.245\textwidth}{(j)}
          \fig{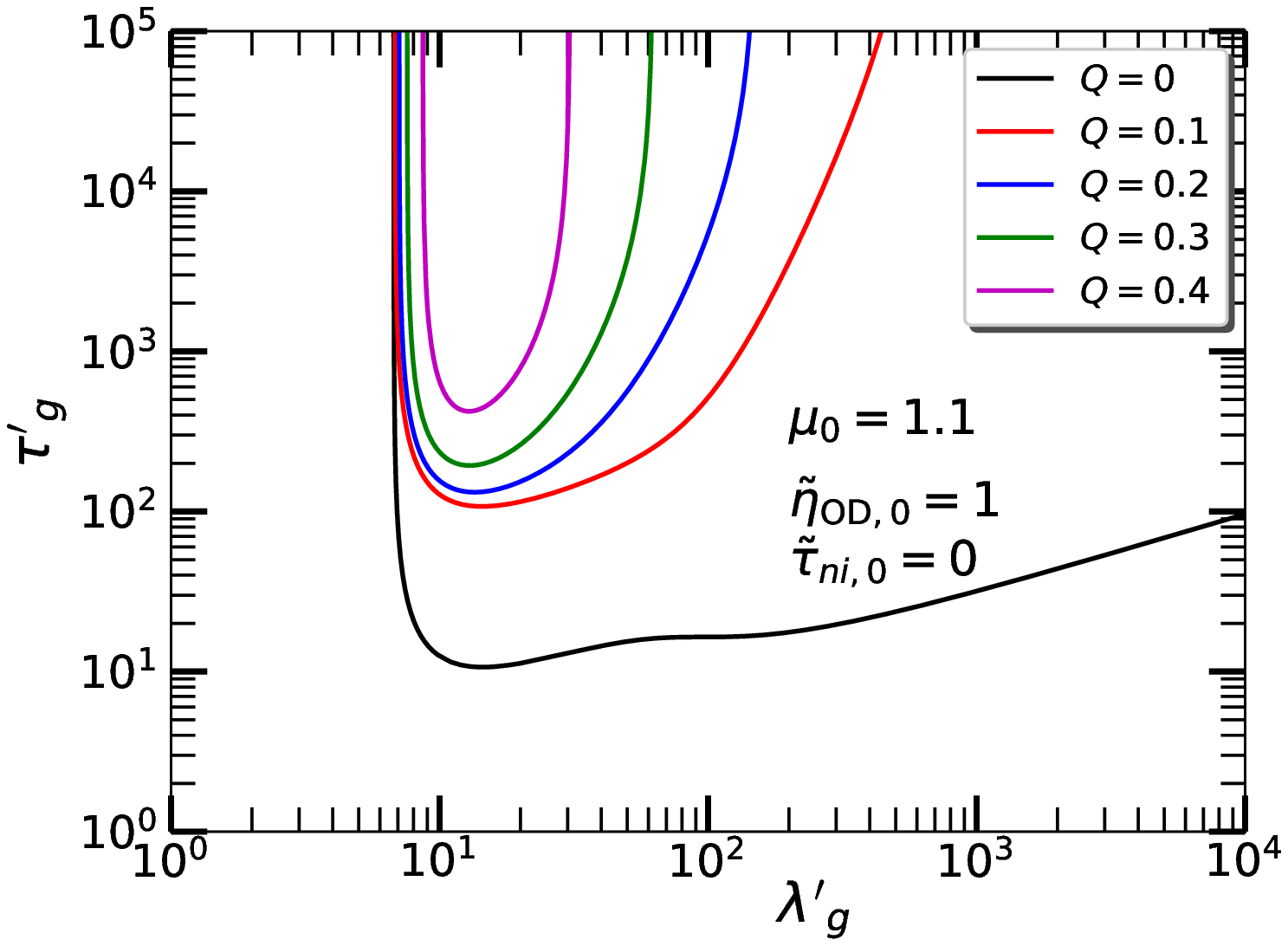}{0.245\textwidth}{(k)}
          \fig{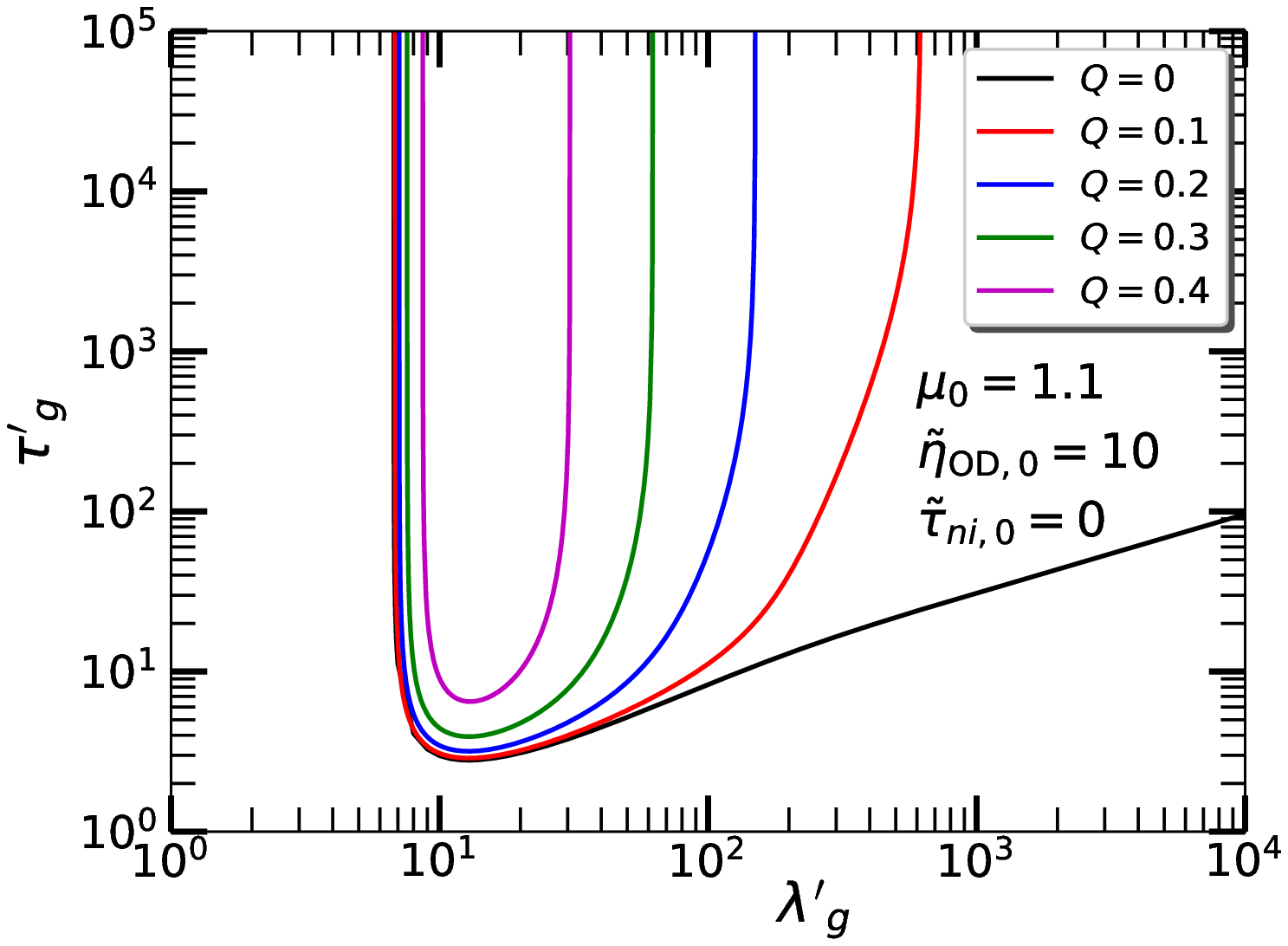}{0.245\textwidth}{(l)}
          }
\gridline{\fig{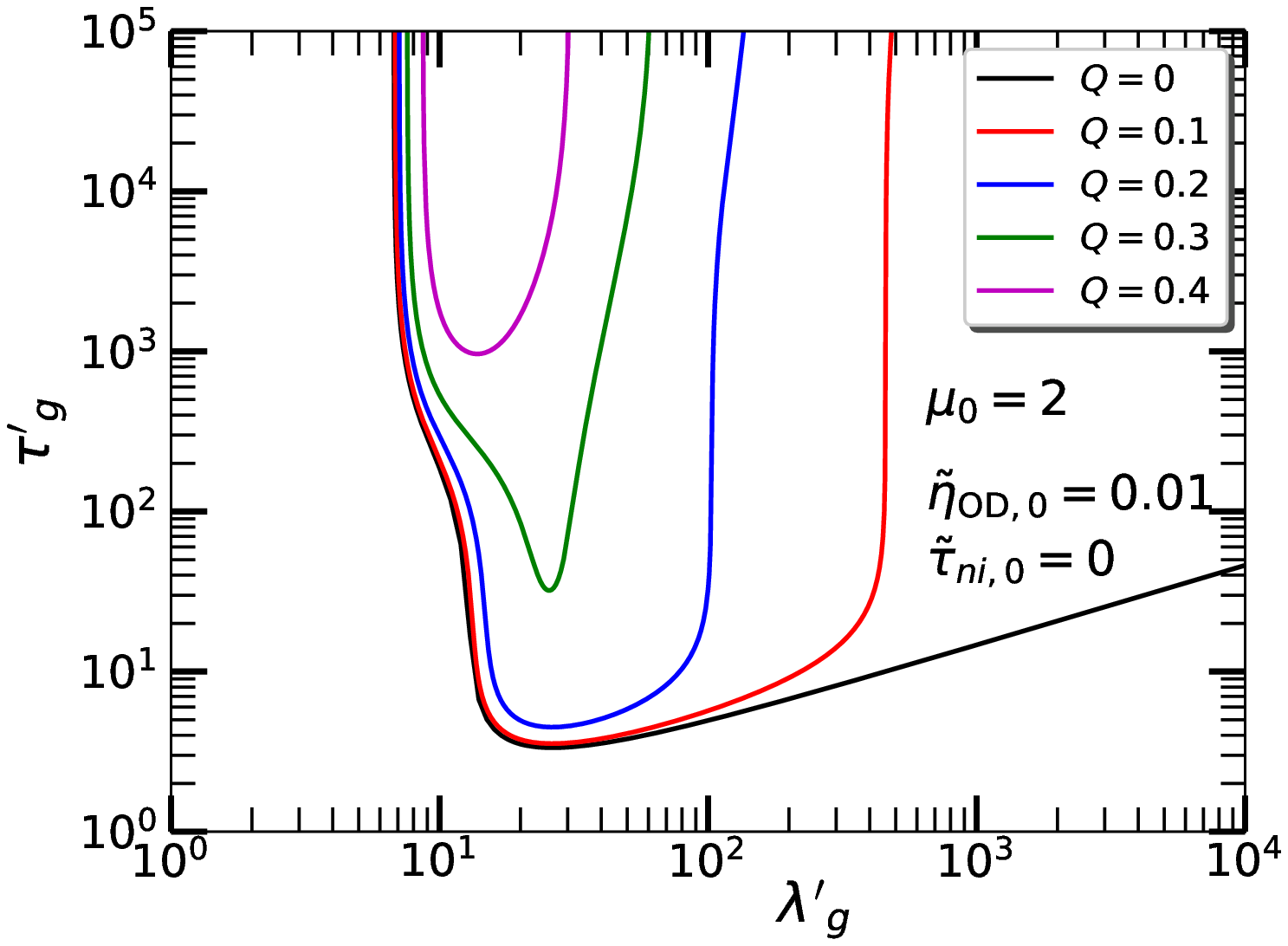}{0.245\textwidth}{(m)}
          \fig{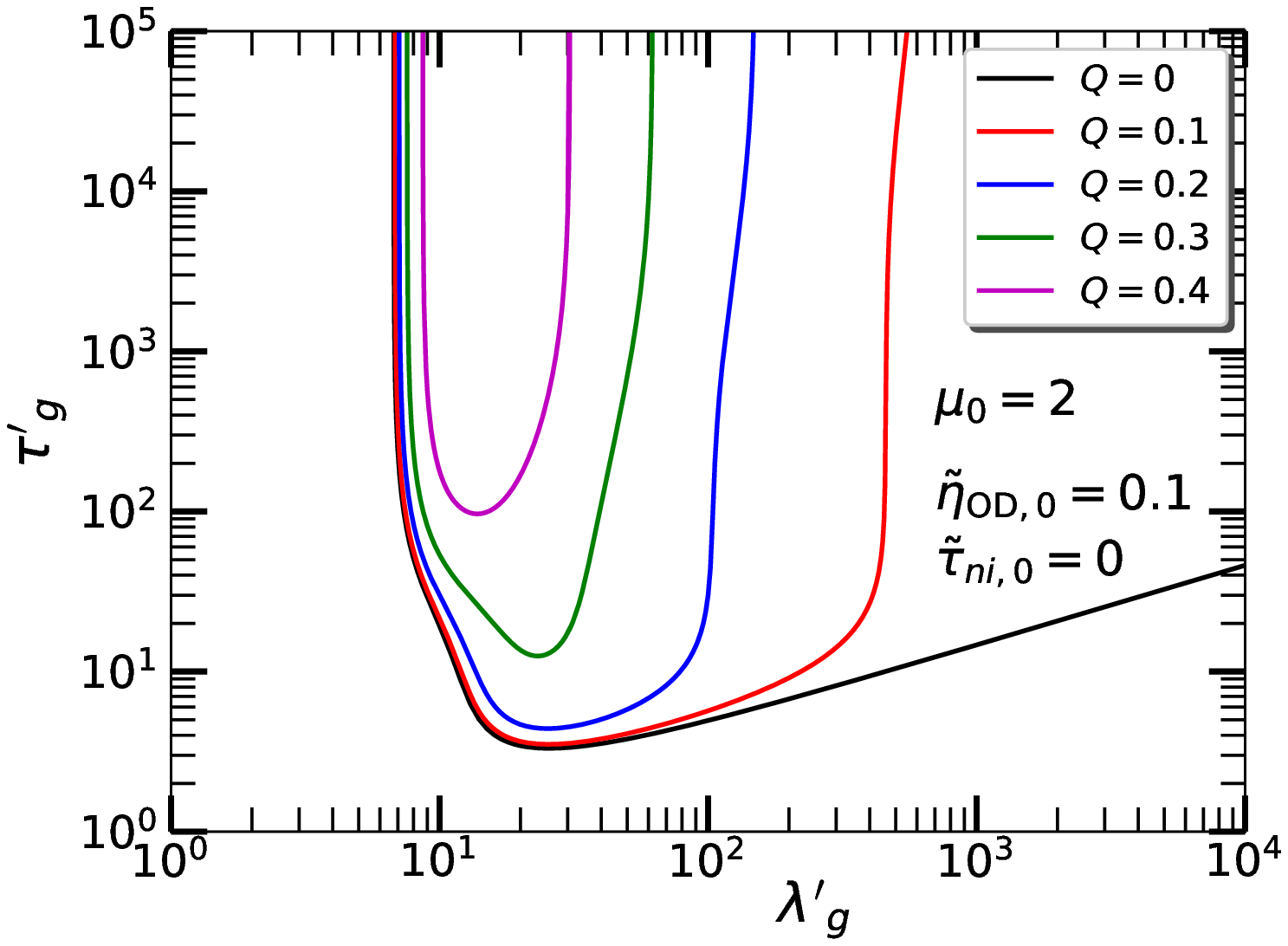}{0.245\textwidth}{(n)}
          \fig{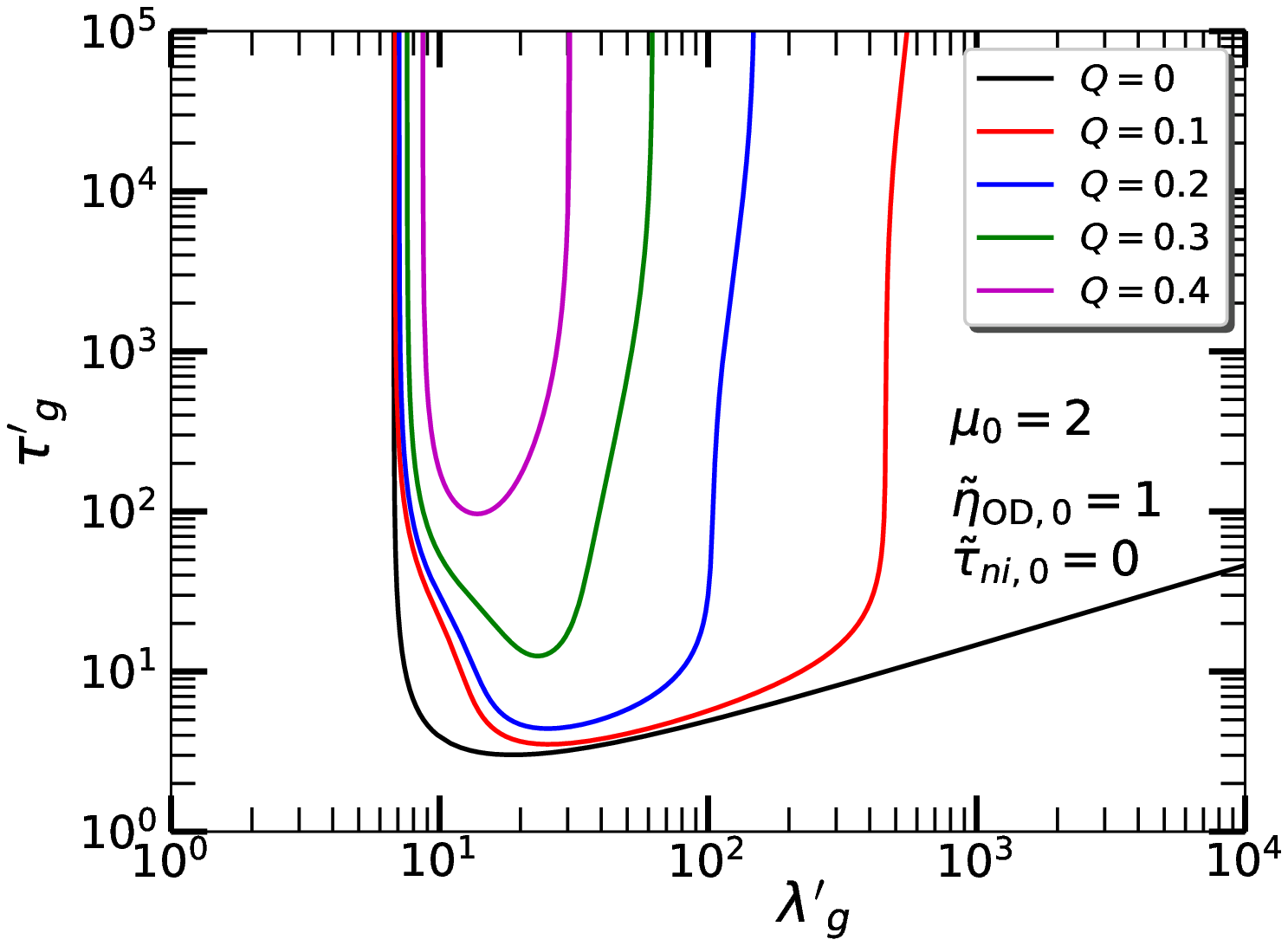}{0.245\textwidth}{(o)}
          \fig{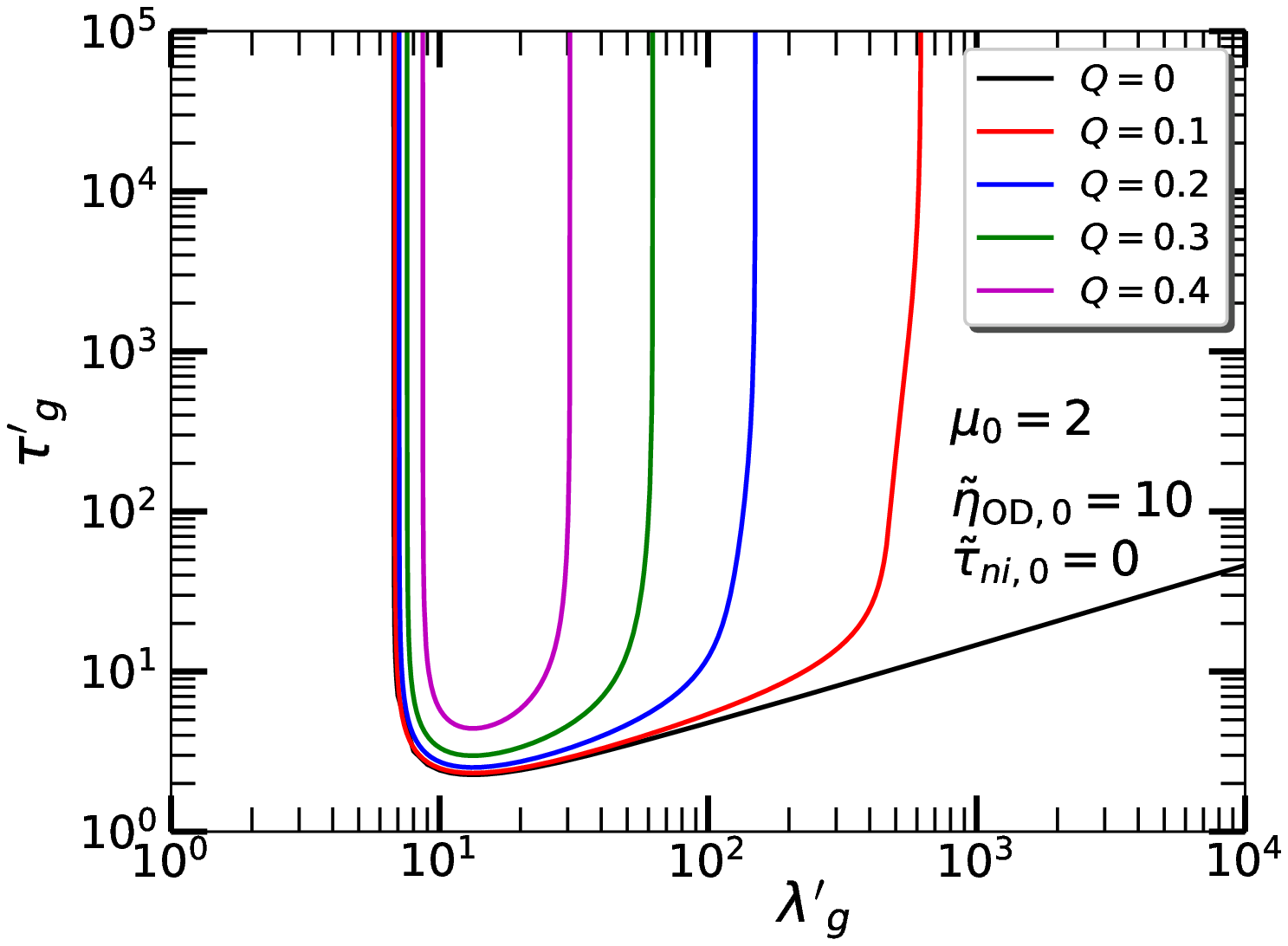}{0.245\textwidth}{(p)}
          }
\gridline{\fig{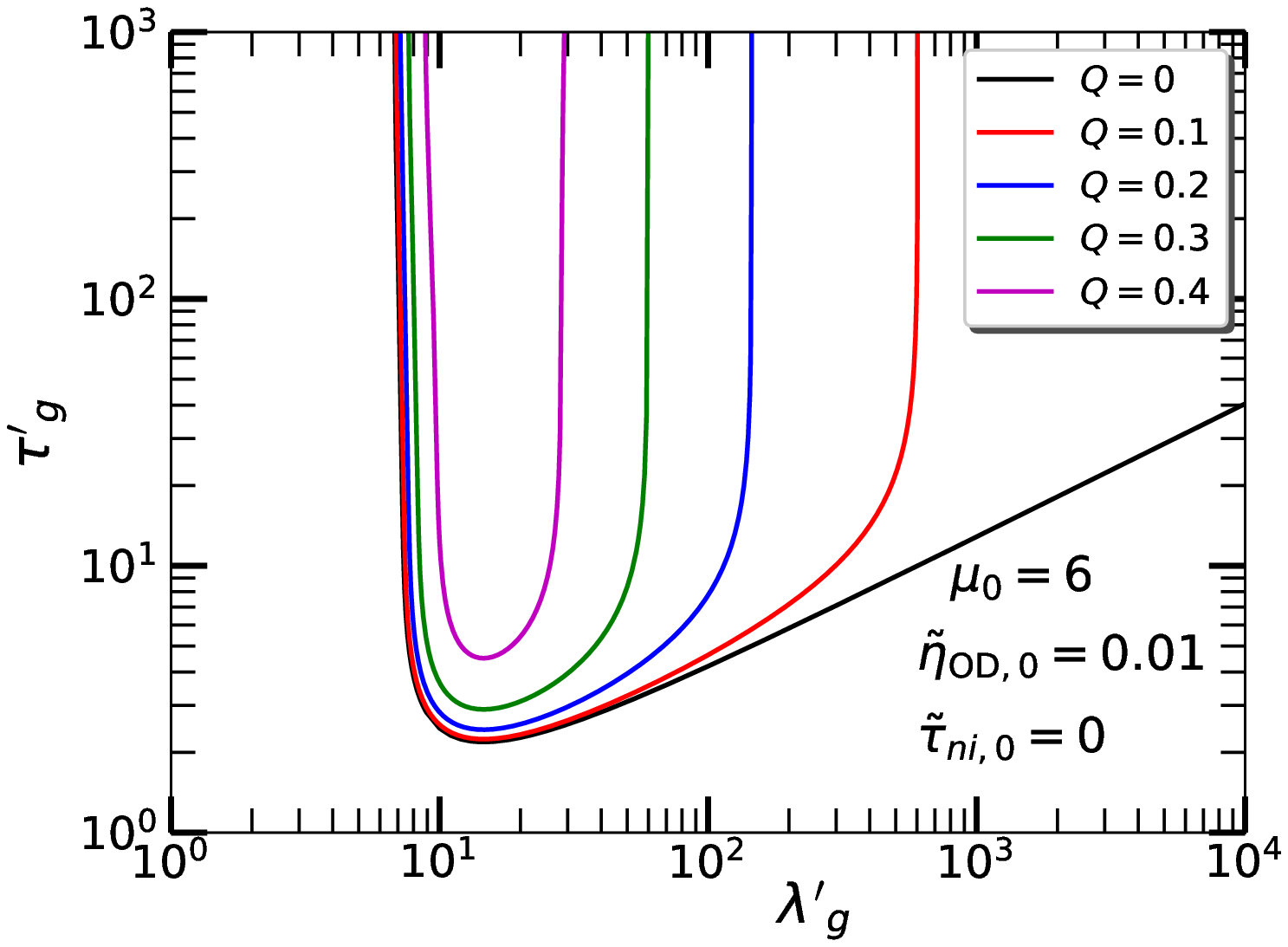}{0.245\textwidth}{(q)}
          \fig{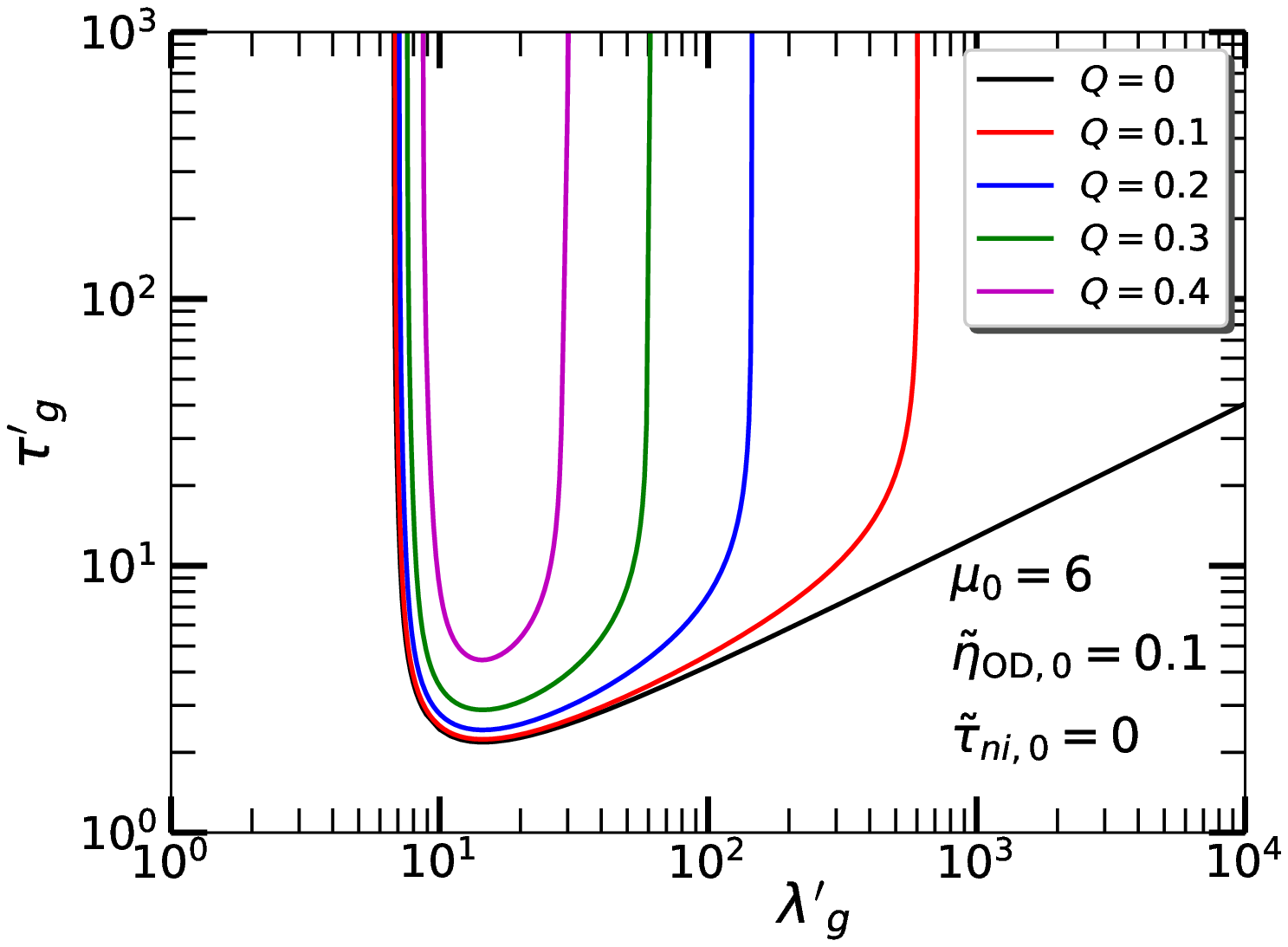}{0.245\textwidth}{(r)}
          \fig{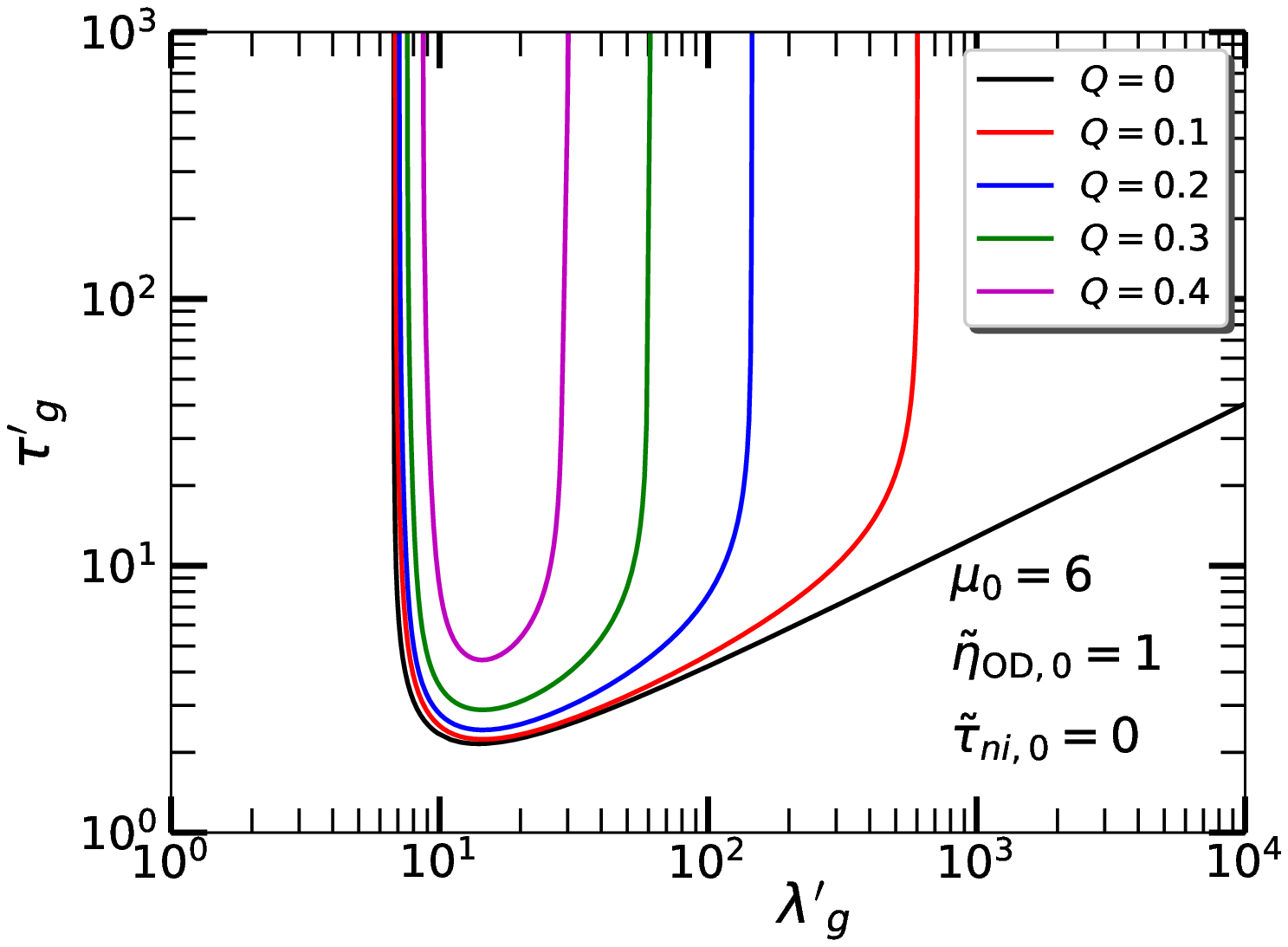}{0.245\textwidth}{(s)}
          \fig{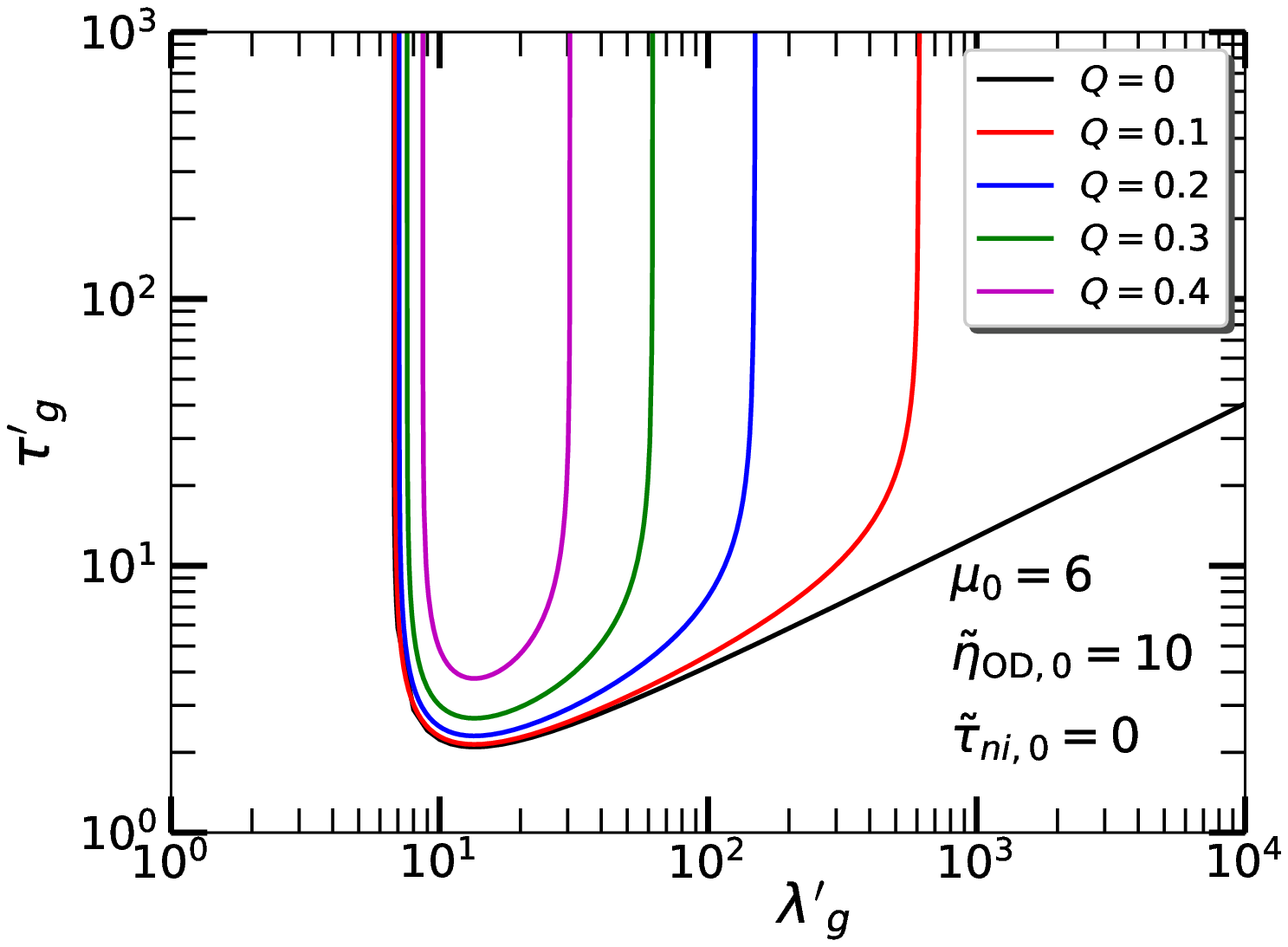}{0.245\textwidth}{(t)}
          }
\caption{Normalized growth time $\tau'_{g}=\tau_{g}/t_0$ of gravitationally unstable mode as a function of the normalized wavelength $\lambda'_{g}=\lambda_{g}/L_0$.
Left to right: For models with different normalized Ohmic diffusivities $\etaODt = 0.01$ (1st column), $0.1$ (2nd column), $ 1$ (3rd column), $10$ (4th column).
Top to bottom: For models with different normalized mass-to-flux-ratio $\mu_0 =$ 0.5 [1st row: a, b, c, d], $\mu_0 = 1$ [2nd row: e, f, g, h], $\mu_0 = 1.1$ [3rd row: i, j, k, l], $\mu_0 = 2$ [4th row: m, n, o, p], $\mu_0 = 6$ [5th row: q, r, s, t]. Each figure shows timescale curves for models with different normalized rotation $Q$ = 0 (black), 0.1 (red), 0.2 (blue), 0.3 (green), and 0.4 (magenta).}
\label{fig:fig1muall}
\end{figure*}   

We evaluate the growth timescale and length scale of gravitational instability with nonideal MHD effects. 
The larger the Ohmic diffusivity ($\etaODt$) and/or the neutral-ion collision time ($\tilde{\tau}_{ni,0}$), the greater are the effects of Ohmic dissipation (OD) and ambipolar diffusion (AD), respectively.

\autoref{fig:fig1AD} presents the instability growth time $\tau'_g$ ($=\tau_g/t_0$) as a function of the wavelength $\lambda'_g$ ($=\lambda_g/L_0$) for different cases of OD and AD (see also \cite{ciolek06} for AD).  
Here, the three different cases in the upper panel represent various Ohmic diffusivities: $\etaODt =0.1, 1, 10$.
The lower panel represents various areas within a molecular cloud: diffuse regions with high ionization fractions ($\tilde{\tau}_{ni,0} = 0.001$), dense core forming regions with low ionization fractions ($\tilde{\tau}_{ni,0} =0.2$) and an intermediate region ($\tilde{\tau}_{ni,0} =0.04$). 
Each panel shows the dependence for several labeled values of $\mu_0$ $(= 1/\tilde{B}_{\rm{ref}})$. Here, $\mu_0 = 0.5$ is a subcritical cloud, $\mu_0 = 1$ is a transcritical cloud, $\mu_0 = 1.1$ is slightly supercritical, $\mu_0 = 2$ is somewhat supercritical, and $\mu_0 = 6$ is highly supercritical. We see that the growth time decreases with greater $\etaODt$ and $\tilde{\tau}_{ni,0}$. 
In the limit of very large length scale, the normalized timescale ($\tau'_g$) for OD and AD asymptotically varies as $\lambda'^2_g$ and $\lambda'_g$, respectively for $\mu_0<1$, as derived from \autoref{eq:OD_taug_statB_asymplim} and \autoref{eq:AD_taug_statB_asymplim}. Whereas, for the supercritical region, $\tau'_g$ asymptotically varies as $\lambda'^{1/2}_g$, as derived in \autoref{eq:normtaug_asymplim}. Hence for this case, the minima of $\tau'_g$ vs $\lambda'_g$ curves look shallower as compared to the subcritical cases. The diffusive-driven instabilities for the subcritical clouds have a sharper minimum (peak) in the growth time.

\autoref{fig:fig1muall} shows the instability growth timescale and length scale with OD as the only nonideal MHD effect.  The first, second, third, and fourth column (from left to right) show the cases for $\etaODt = 0.01$, $\etaODt = 0.1$, $\etaODt = 1$, and $\etaODt = 10$, respectively. 
Each column shows five different normalized mass-to-flux-ratios ($\mu_0 = 0.5, 1, 1.1, 2, 6$) and each panel shows five different rotation levels ($Q = 0, 0.1, 0.2, 0.3, 0.4$). For the subcritical case ($\mu_0 = 0.5$, note first row), as the Ohmic diffusivity ($\etaODt$) increases by each factor of 10, the instability growth time significantly gets reduced. Since the magnetic flux is being dissipated at a faster rate, it shortens the growth timescale. Changing $\etaODt$ from $0.01$ to $10$, the timescale gets smaller by a factor of $10^3$. Also, for the transcritical  ($\mu_0 = 1$, note second row) and slightly transcritical ($\mu_0 = 1.1$, note third row) clouds, the growth timescale is lowered down by a similar magnitude when moving from $\etaODt = 0.01$ to $10$. For the mildly supercritical case ($\mu_0 = 2$, note fourth row), a gradual reduction in the growth timescale is more prominent for the modes with higher rotation. This signifies that in the regime of OD, gravitational collapse is likely to be faster even with the higher rotation speed. Lastly, for the highly supercritical case ($\mu_0 = 6$, note fifth row), since the inward gravitational pull is extremely dominant over the magnetic field and rotation, there is not much appreciable change in the growth modes with the variation of $\etaODt$.

Earlier, for the flux-frozen case, we mentioned that there is no unstable, gravitationally collapsing mode for $\mu_0 < 1$, implying that only initially supercritical clouds can collapse. However, the addition of AD and OD (see \autoref{fig:fig1AD} and \autoref{fig:fig1muall}) allows for unstable, gravitationally collapsing modes to exist for both subcritical ($\mu_0 < 1$) and supercritical ($\mu_0 > 1$) regimes. See also \autoref{sec:app_fig1ADODrot} for the combined effects of OD and AD. For all these plots and for each case of $\mu_0$ shown, we notice that each curve has a distinct minimum. This minimum represents the shortest growth time (fastest growth rate) and a corresponding preferred length scale for gravitational instability.

\begin{figure*}[ht!]
\gridline{\fig{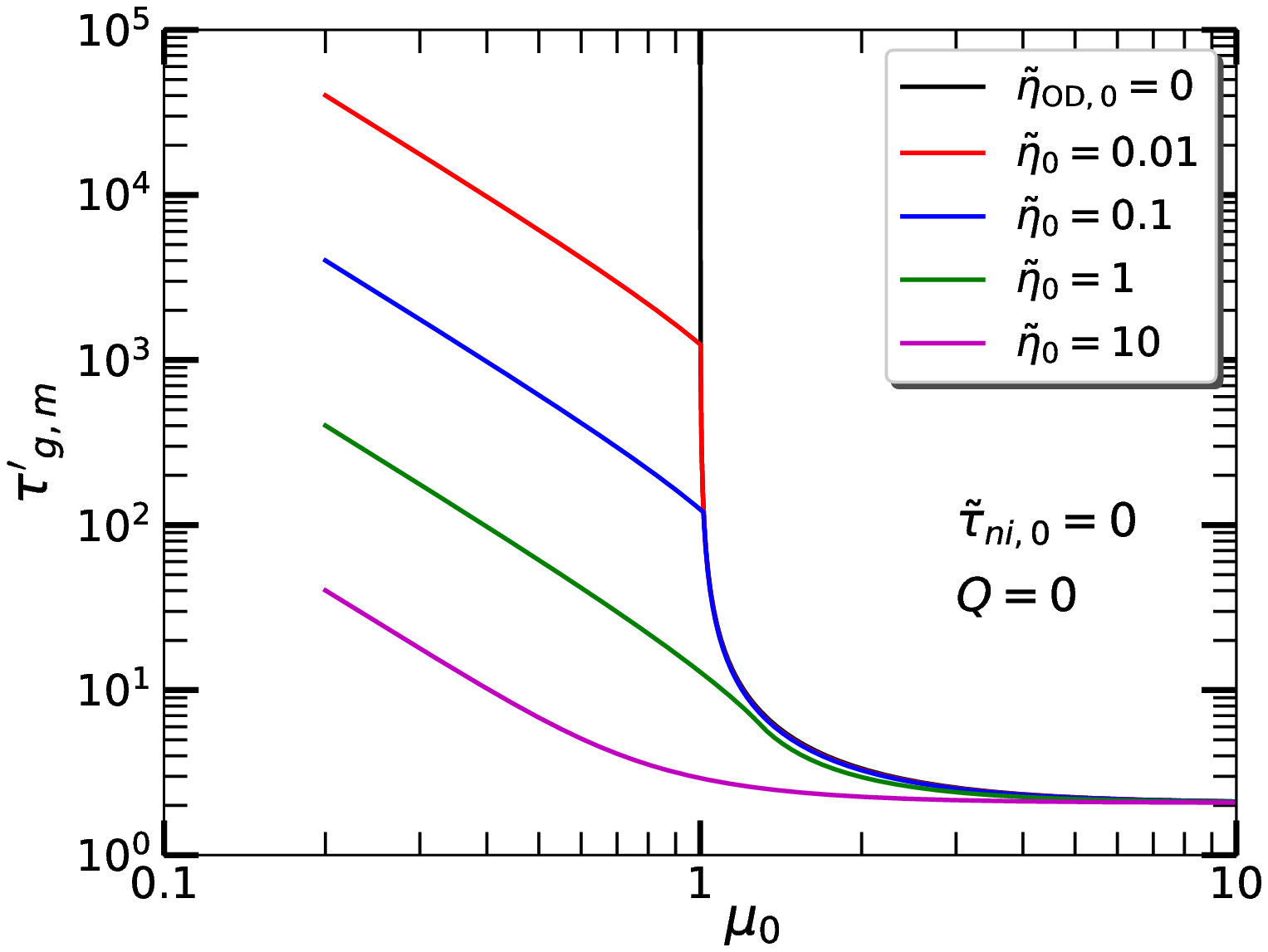}{0.45\textwidth}{(a)}
          \fig{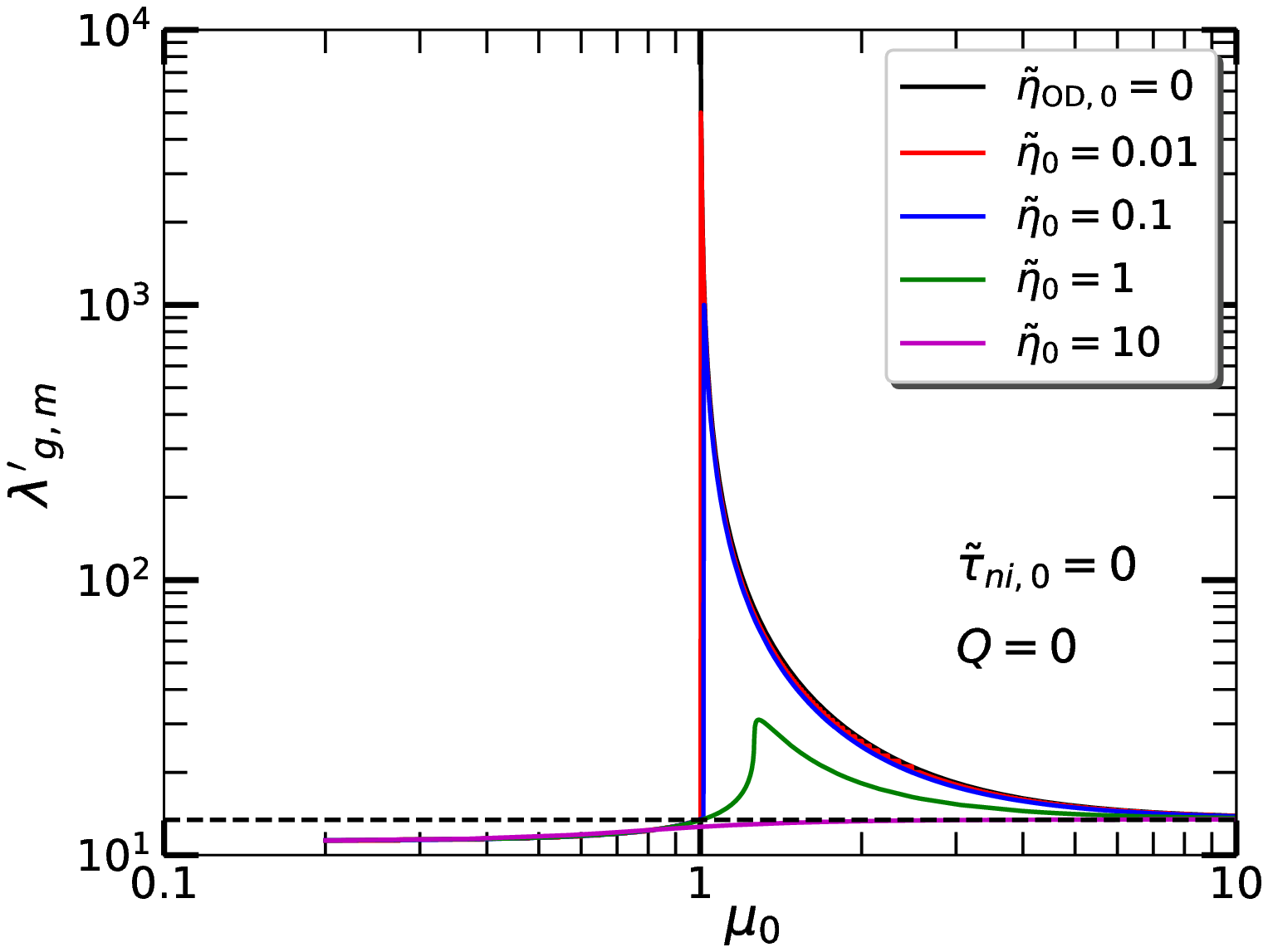}{0.45\textwidth}{(b)}
          } 
\gridline{\fig{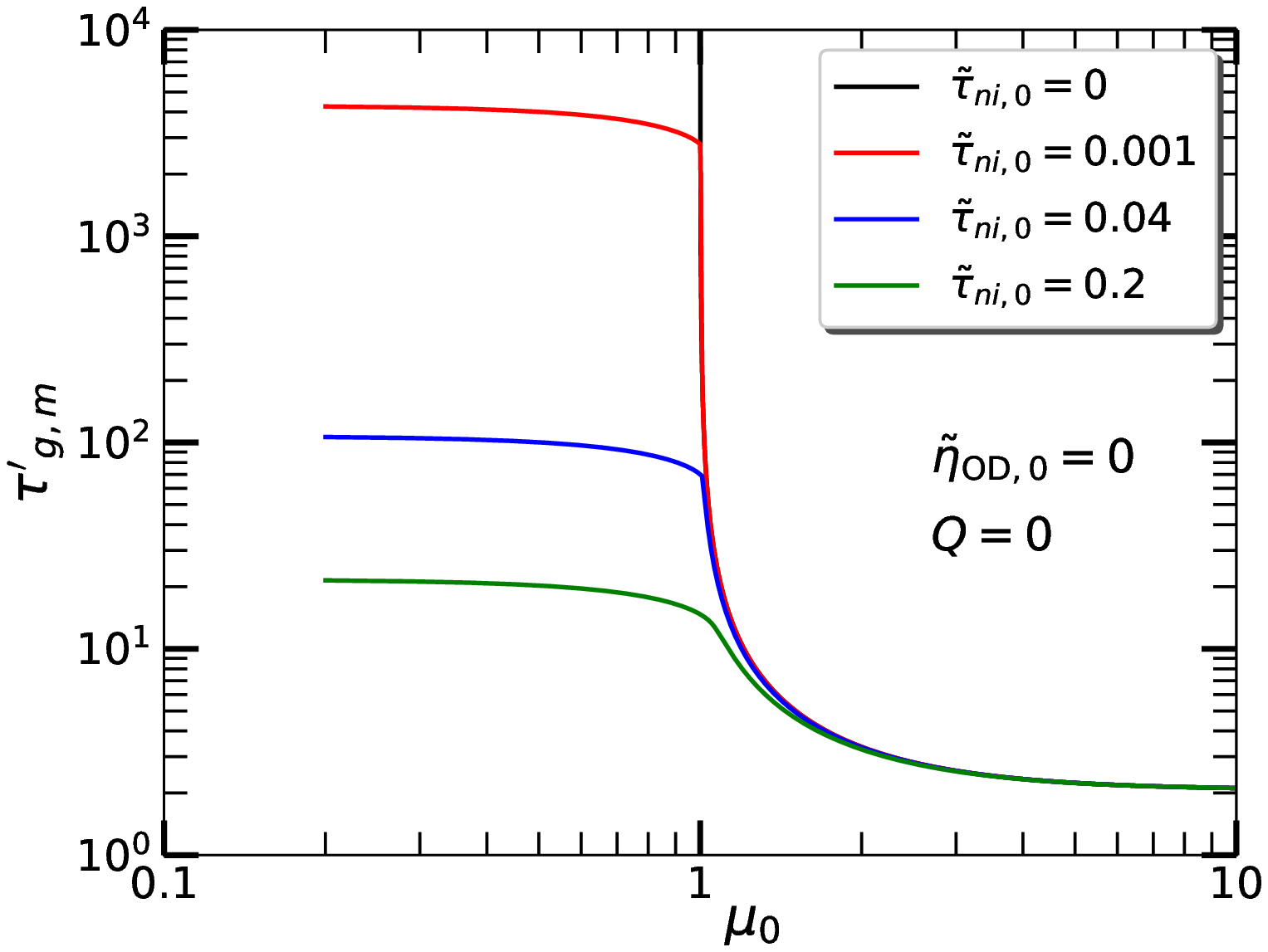}{0.45\textwidth}{(c)}
          \fig{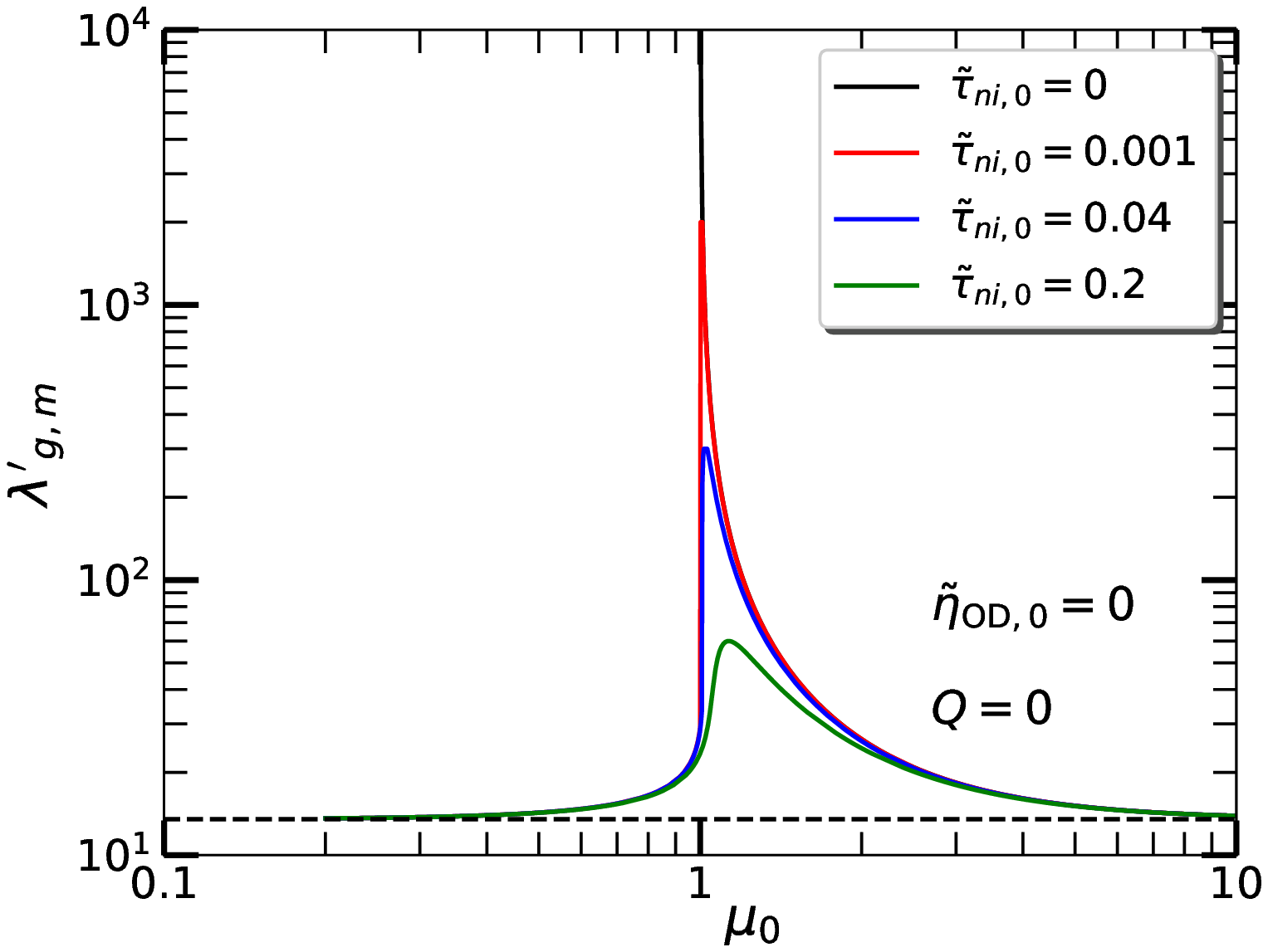}{0.45\textwidth}{(d)}
          }
         
\caption{Normalized shortest growth timescale $\tau'_{g,m}=\tau_{g,m}/t_0$ and preferred length scale $\lambda'_{g,m} = \lambda_{g,m}/L_0$ of the gravitationally unstable mode as a function of the normalized mass-to-flux ratio $\mu_0$. 
Upper panel (a and b) shows the case of Ohmic dissipation for models with normalized Ohmic diffusivities $\etaODt$ = 0 (black), 0.01 (red), 0.1 (blue), 1 (green), and 10 (magenta). Lower panel (c and d) shows the case of ambipolar diffusion for models with normalized neutral-ion collision time $\tilde{\tau}_{ni,0}$ = 0 (black), 0.001 (red), 0.04 (blue), 0.2 (green). The black dashed line in (b) and (d) denotes the value $2 \lambda'_{\rm T}$; $\lambda'_{\rm T}$ is the normalized thermal length scale.}
\label{fig:ADonly_ODonly}
\end{figure*}

\autoref{fig:ADonly_ODonly} shows the normalized minimum growth time of the gravitationally unstable mode $\tau'_{g,m}$ ($=\tau_{g,m}/t_0$) and length scale $\lambda'_{g,m}$ ($=\lambda_{g,m}/L_0$) corresponding to this most unstable mode (which we call the preferred length scale) as a function of $\mu_0$. The upper panel of \autoref{fig:ADonly_ODonly} shows the case with only OD as obtained from our model. 
On the other hand, the lower panel of \autoref{fig:ADonly_ODonly} shows the same for different amounts of AD, as calculated previously by \cite{ciolek06} and \cite{bailey12}. For both nonideal MHD effects, we observe qualitatively similar length scale curves, and timescale curves that are qualitatively similar in the supercritical regime but differ in the subcritical regime. 

In the limit of flux-freezing ($\etaODt=0$, $\taunit=0$), for the supercritical regime ($\mu_0 >1$), each of the flux-freezing curves in \autoref{fig:ADonly_ODonly} shows that the growth time and length scale for instability are short; essentially the timescale and length scale follow the dynamical timescale $(t_{\rm d} = Z_0/c_s)$ and preferred thermal length scale ($\lambda_{{\rm T},m}$). Even with nonideal MHD terms included, the growth times are similar in the supercritical regime, since these modes are dominated by gravity. As the normalized mass-to-flux ratio approaches the transcritical value ($\mu_0 =1$) the growth timescale/length scale for instability becomes infinitely long, since in the flux-frozen case only supercritical clouds can collapse. 
With the addition of either nonideal MHD effect (OD or AD), the growth timescale in the subcritical regime becomes finite. 

In the OD-only regime, \autoref{fig:ADonly_ODonly}(a) shows that an increasing $\etaODt$, which increases the rate of magnetic flux dissipation, makes the growth time tend toward 
that of thermal collapse $\tau'_{g,T,m}$ ($=\tau_{g,T,m}/t_0)$. For $\mu_0 \ll 1$, the preferred length scale attains the thermal length scale as shown in \autoref{eq:lambdagmpref_OD}, hence the corresponding minimum growth timescale goes to infinity as derived in \autoref{eq:taugmpref_OD}.

Similarly, in the AD-only case, for a relatively large $\tilde{\tau}_{ni,0} > 0.2$, the growth timescale of the subcritical regime is decreasing toward that of thermal collapse. However, for each value of $\tilde{\tau}_{ni,0}$, it has a plateau for all values of $\mu_0 \ll 1$. This is a distinguishing characteristic of AD in comparison to OD. As seen in \autoref{eq:etaAD} and \autoref{eq:etaADt}, the ambipolar diffusivity is proportional to the square of the background magnetic field strength, therefore proportional to $\mu_0^{-2}$.
Even as $\mu_0$ decreases in the regime $\mu_0 \ll 1$, $\etaADt$ increases as $\mu_0^{-2}$ and enforces a fixed drift speed of ions and neutrals (see \autoref{sec:perturbation} for more details on the eigenfunctions) and thereby in the growth timescale. 
For a typical normalized neutral-ion collision time as observed in molecular clouds ($\tilde{\tau}_{ni,0} =0.2$), the timescale for collapse of a subcritical region is $\sim 10$ times longer than that of a supercritical region (see \autoref{fig:ADonly_ODonly}(c)). This leads to the often quoted result that the ambipolar diffusion time is $\sim 10$ times the dynamical time. However, note that a transcritical region has a growth time that is intermediate to the two plateau values.

The preferred wavelengths for collapse ($\lambda'_{g,m}$ = $\lambda_{g,m}/L_0$) exhibit an interesting dependence on $\mu_0$ (see \autoref{fig:ADonly_ODonly}(b) and (d)).
For a nonzero Ohmic diffusivity or neutral-ion collision time, the wavelength with the minimum growth time remains close to the flux-freezing value for decidedly supercritical clouds, since these are gravity-dominated modes that collapse quickly with little time for significant magnetic diffusion. For transcritical but slightly supercritical clouds, there is a sharp rise in the preferred wavelength, similar to what happens in the flux-frozen case. However, the preferred wavelength does not diverge at $\mu_0=1$ as in the flux-frozen case. Instead, the magnetic diffusion caps the preferred wavelength at a finite, but potentially large value that depends on the level of diffusivity. In these hybrid transcritical modes, there is enough magnetic field dragging to create an hourglass shape with a strong curvature force that resists the collapse, so that larger perturbations with more mass can more easily overcome the magnetic support. For subcritical regions, where flux-freezing would allow no instability, modes of diffusion-driven contraction now appear. These modes have very little magnetic field enhancement in the perturbed region (see \autoref{sec:perturbation}). Given the withdrawal of magnetic support by OD or AD, the preferred scale converges back toward the thermal length scale. The black dashed line in \autoref{fig:ADonly_ODonly}(b) and (d) denotes the value $2 \lambda'_{\rm T}$. Specifically, for OD it converges to $\lambda'_{\rm T}$, the critical thermal length scale, and for AD it converges to $2 \lambda'_{\rm T}$ as would be found in the hydrodynamic case.
These limits were shown in \autoref{eq:lambdagmpref_OD} and
\autoref{eq:lambdagmpref_AD} and the difference can be attributed to the stronger wavenumber dependence of the OD term (see \autoref{sec:etak_dis}).

\subsection{{Results for a Protostellar Disk}}\label{sec:disk}

In this section we focus on the region of parameter space that is most applicable to protostellar disks, i.e., models with significant nonzero values of $\etaODt$ and $Q$, and potentially $\Tilde{\tau}_{ni,0}$ as well.

\begin{figure*}[htb!]
\gridline{\fig{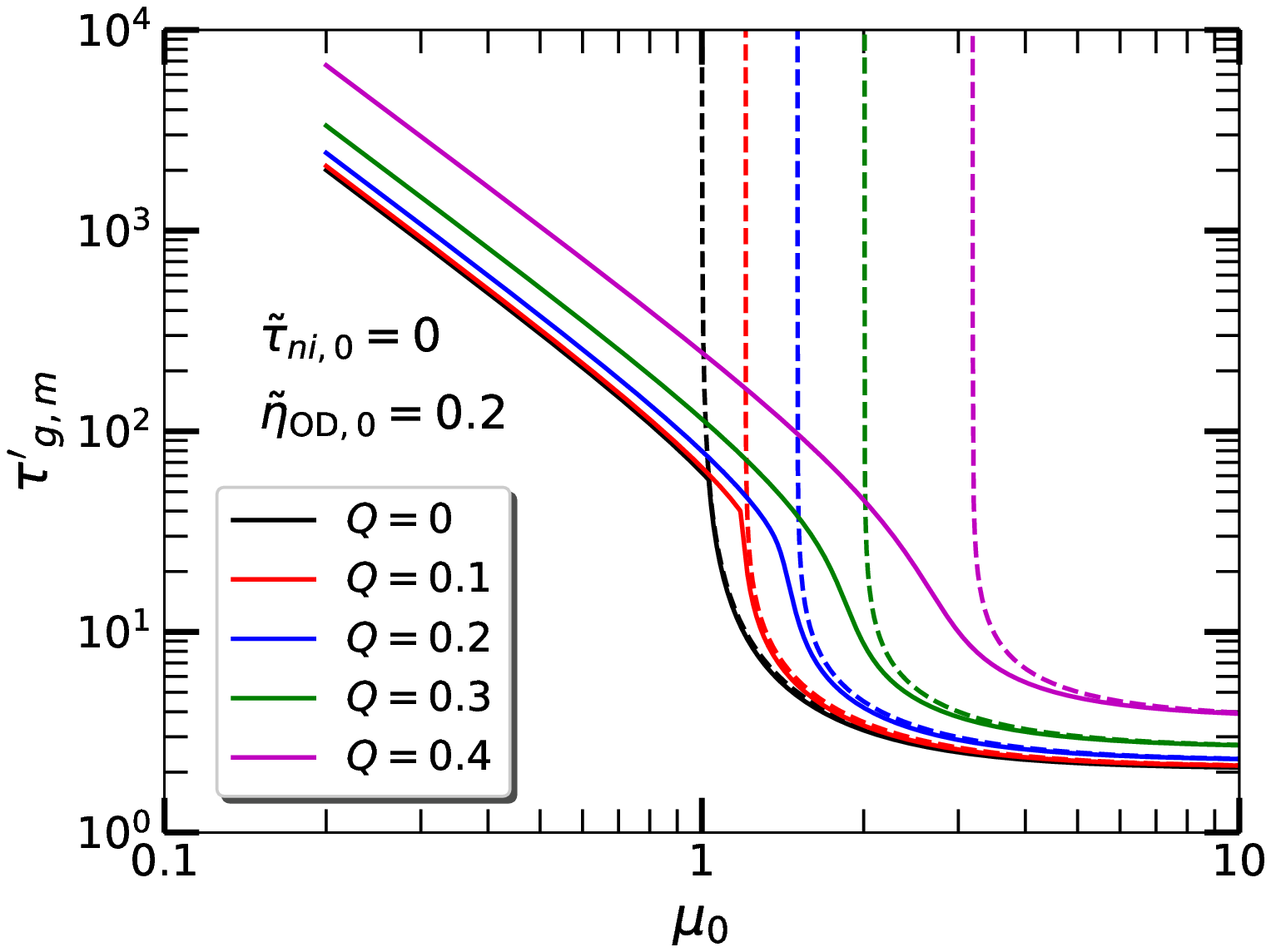}{0.45\textwidth}{(a)}
          \fig{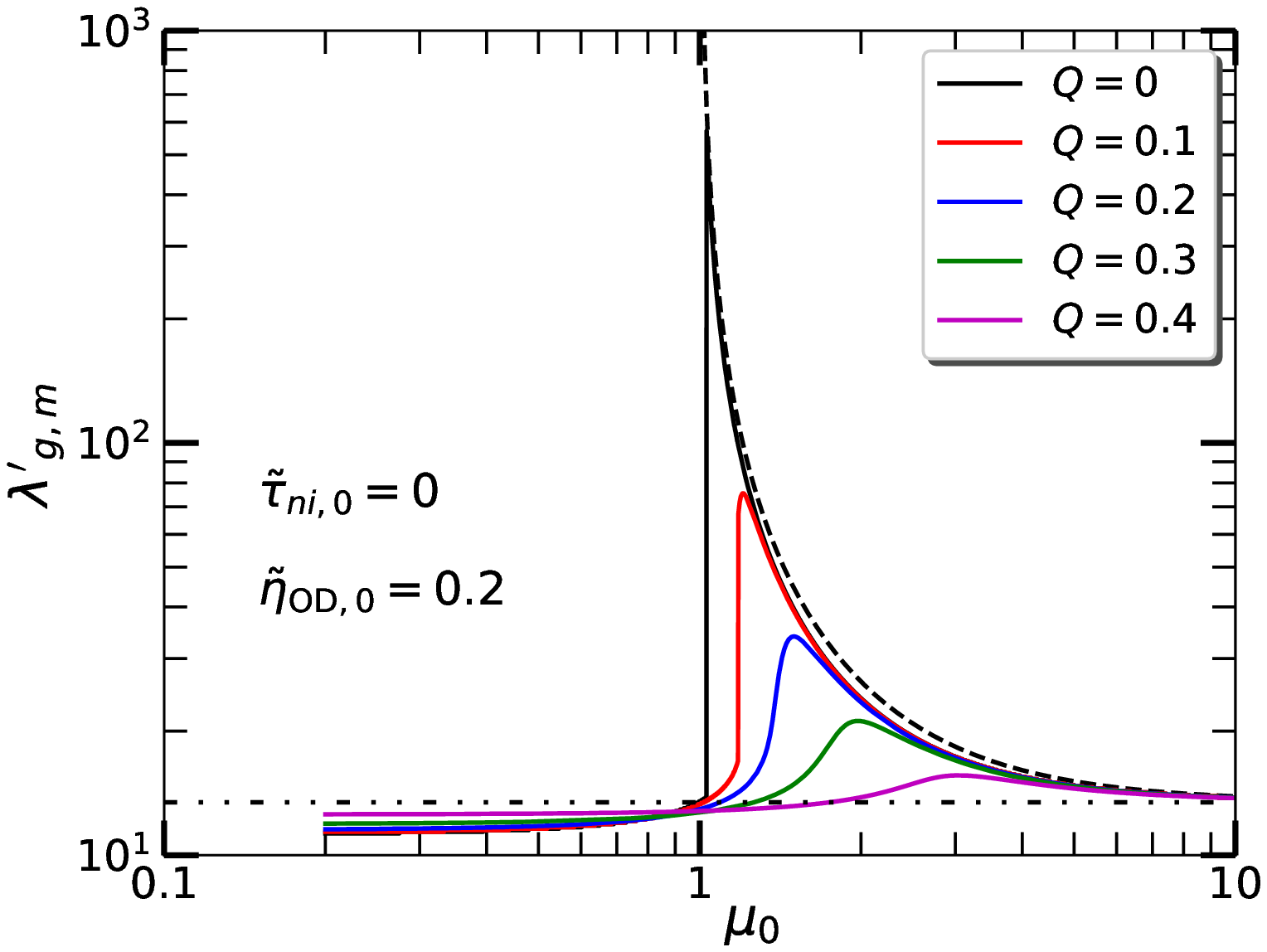}{0.45\textwidth}{(b)}
          }
\caption{Normalized shortest growth timescale ($\tau'_{g,m}=\tau_{g,m}/t_0$), preferred length scale ($\lambda'_{g,m} = \lambda_{g,m}/L_0$) of the most unstable mode as a function of normalized mass-to-flux ratio ($\mu_0 $). 
This represents a model with a fixed normalized Ohmic diffusivity $\etaODt = 0.2$ and without ambipolar diffusion ($\Tilde{\tau}_{ni,0} =0$). Each panel shows preferred timescale and length curves for models with normalized rotation $Q$ = 0 (black),  0.1 (red), 0.2 (blue), 0.3 (green), and 0.4 (magenta). The dashed lines in (a) show the corresponding timescale curves for different $Q$ under the limit of flux-freezing. In (b), the dashed line shows the corresponding length scale curves for all $Q$ under flux-freezing. In (b), the dash-dotted line shows the value $2 \lambda'_{\rm T}$.}
\label{fig:ODandQ}
\end{figure*}


\begin{figure*}[htb!]
\gridline{\fig{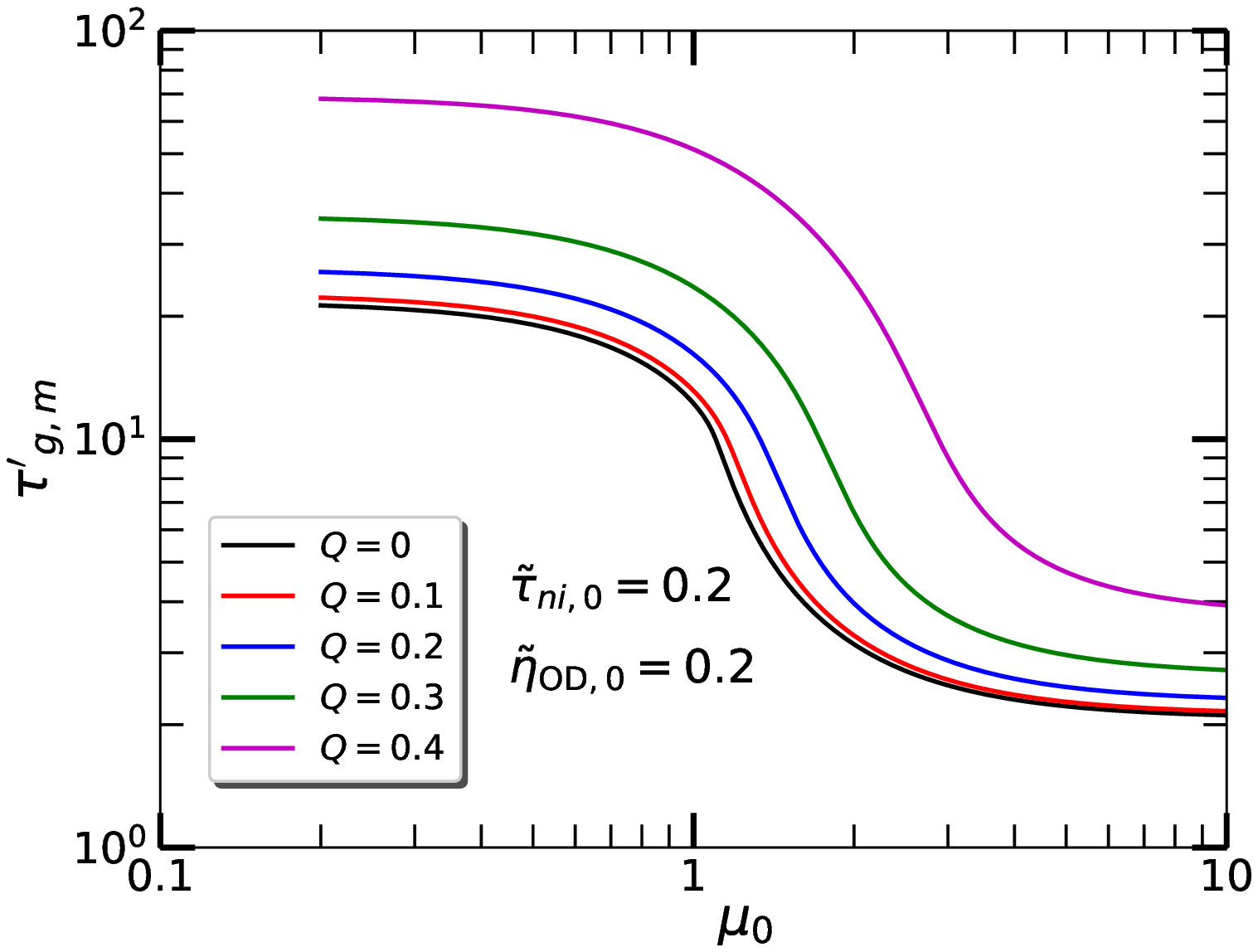}{0.33\textwidth}{(a)}
          \fig{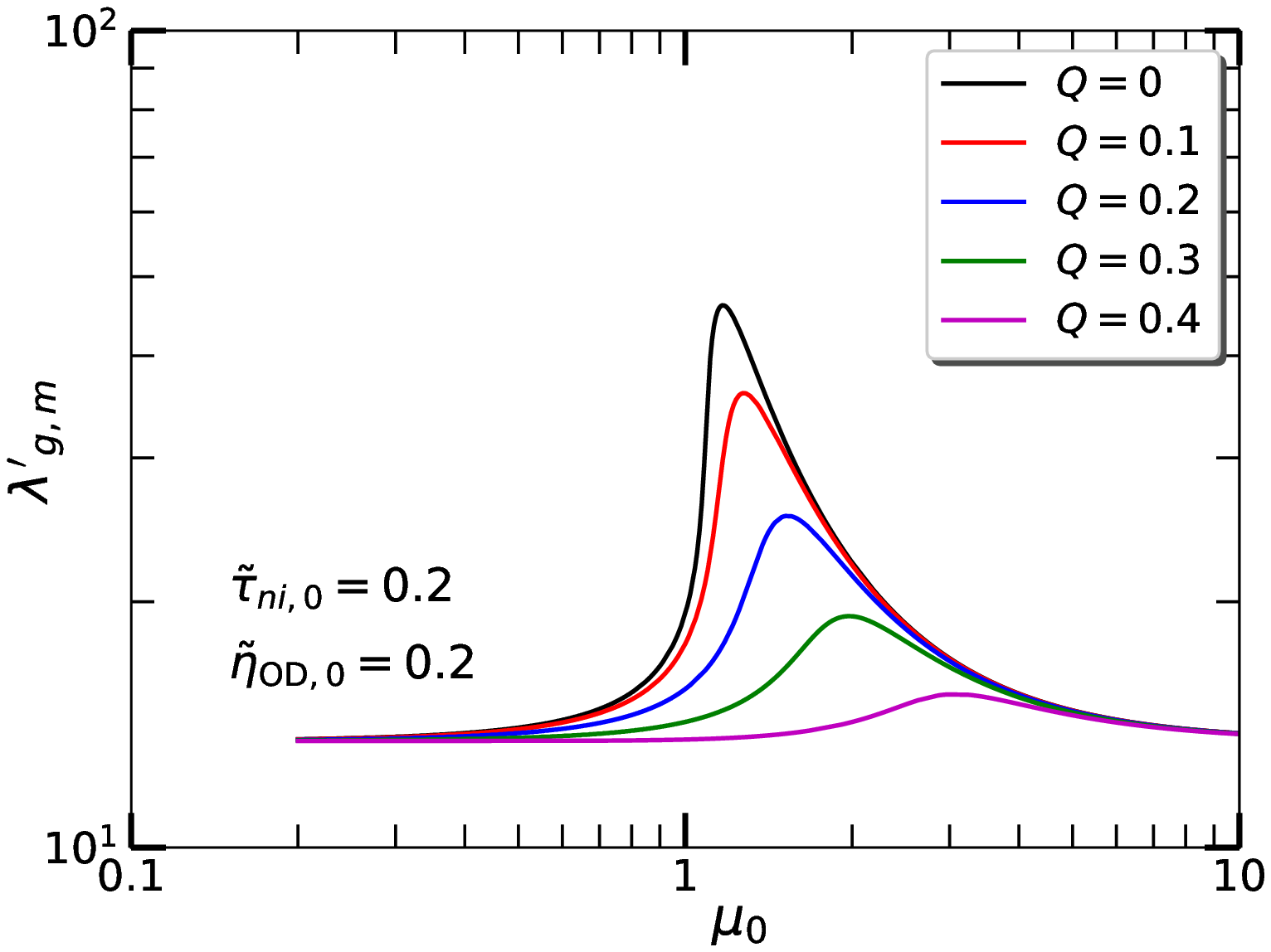}{0.33\textwidth}{(b)}
          \fig{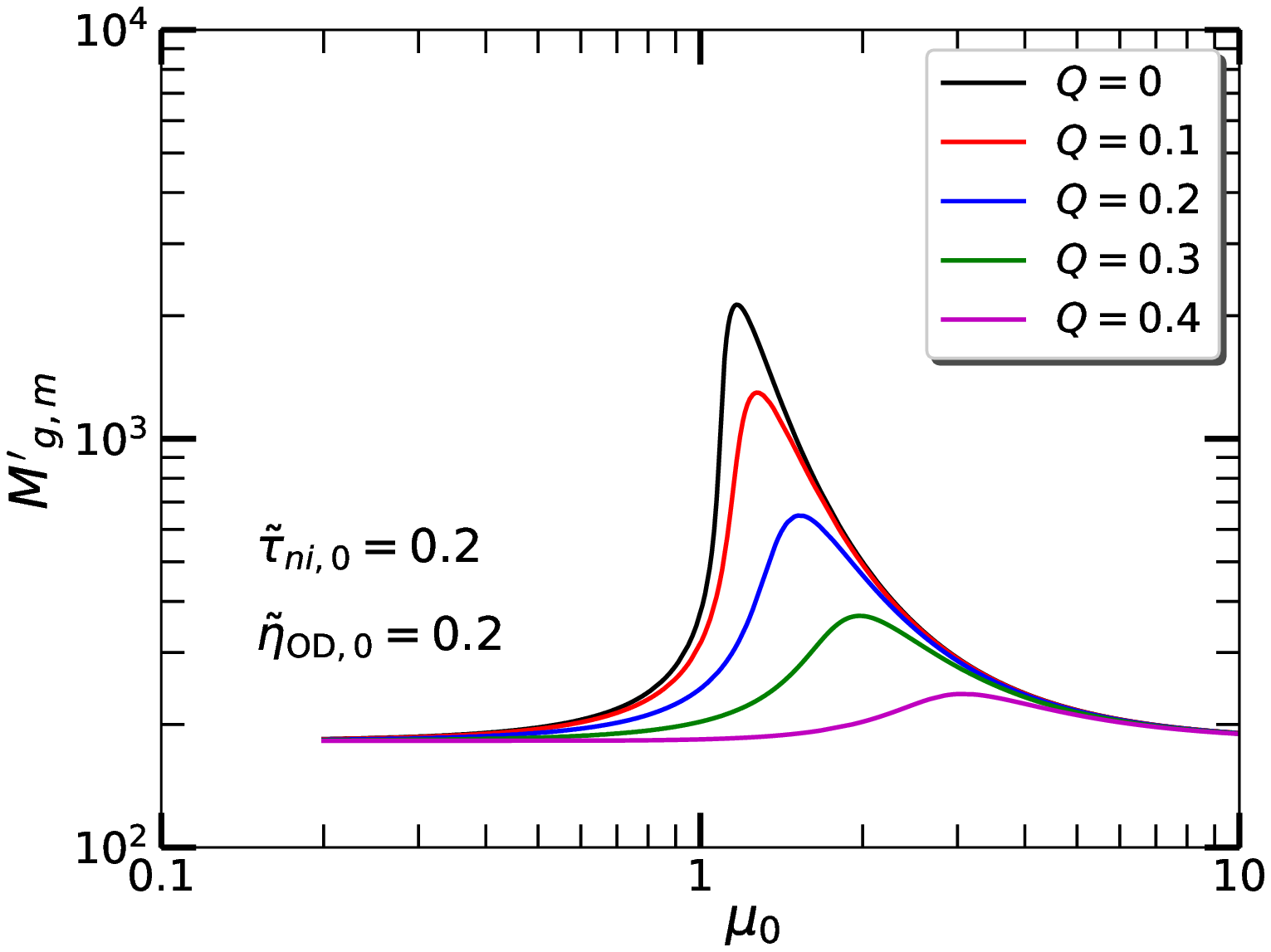}{0.33\textwidth}{(c)}
          }
\caption{Normalized shortest growth timescale ($\tau'_{g,m}=\tau_{g,m}/t_0$), preferred length scale ($\lambda'_{g,m} = \lambda_{g,m}/L_0$) and preferred fragmentation mass ($M'_{g,m} = M_{g,m}/M_0$) of the most unstable mode as a function of normalized mass-to-flux ratio ($\mu_0 $).
This model is shown for a fixed normalized Ohmic diffusivity ($\etaODt = 0.2$) and normalized neutral-ion collision time ($\tilde{\tau}_{ni,0} = 0.2$) corresponding to $n_{n,0} = 10^{11} \> \rm{cm}^{-3}$ and $T=30\, {\rm K}$. Each panel shows instability curves for different normalized rotation $Q$ = 0 (black), 0.1 (red), 0.2 (blue), 0.3 (green), and 0.4 (magenta).}
\label{fig:AD_OD_Q}
\end{figure*}

\autoref{fig:ODandQ} shows the shortest growth timescale of the gravitationally unstable mode and corresponding length scale as a function of the critical mass-to-flux ratio ($\mu_0$) for a rotationally-supported protostellar disk in a regime with Ohmic dissipation (OD) only. We study the case of $\etaODt = 0.2$ corresponding to neutral number density  $n_{n,0} = 10^{11} \> \rm{cm}^{-3}$. In \autoref{fig:ODandQ}(a), we see that the minimum growth timescale of the disk becomes longer with higher rotation. This indicates that rotation is providing more support together with the magnetic field and thermal pressure against the inward gravitational pull. Overall, the reasoning behind this kind of trend in the timescale and the length scale curves has been explained while discussing \autoref{fig:ADonly_ODonly}. We find that adding rotation to the flux-freezing case yields different minimum growth timescale curves for each different rotation rate as shown by the dotted lines in \autoref{fig:ODandQ}(a). In the highly supercritical regime, the growth timescale for each different rotation rate belongs to a different thermal collapse time for each different rotation. See \autoref{eq:normtaug} and \ref{eq:normtauthermal} for the calculation. The dynamical time obtained with a higher rotation is longer than that with smaller rotation.

\autoref{fig:ODandQ}(b) shows that the preferred wavelength becomes smaller with higher rotation, since the rotation stabilizes the longer length scales. Note that as rotation increases, each respective peak preferred wavelength is gradually shifted to a larger $\mu_0$. This is because for an increased $Q$, the disk attains more support from rotation and becomes more stable against the self-gravitational collapse. The field lines are not dragged in as much, and the (restorative) effect of magnetic field curvature is maximized at progressively greater $\mu_0$, where gravity is more dominant. This causes the peak of $\lambda'_{g,m}$ to move to greater values of $\mu_0$, but have decreased value, as $Q$ increases.  
Furthermore, we see that for higher rotation $\lambda'_{g,m}$ becomes larger than that for smaller rotation in the regime $\mu_0<1$, in contrast to its trend in the regime $\mu_0 \geq 1$. 
Since rotation helps to stabilize the smaller length scales, an increment in rotation pushes the lower limit of unstable wavelengths to a larger value. So, the shortest growth time occurs at a relatively larger wavelength for a higher $Q$ in the regime $\mu_0 < 1$. 
We find that adding rotation to the flux-freezing case yields the exactly same preferred length scale curve for each different rotation rate as shown by the black dotted line in \autoref{fig:ODandQ}(b), which is the same as for the $Q=0$ case. 
We found that the preferred wavelength is independent of $Q$ for the flux-frozen case (see \autoref{eq:normlambdag} and \ref{eq:normlambdathermal}).
The black dash-dotted line in \autoref{fig:ODandQ}(b) shows $2 \lambda'_{\rm T}$, as discussed in \autoref{fig:ADonly_ODonly}.

In \autoref{fig:AD_OD_Q}, we present a more realistic case of a rotationally-supported protostellar disk in the hybrid regime where OD and AD are both active. \autoref{fig:AD_OD_Q}(a) and \ref{fig:AD_OD_Q}(b) show the minimum growth time of the gravitationally unstable mode $\tau'_{g,m}$ and the corresponding length scale, $\lambda'_{g,m}$, respectively, as a function of $\mu_0$. These are shown for the density $n_{n,0} = 10^{11} \> \rm{cm}^{-3}$, with specific values of normalized Ohmic diffusivity $\etaODt = 0.2$ and normalized neutral-ion collision time $\tilde{\tau}_{ni,0} = 0.2$, and adopting different rotational speeds. Interestingly, in such high density regions, the effects of AD are still present and cannot be neglected. Because of the AD, the minimum growth timescale curves get plateaued in the subcritical regime, which are indefinitely long for OD only, as seen in \autoref{fig:ODandQ}(a). On the whole, the behavior of $\lambda'_{g,m}$, as shown in \autoref{fig:AD_OD_Q}(b), looks similar to the previously discussed length scale plots. The presence of these two nonideal MHD effects together reduces the preferred length scale by an order of 10 as compared to \autoref{fig:ODandQ}(b) for the case of no rotation ($Q=0$). In the subcritical regime, as the magnetic diffusion becomes strong in the presence of AD and OD together, the shortest growth times occur at the same preferred wavelength (that is similar to that of thermal collapse) regardless of the level of rotation.

Continuing with these parameters, \autoref{fig:AD_OD_Q}(c) shows the normalized preferred mass $M'_{g,m} = M_{g,m}/M_0$ corresponding to fastest growing mode as a function of normalized mass-to-flux ratio ($\mu_0$) for different values of rotation. We see that the preferred mass for collapse exceeds the Jeans mass by a factor of up to 10 when including OD and/or AD. The influence of the magnetic field on the preferred mass of the most unstable mode can essentially lead to the concept of a modified threshold for the fragmentation mass, as opposed to the Jeans mass alone. This can allow a step forward to the understanding for the formation of clumps within a protostellar disk in the early embedded phase.




\begin{figure*}[ht!]
\gridline{\fig{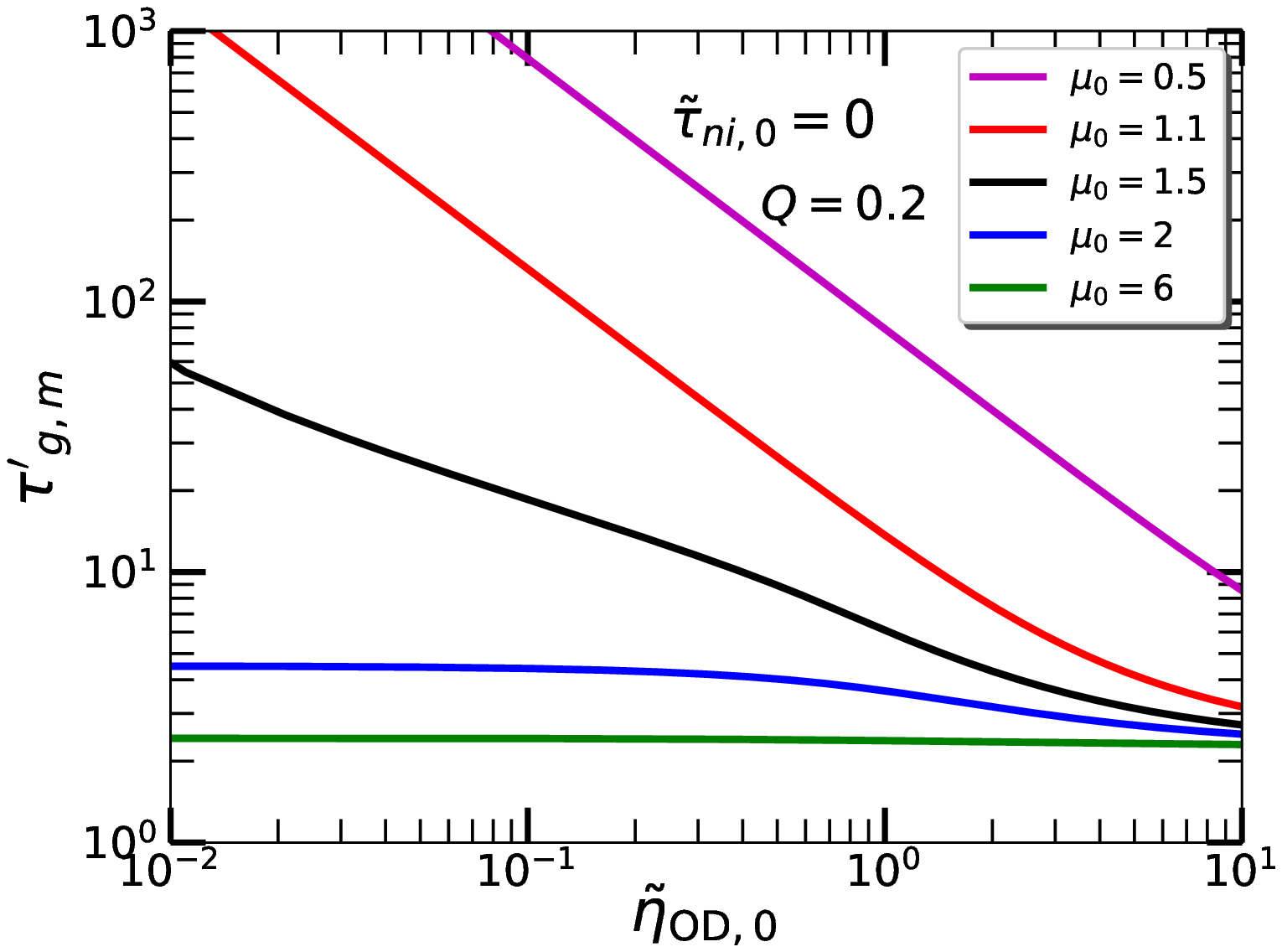}{0.245\textwidth}{(a)}
          \fig{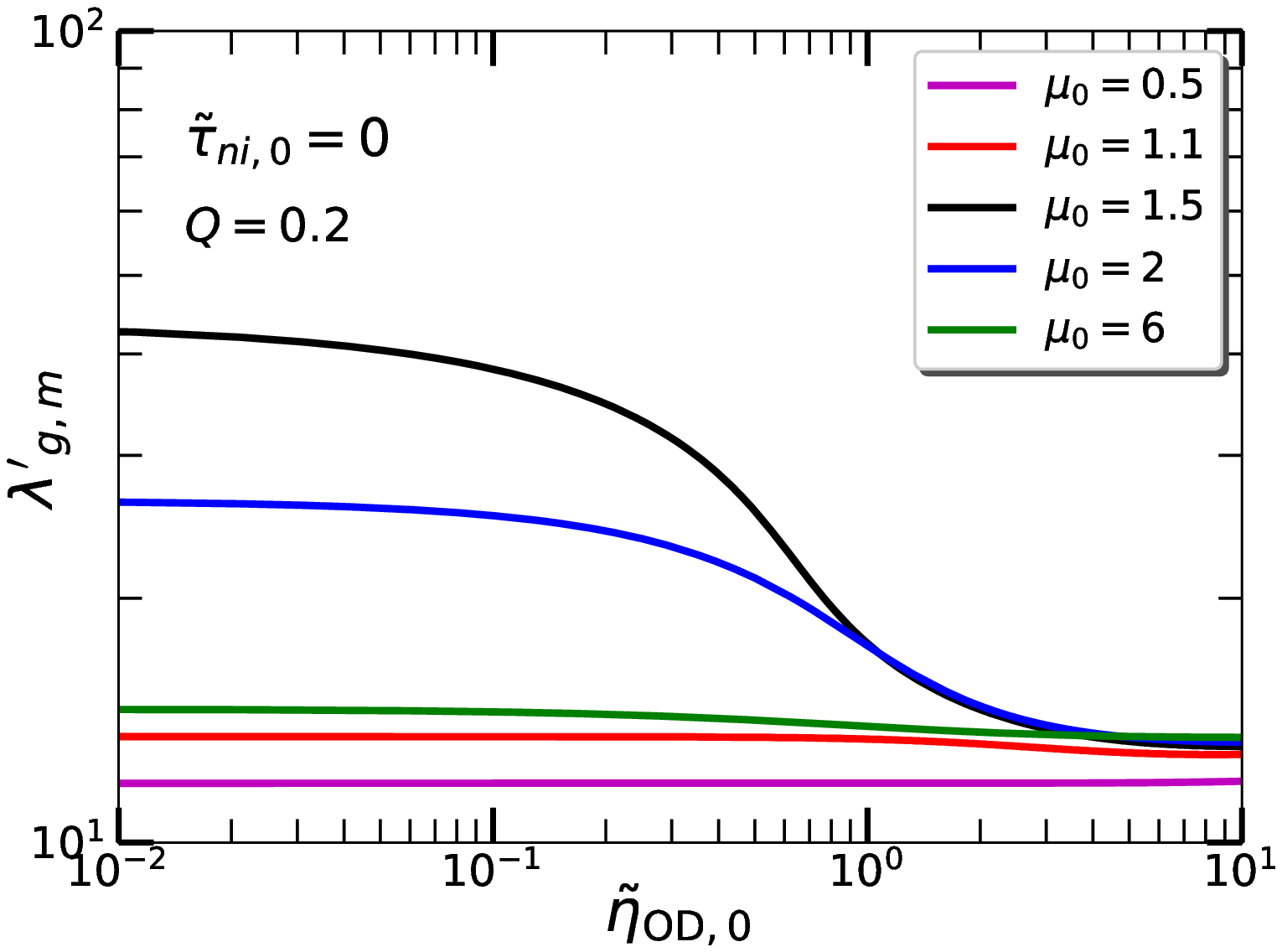}{0.245\textwidth}{(b)}
          \fig{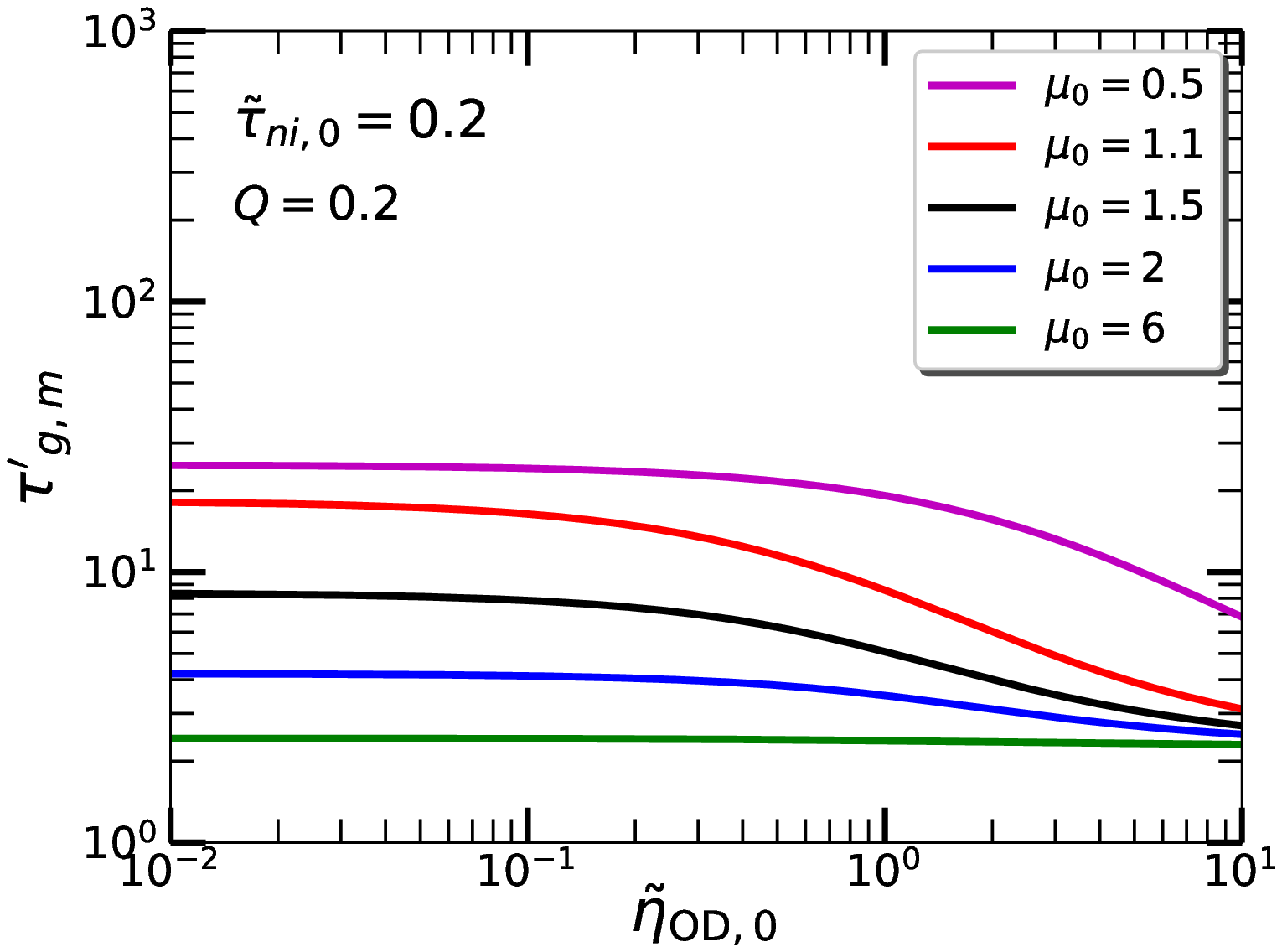}{0.245\textwidth}{(c)}
          \fig{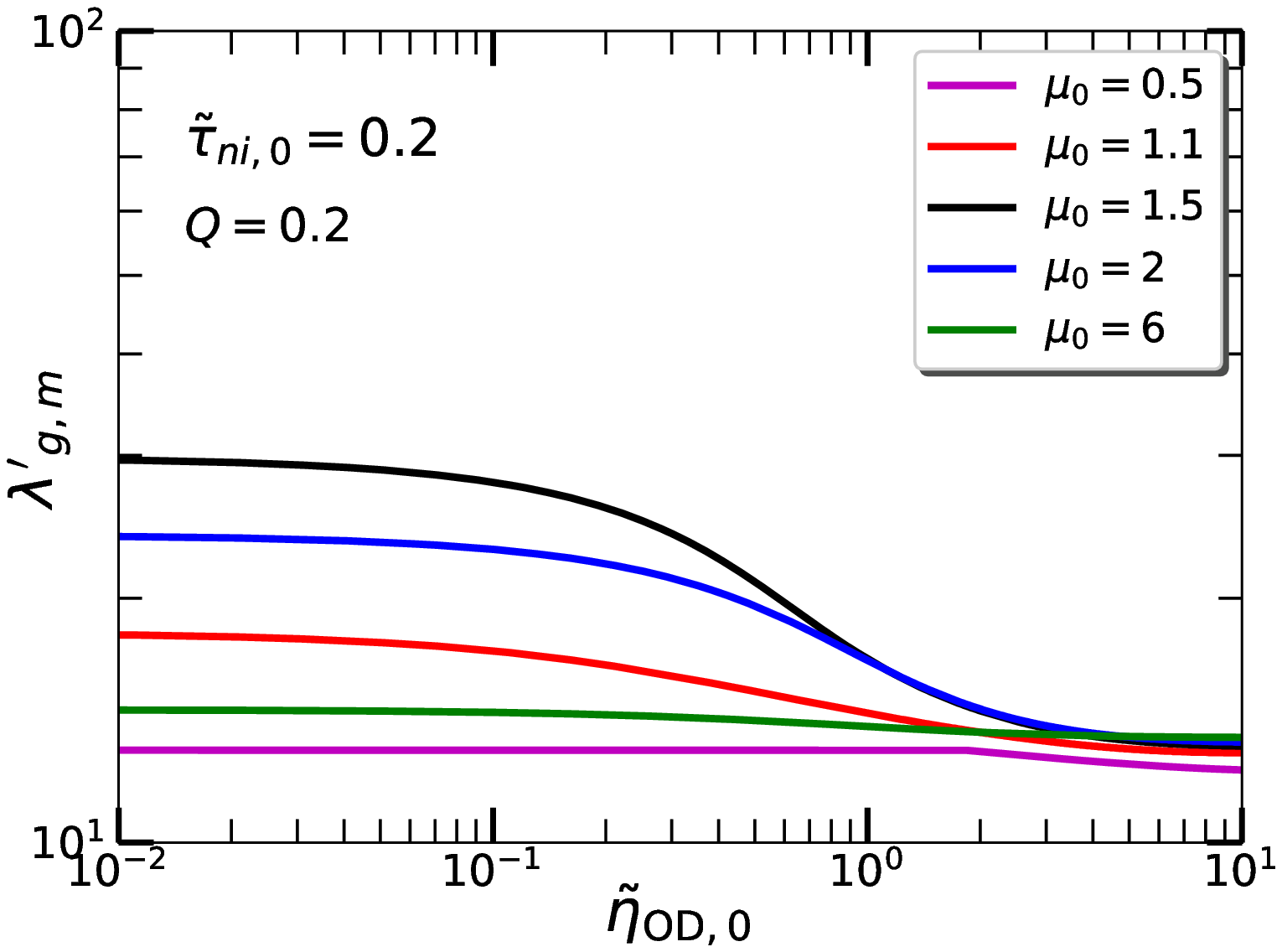}{0.245\textwidth}{(d)}
          }
\caption{Normalized shortest growth time of gravitationally unstable mode ($\tau'_{g,m} = \tau_{g,m}/t_0$) and normalized preferred length scale of most unstable mode ($\lambda'_{g,m} = \lambda_{g,m}/L_0$) as a function of normalized Ohmic diffusivity ($\etaODt$) for a fixed normalized rotation $Q=0.2$.
Fig (a) and (b) show the model for (\romannumeral 1) $\tilde{\tau}_{ni,0} = 0$. 
Fig (c) and (d) show the model for (\romannumeral 2) $\tilde{\tau}_{ni,0} = 0.2$.}
\label{fig:functionofeta}
\end{figure*}



\begin{figure*}[ht!]      
\gridline{\fig{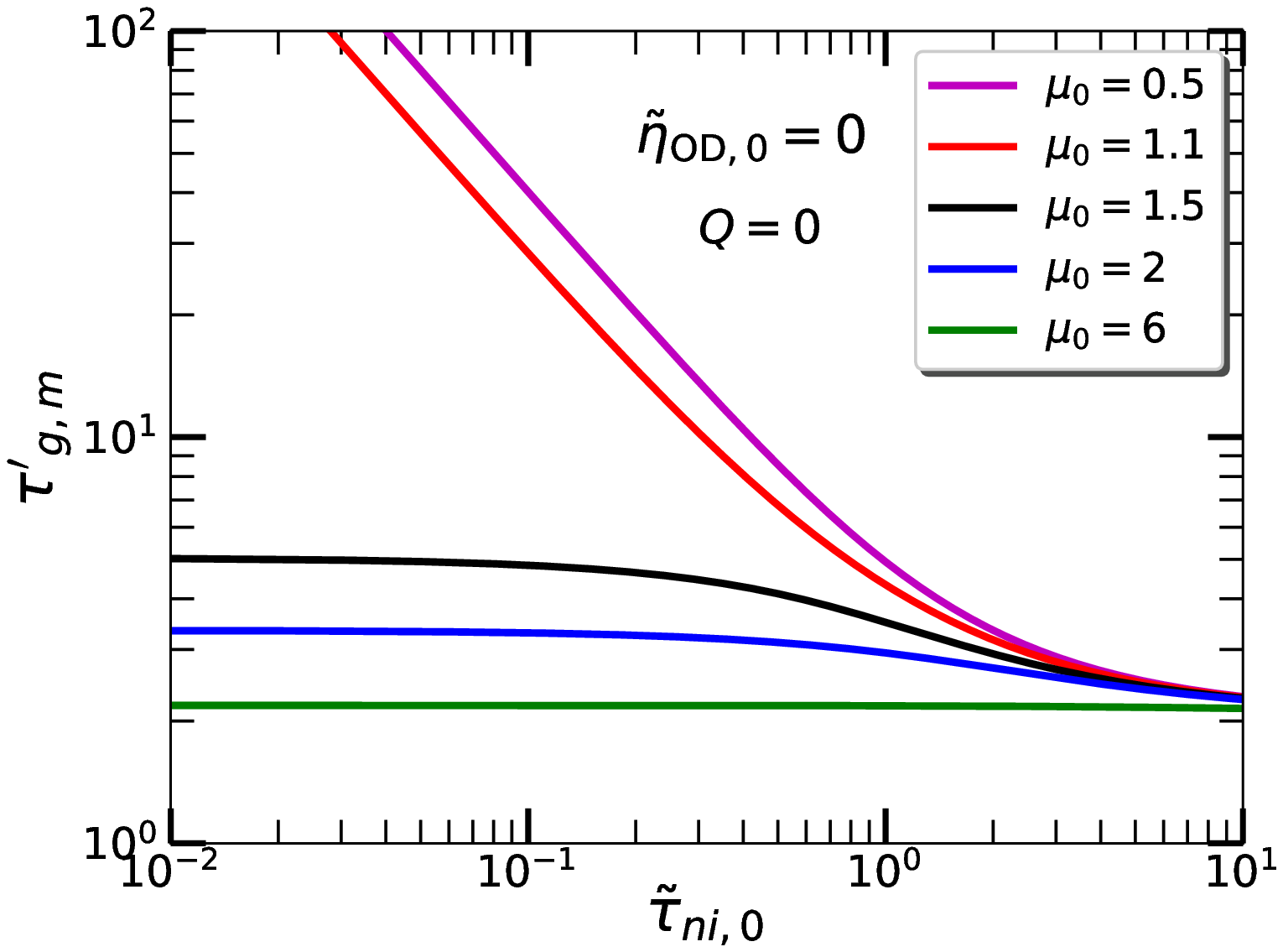}{0.245\linewidth}{(a)}
          \fig{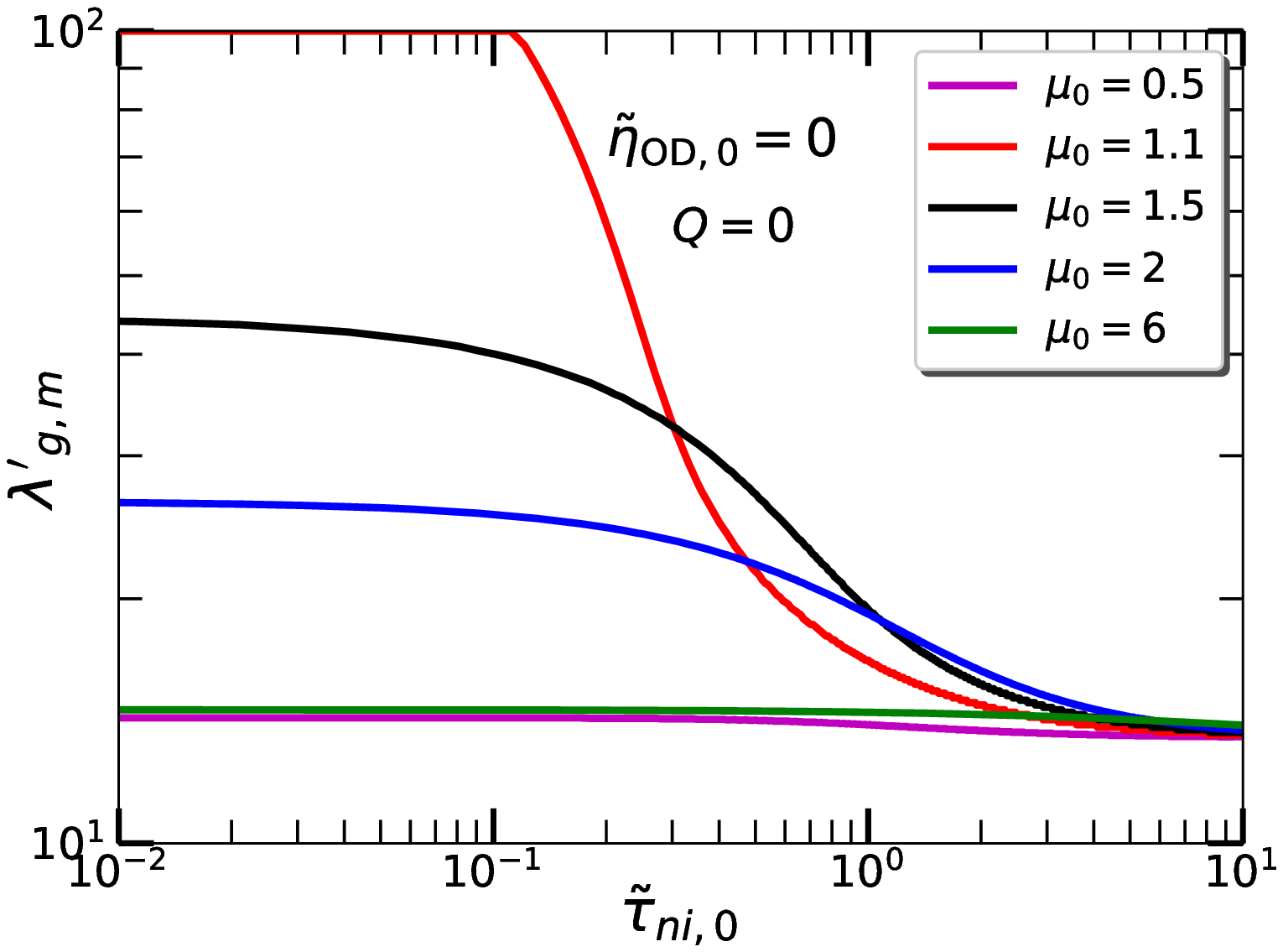}{0.245\linewidth}{(b)}
          \fig{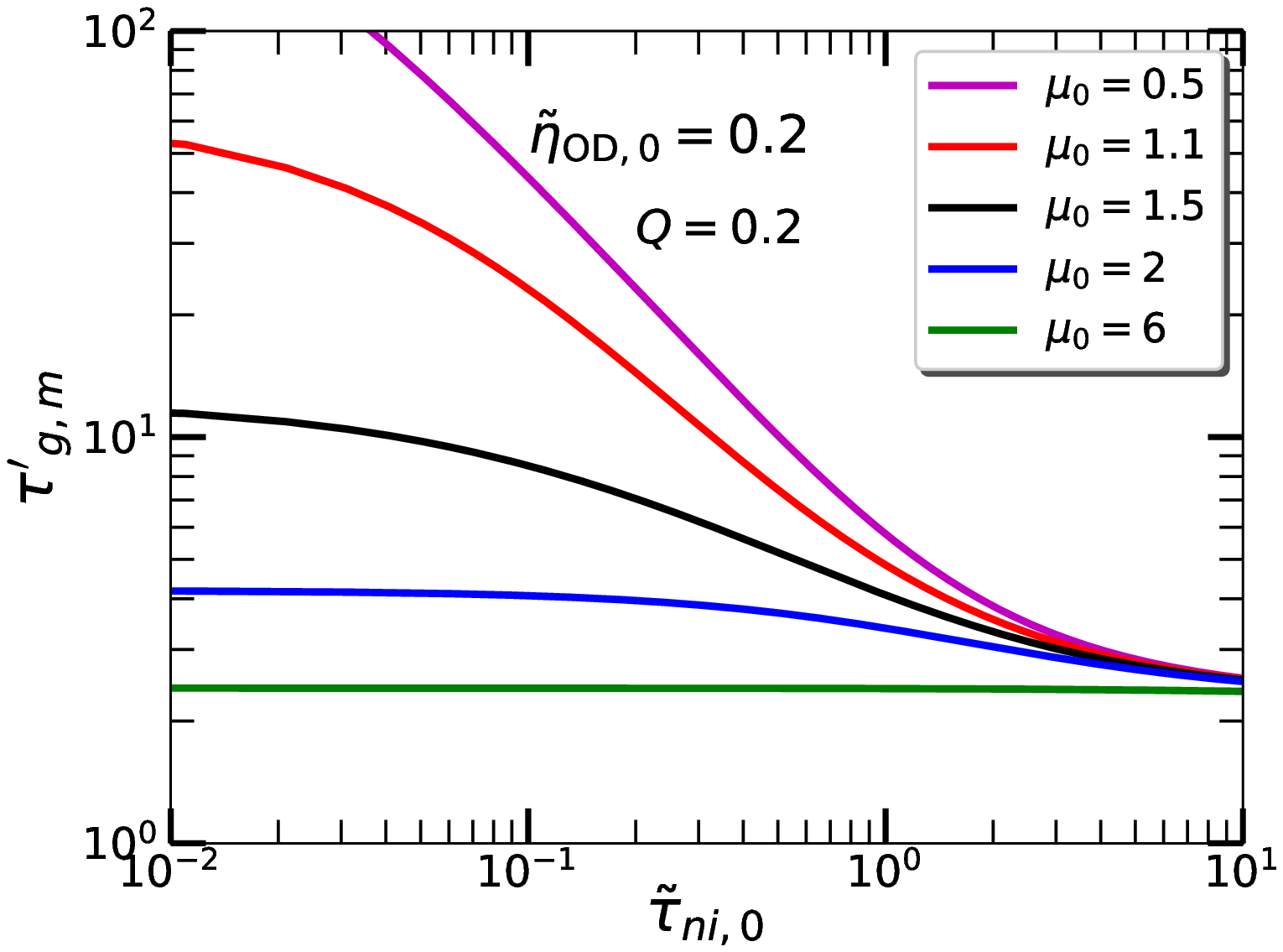}{0.245\textwidth}{(c)}
          \fig{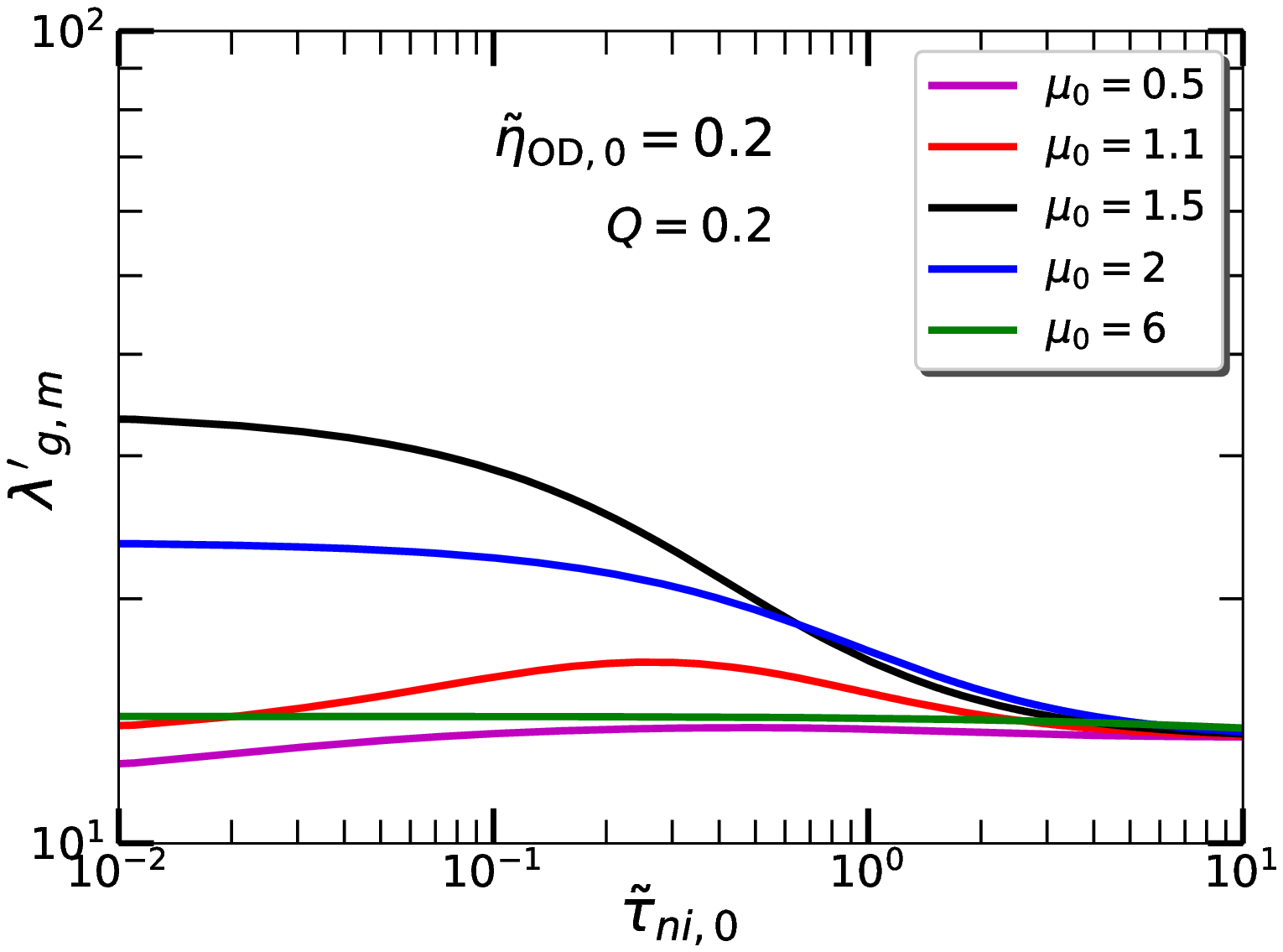}{0.245\textwidth}{(d)}
          }
\caption{Normalized shortest growth time of gravitationally unstable mode ($\tau'_{g,m} = \tau_{g,m}/t_0$) and normalized preferred length scale of most unstable mode ($\lambda'_{g,m} = \lambda_{g,m}/L_0$) as a function of normalized neutral-ion collision time ($\tilde{\tau}_{ni,0}$). Fig (a) and (b) show the model for (\romannumeral 1) $\etaODt = 0$, $Q = 0$.
Fig (c) and (d) show the models for (\romannumeral 2) $\etaODt = 0.2$, $Q = 0.2$.}
\label{fig:functionoftaoni} 
\end{figure*}

\subsection{$\tau'_{g,m}$ and $\>$ $\lambda'_{g,m}$ as Functions of the Diffusion Parameters}

We present an alternative way to look at the minimum growth timescale ($\tau'_{g,m}$) and length scale ($\lambda'_{g,m}$) by studying them as a function of diffusion parameters, i.e., Ohmic diffusivity ($\etaODt$) and neutral-ion collision time ($\taunit$), for fixed $\mu_0$.
\autoref{fig:functionofeta} shows $\tau'_{g,m}$ and $\lambda'_{g,m}$ as a function of $\etaODt$ with a finite rotation $Q=0.2$. \autoref{fig:functionofeta}(a) and (b) represent the case of only Ohmic dissipation (OD) over a range of $\etaODt$ from 0.01 to 10. 
\autoref{fig:functionofeta}(a) shows that for a subcritical cloud with $\mu_0 = 0.5$, $\tau'_{g,m}$ falls off with $\etaODt$ in almost a linear fashion. For $\mu_0 \approx 1.1$, the growth time decreases at a much faster rate up until $\etaODt \approx 1$, after which it settles down to the thermal collapse time. For $\mu_0 =2$, the timescale is plateaued for highly ionized regions where the collapse time is longer, while for low ionization fractions the collapse time again attains the thermal collapse time. For a highly supercritical case ($\mu_0 =6$), the cloud is unstable on the thermal timescale since gravity predominates. 

Now, coming to \autoref{fig:functionofeta}(b), for $\etaODt \lesssim 1$, we see that as $\mu_0$ increases from below unity, $\lambda'_{g,m}$ increases from the thermal wavelength ($\lambda'_{\rm T}$) and becomes maximum at $\mu_0 \gtrsim 1$, and then goes back toward $\lambda'_{\rm T}$ for greater values of $\mu_0$. This is due to a sharp resonant-like peak in $\lambda'_{g,m}$ at $\mu_0 \gtrsim 1$ (discussed earlier in \autoref{sec:nonideal_theory_results}).
On the other hand, for $\etaODt \gg 1$, $\lambda'_{g,m}$ drops down toward $\lambda'_{\rm T}$ as the preferred mode is dominated by OD. 

The addition of AD causes a significant reduction in the timescale and length scale curves, as shown in \autoref{fig:functionofeta}(c) and (d). Overall, it depicts the interaction of the field lines with two different magnetic diffusion mechanisms and self-gravity. 
For $\mu_0 \lesssim 1$, \autoref{fig:functionofeta}(c) shows that the growth time becomes shorter by an order of about 100. In contrast, for $\mu_0 = 1.5$ (slightly supercritical cloud) this reduction in timescale is relatively smaller and for $\mu_0 = 2$ it is smaller still. As $\mu_0$ increases to a highly supercritical  value ($\mu_0 =6$), self-gravity dominates and the growth time tends to the thermal timescale. \autoref{fig:functionofeta}(d) shows that for $\etaODt \lesssim 1$ and $\mu_0 =1.5$ or 2, the $\lambda'_{g,m}$ is shortened by a factor of about $1-2$ compared to the case without AD. Whereas, for $\etaODt \gtrsim 1$, all the  $\lambda'_{g,m}$ corresponding to different $\mu_0$ become indistinguishable from each other and merge to the thermal scale. 

\autoref{fig:functionoftaoni} shows the variation of the growth time and wavelength as a function of the neutral-ion collision time $\taunit$. The general trend of timescale and length scale curves shown in \autoref{fig:functionoftaoni}  behaves qualitatively in the same way as it does with regards to $\etaODt$.
However, from a closer look some subtle differences can be seen. For $\mu_0=0.5 \; \rm{and} \> 1$, the timescale curves decrease almost linearly until they reach $\taunit \sim 1$, as seen in \autoref{fig:functionoftaoni}(a) \citep[see also][]{bailey12}. But in \autoref{fig:functionoftaoni}(c), we see that the timescale curves corresponding to these $\mu_0$ values attain the plateau at a much faster rate as they approach toward smaller $\taunit$. Moreover, because of a nonzero rotation, the growth timescale for $\mu_0=1.5$ and 2 becomes a little longer in comparison to that seen in \autoref{fig:functionoftaoni}(a).

Moving to \autoref{fig:functionoftaoni}(d), we notice that the maximum wavelength occurs at $\mu_0= 1.5$ as compared to \autoref{fig:functionoftaoni}(b) where $\mu_0 =1.1$ corresponds to the maximum wavelength. This again shows that rotation provides an enhanced support even in a nearly transcritical regime, because of which the peak preferred length scale is shifted toward a slightly more supercritical region. Furthermore, on the side of high ionization fractions (i.e., $\taunit \approx 0.01$), the length scale curves for $\mu_0 \lesssim 1$ go to the thermal length scale (as discussed earlier in \autoref{sec:statBlimit}). 
Also, the length scale curve for $\mu_0 =0.5$ continues to decrease more rapidly than that for $\mu_0=1.1$. 
This happens entirely because of stronger magnetic diffusion that is essentially lowering down the length scale toward the thermal length scale.
Lastly, coming to $\mu_0=6$; being highly supercritical it evolves on the thermal length scale and timescale irrespective of any magnetic effects and the adopted rotation.


\subsection{Critical Limit of the Generalized Toomre Criterion}\label{sec:dis}


\begin{figure*}[ht!]
\gridline{\fig{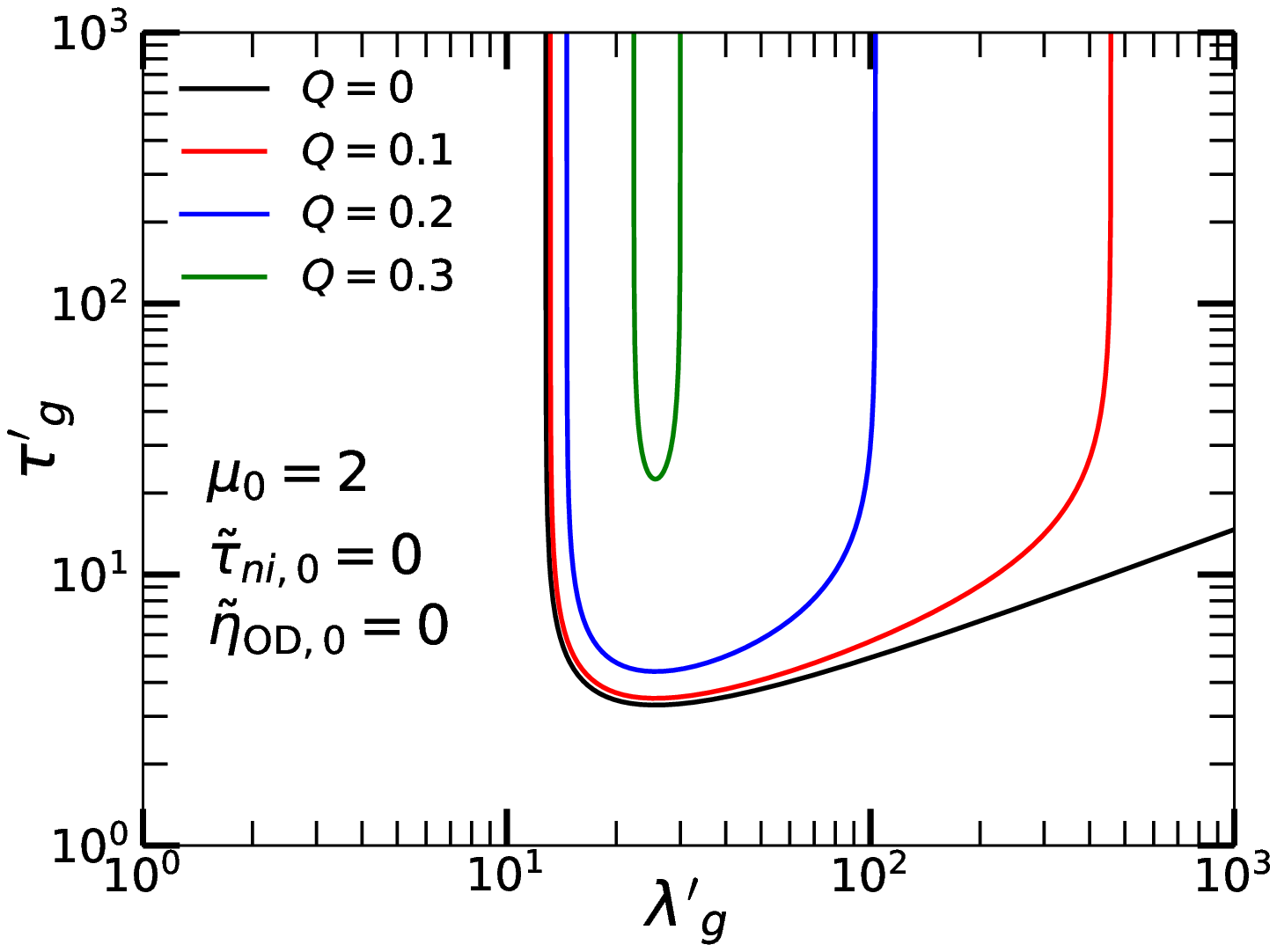}{0.245\textwidth}{(a)}
          \fig{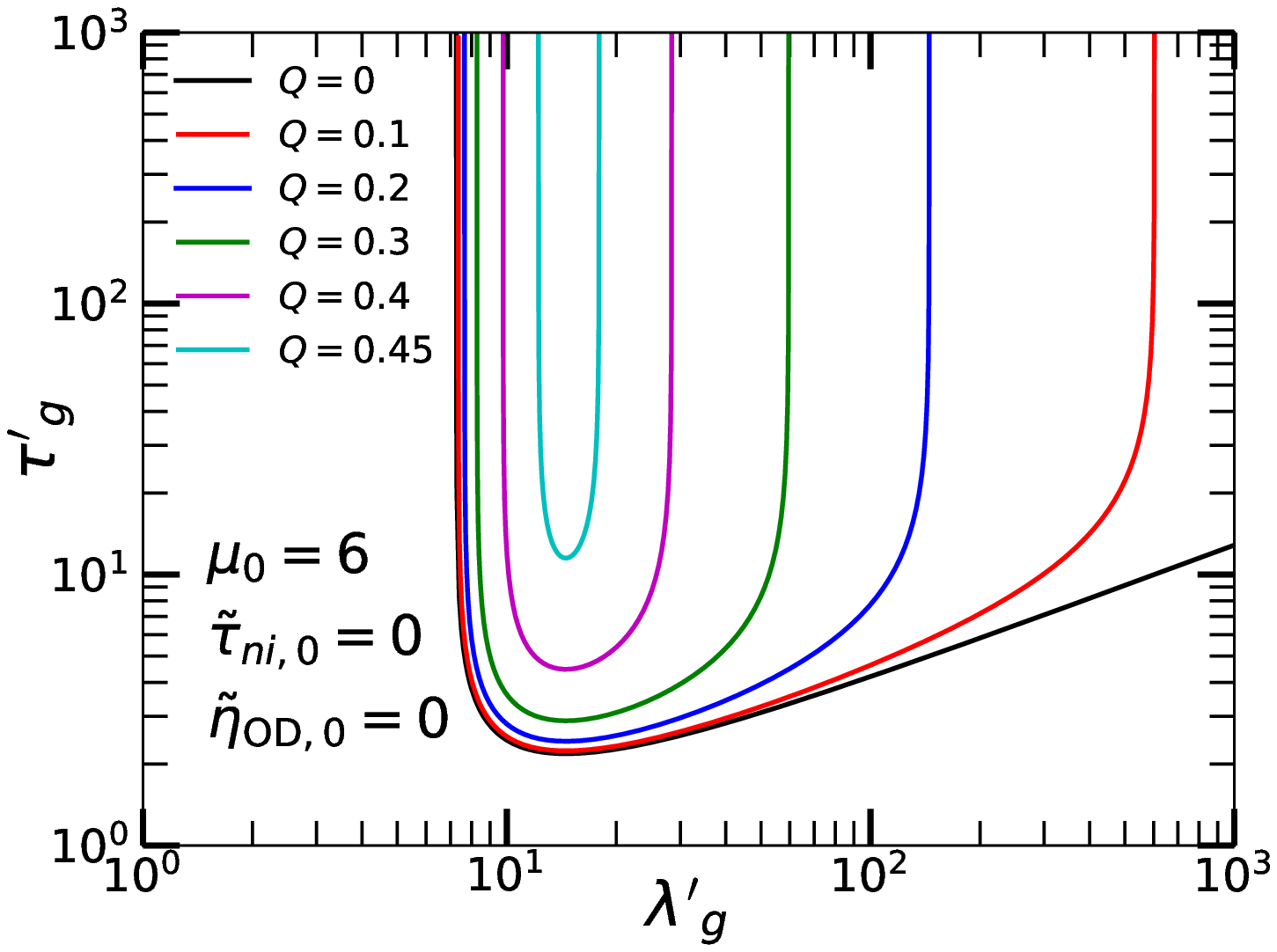}{0.245\textwidth}{(b)}
          \fig{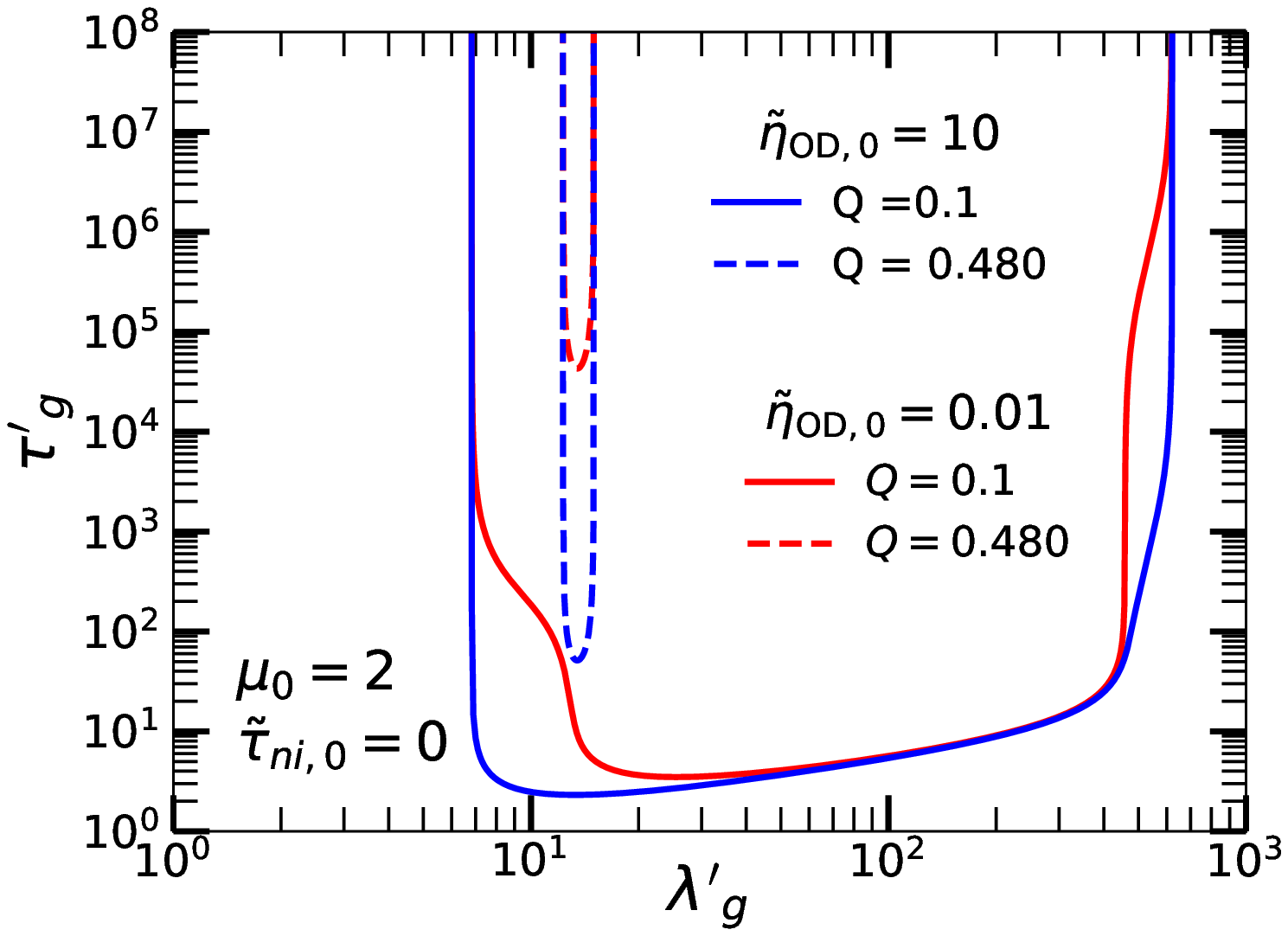}{0.245\textwidth}{(c)}
          \fig{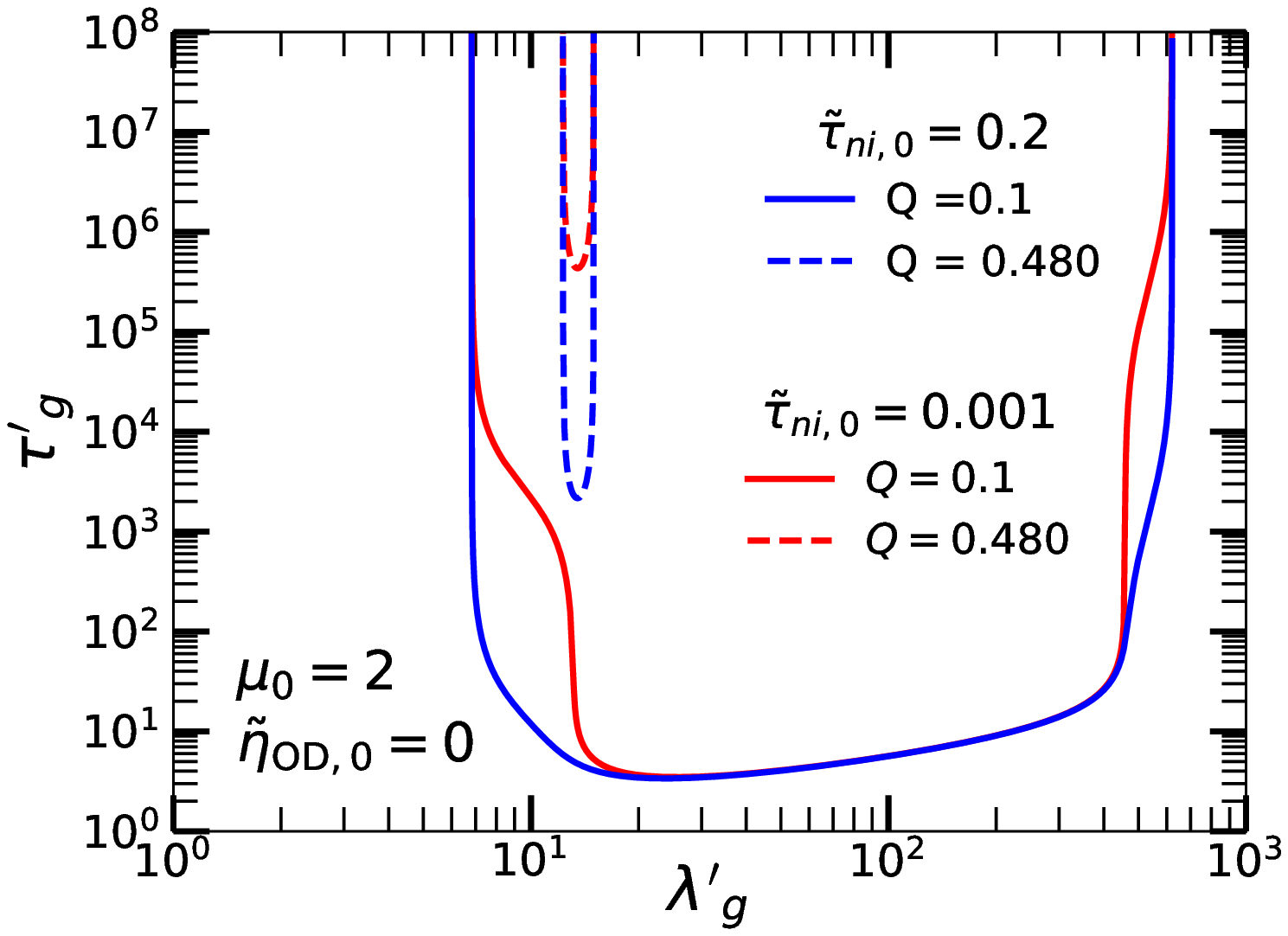}{0.245\textwidth}{(d)}
          }
\caption{Normalized growth time of gravitationally unstable mode ($\tau'_{g}= \tau_{g}/t_0$) as a function of normalized wavelength ($\lambda'_{g}=\lambda_{g}/L_0$) for the following models. Fig (a) and (b) shows flux-frozen model with fixed normalized mass-to-flux-ratio $\mu_0 =2$ and $\mu_0=6$ respectively. In (a) and (b), the timescale curves are shown for models with normalized rotation $Q$ = 0 (black), 0.1 (blue), 0.2 (red), 0.3 (green), and 0.4 (magenta), 0.45 (cyan).
Fig (c) shows timescale curves for $\mu_0 =2$ solely with Ohmic dissipation and $\etaODt = 0.01, 10$ (red and blue, respectively). Fig (d) shows timescale curves for $\mu_0 =2$ solely with ambipolar diffusion [$\tau_{ni,0} = 0.001, 0.2$ (red and blue, respectively)]. Each one of (c) and (d) shows timescale scale curves for $Q = 0.1, 0.480$ (solid line and dashed line for respective color). See also \autoref{tab:Qcritical}.}
\label{fig:Qcritical}
\end{figure*}

\begin{table}[ht!]
\caption{Feasible Range of $Q$ in Different Regimes}
\flushleft
\begin{tabular}{|cc|c|c|c|c|}
\hline 
$\Tilde{P}_{\rm{ext}}$ & $\tilde{C}_{\rm{eff}}$ & Regime & Nonideal  & $\mu_0 $ & $Q_{\rm crit, m}$ \\
& & & MHD  & $=( 1/\tilde{B}_{\rm{ref}})$ & \\
& & & parameters & & \\
\hline 
0 & 1 & HD & - & $\infty$ & 0.5 \\ 
\hline
0.1 & 1.037 & HD & - & $\infty$ & 0.482 \\ \cline{3-6}
 & & FF & - & 6 & 0.459 \\
 & & & - & 5 & 0.449 \\
 & & & - & 4 & 0.44 \\
 & & & - & 3 & 0.40  \\
 & & & - & 2 & 0.304 \\
 & & & - & 1.1 & 0.054 \\ \cline{3-6}
 & & OD & $\etaODt = [0.01,$ & all & 0.482 \\
 & & & $0.1,1,10]$ & & \\
  & & AD & $\tilde{\tau}_{ni,0} = [0.001,$ & all & 0.482 \\
  & & & $0.04,0.2]$ & & \\
  \hline
\end{tabular}
\label{tab:Qcritical}
\tablecomments{Generalized Toomre Instability criterion ($Q < Q_{\rm crit,m} $), see \autoref{fig:Qcrit}, \autoref{eq:Q_crit_mag}. see also \autoref{fig:Qcritical} for nonideal MHD cases. \\
Here, HD, FF, OD, and AD present Hydrodynamic, Flux-frozen, ohmic dissipation, and ambipolar diffusion, respectively.}
\end{table}

We introduced the effect of the magnetic field on Toomre's instability criterion and obtained an analytic expression in the flux-freezing limit (see \autoref{sec:toomreQeff}). Here, we investigate the dependence of the nonideal MHD effects on the rotation parameter by analyzing our numerically obtained results. 

\autoref{fig:Qcritical} shows normalized growth timescale as a function of length scales in different MHD regimes (ideal and nonideal). In \autoref{fig:Qcritical}(a) and (b), under flux-freezing condition, we see unstable growing modes can be obtained for a larger $Q$ with $\mu_0=6$ than that with $\mu_0=2$.
As the cloud becomes more supercritical, the feasible instability range of $Q$ expands with the increase of $\mu_0$ until it merges with that of the hydrodynamic case (see \autoref{tab:Qcritical} and \autoref{fig:Qcrit}). 
When we incorporate any of the nonideal MHD effects (either $\etaODt$ or $\taunit$), growing modes can be obtained almost for the entire feasible range of $Q$ as seen in \autoref{fig:Qcritical}(c) and (d). These two cases are shown for a slightly supercritical value, $\mu_0 =2$, to provide a better comparison with respect to \autoref{fig:Qcritical}(a).
After adding a very small Ohmic diffusivity ($\etaODt=0.01$) or neutral-ion collision time ($\taunit=0.001$), we obtain an unstable mode even for $Q=0.480$ when $\mu_0=2$, unlike the flux-frozen case in which $Q=0.480$ corresponds to stable modes. For a fixed value of $Q$ at which an unstable mode exists in the flux-frozen case, introducing a non-zero $\etaODt$ or $\taunit$ reduces the growth time of the instability as compared to the flux-frozen case. 
In our model, an unstable mode occurs for $Q < 0.482$ ($=1/(2 \tilde{C}_{\rm{eff},0})$) when we take $\tilde{P}_{\rm{ext}} =0.1$, and for $Q < 0.5$ when $\tilde{P}_{\rm{ext}} =0$.

Therefore, in the supercritical regime with magnetic diffusion, the critical instability limit of $Q$ reverts back to the hydrodynamic value.
The growth times and wavelengths of preferred unstable modes for low values of diffusivity are however much longer than for higher values of diffusivity.

\section{Discussion}\label{sec:discussion}

We calculate some typical numbers based on our model for a rotationally-supported protostellar disk. In the high density environment of a protostellar disk, both Ohmic dissipation (OD) and ambipolar diffusion (AD) are expected to be active. Based on the parameters discussed in \autoref{sec:app_params}, we estimate the nonideal MHD parameters using
\begin{equation}
    \tau_{ni, 0}= 3.74 \times 10^4 \> \left(\frac{T}{10 \, \rm{K}}\right)  \left(\frac{0.01 \> \rm{g}\> \rm{cm}^{-2}}{\sigma_{n,0}}\right)^2 \; \left(\frac{10^{-7}}{\chi_{i,0}}\right) \left(1+\Tilde{P}_{\rm{ext}}\right)^{-1} \> \> \rm{yr},
    \label{tauni}
\end{equation}
where the ionization fraction is given by the approximate relation
\begin{equation}
    \chi_{i,0} =  10^{-7} \Bigg(\frac{n_{n,0}}{10^4 \; \rm{cm^{-3}}}\Bigg)^{-1/2},
\end{equation}
and for OD we use
\begin{equation}
\label{etaOD}
\begin{aligned}
    \etaOD = & C_1 \> 1.3 \times 10^{18}  {}  \left(\frac{n_{n,0}}{10^{12}\, \rm{cm}^{-3}}\right) \> \left(\frac{T}{10 \> \rm{K}}\right)^{1/2}\\
                                        & \times \left[1-\tanh \left(\frac{n_{n,0}}{10^{15}\, \rm{cm}^{-3}}\right)\right] \> \> \rm{cm}^2 \> \rm{s}^{-1},
\end{aligned}               
\end{equation}
as used by \cite{machida07} based on calculations by \cite{nakano2002}.
The term within the square brackets of \autoref{etaOD} is a cutoff representing the restoration of flux-freezing at high densities due to thermal ionization. The uncertainties in $C_1$ (adopting a dimensionless scaling parameter whose standard value is unity) hinge largely on the grain properties \citep[e.g.,][]{dapp10}.

The magnetic field is expected to be largely dragged in by the collapse of a prestellar core and there should be significant magnetization at early times. Three-dimensional nonideal MHD simulations that start from a prestellar core show that in the very early embedded phase the disk is only mildly supercritical with normalized mass-to-flux ratio $\mu_0 \gtrsim 3$ if the prestellar core has $\mu_0=3$ \citep{Hirano2020}. Observationally, the magnetic field is difficult to detect directly through the Zeeman effect in small-scale low mass objects like disks \citep{brauer17}. Furthermore, the strength of polarized emission from embedded dust cannot be related to a field strength, but an indirect means such as the DCF method could work in principle. However, this has also proven to be challenging, since the spatially resolved polarization vectors detected at mm wavelengths tend to be dominated by dust scattering instead \citep{kataoka15,kataoka16,yang16,yang16b}. 
Observations of masers in compact high density regions near massive protostars often reveal polarization that can be used to infer the magnetic field strength through the Zeeman effect. \cite{vlemmings2010} used methanol (CH$_3$OH) maser emission around the massive protostar Cepheus A HW2 to infer a large-scale magnetic field of strength $\approx 23$ mG in the $\approx 1000$ AU circumstellar disk. The field direction was also estimated to be nearly perpendicular to the disk elongation and nearly parallel to the observed outflow. The inferred mass-to-flux ratio was $\mu \approx 1.7$. In regions of low-mass star formation, \cite{goncalves08} and \cite{PhilMyers2020} used the indirect means of fitting the magnetic field morphology to determine the mass-to-flux ratios of the protostellar envelopes on $\approx 1000$ AU scales, yielding $\mu \approx 1.7$ and $\mu \approx 1.5$ for NGC 1333 IRAS 4A and BHR71 IRS1, respectively. 

To gain insight into the values of the nonideal MHD coefficients, we refer to figure 4 of \cite{dapp12}, which shows the relative contribution from AD and OD to their respective diffusivity coefficients. 
Their simulation shows that the diffusion coefficients for AD and OD are nearly the same on the scale of the first core at $r \approx 1 \; \rm{AU}$. 
The contribution of OD continues to increase sharply at higher densities ($n_{n,0} \gtrsim 10^{11}\; \rm{cm}^{-3}$) and significantly exceeds the contribution from AD. 
At lower densities, the AD coefficient dominates that due to OD but this does not cause a large flux loss since the dynamical time is less than the diffusion time associated with AD during the runaway collapse phase. 

Based on the simulations of \citet[][see their figure 4]{Vorobyov06} that follow the self-consistent formation of disks from the collapse of prestellar cores, we infer a typical mid-range neutral number density $n_{n,0} \approx 10^{11} \; \rm{cm}^{-3}$ during the early embedded phase of the disk that is characterized by recurrent GI.
This is also the density at which both OD and AD contribute significantly to magnetic dissipation, so it makes an interesting reference point to study. Based also on these simulations \citep{Vorobyov06,vorobyov07}, we estimate a typical disk temperature
$T=30 \; \rm{K}$. These numbers lead to $\rho_{n,0} = 3.90 \times 10^{-13}\, \rm{g} \, \rm{cm}^{-3}$, $\sigma_{n,0} = 59.95 \, \rm{g} \, \rm{cm}^{-2}$, $L_0 = 2.80$  AU, $t_0 = 41.74 \, \rm{yr}$, $M_0 = 5.32 \times 10^{-5}\, M_{\odot}$, and $\chi_i = 3.16 \times 10^{-11}$. See figure 5 of \cite{dapp12} for a comparison of ionization fraction for large number density. 
Using equations (\ref{tauni}) and (\ref{etaOD}), we then find that $\etaOD = 2.25 \times 10^{17} \> \rm{cm}^2 \> \rm{s}^{-1}$ and ${\tau}_{ni,0} = 8.98 \; \rm{yr}$, leading to $\etaODt = 0.2$ and  $\tilde{\tau}_{ni,0} = 0.2$, respectively \citep[see figure 2 of][]{dapp12}.

The adopted surface density $\sigma_{n,0} = 59.95 \, \rm{g} \, \rm{cm}^{-2}$ is congruent with a typical value in simulations of the early embedded class 0 phase \citep[][see their figure 5]{Vorobyov06}. Observationally, \cite{perez2016} estimated a 
surface density $\approx 5 \, \rm{g}\; \rm{cm}^{-2}$ for the disk surrounding Elias 2--27, however that is a class II object representing a later stage of evolution. Our estimated $\sigma_{n,0}$ does not include the inward pressure of an extra vertical squeezing $W_*$ due to the gravity of the central star. This can reduce the value of the surface density for a given volume density. The effect of $W_*$ is calculated quantitatively in \autoref{sec:w_star_app}, and we do not deal with it further here as we are making order of magnitude estimates.

We refer the reader back to \autoref{fig:AD_OD_Q}(c) and note that for the values $Q = 0, 0.1, 0.2, 0.3, 0.4$, the peak preferred modes with minimum growth time occur at $\mu_0 = 1.17, 1.27, 1.52, 1.96, 3.05$, respectively. These correspond to magnetic field values $B_{\rm{ref}} =  83.29, 76.56, 64.06, 49.72, 31.91 \; \rm{mG}$, respectively.  
The normalized ambipolar diffusivity $\etaADt = \zt \mu_0^{-2}\tilde{\tau}_{ni,0} = 0.29, 0.25, 0.17, 0.10, 0.04$, respectively,  for the above mentioned values of $\mu_0$.
These arise from $\etaAD = V^2_{A,0} \tau_{ni,0} = 3.95, 3.35, 2.34, 1.41, 0.58$, in units of $10^{17}\, \rm{cm}^2 \;\rm{s}^{-1}$ for $\mu_0 = 1.17, 1.27, 1.52, 1.96, 3.05$. 
Meanwhile, $\etaODt$ does not explicitly depend on the magnetic field strength.
Finally then, from \autoref{fig:AD_OD_Q}(c), we obtain the peak preferred fragmentation mass $M_{g,m}$ to be $93.13, 56.78, 28.39, 16.13, 10.37$ in units of  $M_{\rm Jup}$
for $Q=0, 0.1, 0.2, 0.3, 0.4$, respectively. 
Here, $M_{g,m} = M'_{g,m} M_c$ and $M_c = \pi \sigma_{n,0} (L_{0}/2)^{2}$ as the perturbation is taken to be circular with radius $L_0/2$. 
For a typical disk temperature $T=30 \, \rm{K}$ and neutral number density $n_{n,0} = 10^{11} \, \rm{cm}^{-3}$, $M_c = 4.18 \times 10^{-5}\, M_{\odot}$.

Protostellar disks in the early embedded class 0 phase can be prone to GI, especially while they are still accreting matter from their surrounding envelope \citep{Vorobyov06}. Resistive MHD simulations also show that the magnetic field that is dragged in from the core collapse leads to mildly supercritical disks in which magnetic dissipation mechanisms are active \citep{Hirano2020}. Hydrodynamic simulations of global disk evolution have established that the $Q$ parameter, although initially derived through a local analysis, has wide ranging applicability to understanding global nonlinear disk evolution \citep[see, e.g.,][]{Vorobyov06,vorobyov07}. For the intermediate regime $1 \lesssim Q \lesssim 2$, small-amplitude fluctuations can persist and lead to meaningful flocculent spiral structure. In the decidedly unstable regime $Q \lesssim 1$, grand design spiral arms are formed and clumps within them can form if the local cooling time is also less than the orbit time \citep{vor10}; this is a criterion on the nonlinear evolution that is not present in an isothermal linear analysis. Giant planet (or other companion) formation by GI can then occur. The effect of the magnetic field on this scenario is just beginning to be explored. Magnetic fields and nonideal MHD lead to a more complex instability criterion including affecting the length scales and timescales of the instabilities, as we have shown in this paper. 
The diffusivities play an important role in setting these quantitatively, and for $\mu_0 > 1$ the preferred modes generally have larger length scales and longer timescales than in the hydrodynamic case. 

Future global simulations of the long term evolution of disks including nonideal MHD will be able to explore the effect of nonideal MHD in clump formation and can potentially use the linear results in this paper as a benchmark. The OD will also introduce important nonlinear effects, since the resistive heating (not present in our linear isothermal analysis) can potentially counteract the surface cooling. In fact, as \cite{lizano2010b} point out, a large amount of OD is required in order to reduce the magnetic flux of disk material to the values inferred from the paleomagnetism of meteorites in our solar system.

\section{\rm{Summary}} \label{sec:summary}
We have studied the effect of ambipolar diffusion (AD) and Ohmic dissipation (OD) on gravitational instability within rotationally-supported protostellar disks, employing a linear analysis. 
Our model clouds are isothermal, partially ionized, thin planar sheets with a finite local vertical half thickness. Here, we highlight several interesting results that emerge. 


We derive generalized criteria of Toomre instability that has a magnetic dependence (see \autoref{sec:toomreQeff}). 
We show that the magnetic field strength influences the critical limit of rotation such that the instability criterion appears as $Q<Q_{\rm crit,m}$. 
In the hydrodynamic limit ($\mu_0 \rightarrow \infty$), $Q_{\rm crit,m}$ reduces to $1/(2 \tilde{C}_{\rm{eff},0})$, which is equivalent to standard Toomre's instability criterion. 
With the magnetic diffusion effects, i.e., AD and/or OD, the value of $Q_{\rm crit,m}$ also reverts back to that of the hydrodynamic case (see \autoref{fig:Qcrit}). 

Subcritical clouds ($\mu_0 <1$)  are stable against gravitational fragmentation in the flux-freezing limit ($\etaODt \rightarrow 0, \tilde{\tau}_{ni,0} \rightarrow 0$). Supercritical clouds ($\mu_0 > 1$) are unable to support themselves against their own gravity and are prone to collapse even in the flux-freezing regime. In that regime, adding rotation helps to stabilize the longer wavelengths to a greater extent (refer to \autoref{fig:fluxfrozen}(b)). However, in the presence of any form of magnetic diffusion (OD or AD), a fastest growing mode of gravitational instability having a minimum growth timescale and an associated preferred length scale can be obtained even for subcritical clouds (\autoref{fig:ADonly_ODonly}). The two nonideal MHD effects reveal qualitatively similar kinds of features in the gravitationally unstable modes, but there are quantitative differences. For highly subcritical clouds the preferred length scale in the AD only case converges to $2\lambda_{\rm T}$, i.e., twice the thermal critical length scale, as in the highly supercritical (i.e., nonmagnetic) limit. For OD it converges to $\lambda_{\rm T}$, the minimum possible wavelength for instability due to the presence of thermal pressure, since the OD-driven modes have stronger affinity for short wavelengths. In this highly subcritical limit, the timescale of the fastest growing OD mode tends to infinity, since the preferred wavelength is converging to $\lambda_{\rm T}$. However, for AD, the diffusivity is proportional to the square of the field strength, and this compensates for the strong magnetic support, and enforces a finite constant drift speed and growth time that is independent of $\mu_0$ for $\mu_0 \ll 1$. In a realistic situation of a partially ionized protostellar disk, OD and AD are simultaneously active, and in this case AD places an upper bound on the timescale of the diffusive-driven instability.

A peak length scale for collapse occurs at transcritical (but slightly supercritical, $\mu_0 \gtrsim 1$) mass-to-flux ratios, but the peak occurs at different values for OD and AD and also depending on the value of the diffusivities. For very high diffusivities, the peak can disappear. The timescale for growth of the transcritical modes is intermediate between the dynamical (free-fall) time and the ambipolar diffusion time. 

The interplay of the effects of two nonideal MHD effects together with rotation in a protostellar disk can be seen in \autoref{fig:AD_OD_Q} (\autoref{sec:disk}). 
Rotation makes the growth timescale longer and the peak preferred length scale becomes shorter because of an additional support from rotation against gravitational collapse. The peak preferred wavelength of instability gradually moves to a larger $\mu_0$ ($\sim 2$) as rotation increases. Furthermore, we find that the peak preferred mass for collapse exceeds the thermal critical (Jeans) mass by a factor of up to 10 when including OD and/or AD.
The peak preferred fragmentation mass is likely to be $\sim 10- 90 \ M_{\rm Jup}$ (see \autoref{sec:discussion}, \autoref{fig:AD_OD_Q}(c)).
This magnetic field dependent mass creates a modified threshold for AD and/or OD driven gravitational fragmentation in the magnetized disks.

The linear analysis we have presented is formally applicable to a local patch within a larger disk-like cloud. The inclusion of rotation, OD, and AD makes the results particularly relevant for protostellar disks. A local analysis of a nonmagnetic rotating cloud yields the usual Toomre criterion, which has proven surprisingly effective in the interpretation of the global evolution of disks that contain significant inhomogeneities. In a similar manner, our results may prove to be useful in the analysis of global nonideal MHD models of disk evolution. Such simulations are in their infancy, and the role of OD and AD in regulating GI and giant planet formation may prove to be crucial. Future simulations have much to explore.

\acknowledgments
We thank the anonymous referee for comments that improved the manuscript. We also thank Sayantan Auddy for his comments. SB is supported by a Discovery Grant from NSERC.


\numberwithin{equation}{section}
\numberwithin{equation}{subsection}
\renewcommand*{\theequation}{%
  \ifnum\value{subsection}=0 %
    \thesection
  \else
    \thesubsection
  \fi
  .\arabic{equation}%
}

\appendix
\section{Units of Defined Parameters} \label{sec:app_params}
The typical values of the units used and other derived quantities are
\begin{equation}
    \sigma_{n,0}= \frac{3.63 \times 10^{-3}}{\left(1+\Tilde{P}_{\rm{ext}}\right)^{1/2}} \> \left(\frac{n_{n,0}}{10^3\, \rm{cm}^{-3}}\right)^{1/2} \> \left(\frac{T}{10\, \rm{K}}\right)^{1/2} \> \>  \rm{g} \> \rm{cm}^{-2}, 
\end{equation}


\begin{equation}
    L_0 = 1.54 \times 10^4\> \left(\frac{T}{10\, \rm{K}}\right)^{1/2} \> \left(\frac{10^3\, \rm{cm}^{-3}}{n_{n,0}}\right)^{1/2}\> \> \left(1+\Tilde{P}_{\rm{ext}}\right)^{1/2} \> \> \rm{AU},
\end{equation}

\begin{equation}
    t_0 = 3.98\>  \times 10^{5} \> \left(\frac{10^3\, \rm{cm}^{-3}}{n_{n,0}}\right)^{1/2} \>\> \left(1+\Tilde{P}_{\rm{ext}}\right)^{1/2} \> \> \rm{yr},
\end{equation}

\begin{equation}
    c_s = 0.188 \> \left(\frac{T}{10\, \rm{K}}\right)^{1/2} \> \> \rm{km} \>\> \rm{s}^{-1}, 
\end{equation}

\begin{equation}
\begin{aligned}
    M_0 = 9.76 \times 10^{-2} \left(\frac{T}{10\, \rm{K}}\right)^{3/2} \; \left(\frac{10^3\,\rm{cm}^{-3}}{n_{n,0}}\right)^{1/2} \left(1+\Tilde{P}_{\rm{ext}}\right)^{1/2}  \> \> \> M_\odot\ ,
    \end{aligned}
\end{equation}

\begin{equation}
\begin{aligned}
    B_{\rm{ref}} = \frac{5.89 \times 10^{-6}}{\mu_0} \left(\frac{n_{n,0}}{10^3\, \rm{cm}^{-3}}\right)^{1/2} \>\left(\frac{T}{10\, \rm{K}}\right)^{1/2} \left(1+\Tilde{P}_{\rm{ext}}\right)^{-1/2}\> \; {\rm G} \ ,
\end{aligned}
\end{equation}

\begin{equation}
    \etaAD = 6.01 \times 10^{21} \frac{\mu_0^{-2}}{(1+ \tilde{P}_{\rm ext})} \left(\frac{T}{10 \> \rm{K}}\right) \left(\frac{10^{-7}}{\chi_{i,0}}\right) \left(\frac{10^3 \, {\rm {cm}^{-3}}}{n_{n,0}} \right) \ \rm{cm}^2 \rm{s}^{-1}.
\end{equation}


\section{collision timescales} \label{sec:coll_time_app}
We use the collision time formula between the different species $s$ and neutrals as computed by \cite{dapp12}, employing the work by \cite{mou1996}. The following expression is the collision time for a charged species $s$ with the neutrals: 
\begin{equation}
    \tau_{sn} = k_{s,\rm{He}} \frac{m_s + m_{\rm{H}_2}}{\rho_n \langle\sigma w \rangle_{s\rm{H}_2}},
\end{equation}
where $\sigma$ is the elastic scattering cross-section for electron-neutral or ion-neutral encounters, and $w$ equals the relative velocity of the charged particle as seen from the rest frame of the neutrals. The angular bracket denotes an average over the velocity distribution function of the charged species. The quantity $k_{s,\rm{He}}$ is a correction factor due to the fact the gas also contains helium.
Helium contributes only a small correction due to its low polarizability as compared to $\rm{H}_2$ \citep[see][]{spitzer1978, mou1996}:

\begin{equation}
\begin{aligned}
    k_{s,\rm{He}} & {} =  1.23 \;\;\; \rm{if} \; s=i, \\
                  & {} =  1.21 \;\;\; \rm{if} \; s=e. \\
\end{aligned}
\end{equation}

The values of the collision rate $\langle \sigma w \rangle_{s\rm{H}_2}$ are \citep{Mott1949, mcdaniel1973}:
\begin{equation}
\begin{aligned}
    \langle \sigma w \rangle_{s\rm{H}_2} & {} = 1.69 \times 10^{-9}\, \rm{cm}^3\;{\rm{s}^{-1}} \;\;\; \rm{if} \; s=i, \\
                                & {} = 1.30 \times  10^{-9}\, \rm{cm}^3\;{\rm{s}^{-1}} \;\;\;\;\; \rm{if} \; s=e. \\
\end{aligned}
\end{equation}


\section{Characteristic diffusion length scales for OD and AD}
\label{sec:etak_dis}
From \autoref{eq:gamma} we see that the characteristic diffusion length scale for Ohmic dissipation (OD) is
\beq
l_{\rm OD} \sim \frac{1}{k}.
\eeq
It corresponds to the typically encountered diffusion rate (\autoref{eq:gamma}) proportional to $k^2$, arising from the application of a resistivity $\etaOD$ within the assumed finite thickness of our model cloud.
Similarly, from \autoref{eq:theta} we see that the characteristic diffusion length scale for ambipolar diffusion (AD) is 
\beq
l_{\rm AD} \sim \left(\frac{k}{Z_0} + k^2 \right)^{-1/2},
\eeq
which contains an additional term $(Z_0/k)^{1/2}$ as compared to $l_{\rm OD}$. The diffusion rate (\autoref{eq:theta}) is the sum of two terms, with
a term proportional to $k$ that comes from the magnetic tension term in the Lorentz force (see \autoref{eq:Fmagequn}) while a term proportional to $k^2$ comes from the magnetic pressure gradient force that acts within the finite thickness region of the cloud. The magnetic tension, arising from a surface stress (see \autoref{eq:stresstensor}), would exist even in the limit of an infinitesimally thin sheet, and illustrates the fact that the relevant length scale for a diffusive process in the limit of an infinitesimally thin sheet is $l \sim (Z_0/k)^{1/2}$, which is the the geometric mean of $Z_0$ and $1/k$ \citep[see discussion in][Appendix]{lizano2010b}. Even though the sheet can be infinitesimally thin, one can still identify an effective length scale $Z_0 \propto c_s^2/(G \sigma_{n,0})$ as a combination of the relevant parameters. The OD term would also attain such a form if the sheet was infinitesimally thin and we only considered the dissipation of surface currents, as shown by \cite{lizano2010b}. In this study we consider the OD of the current inside the finite thickness disk to be the most applicable.

\section{Effective Sound Speed} \label{sec:figA}


\begin{figure}
\epsscale{1}
\plotone{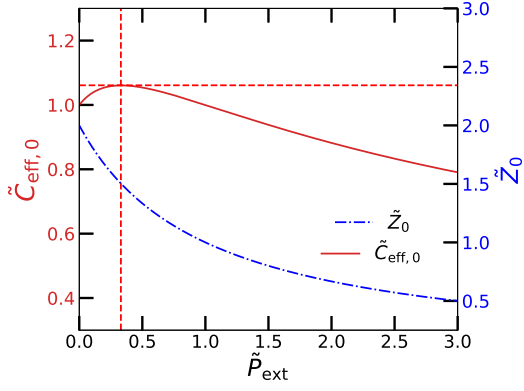}
\caption{Normalized local effective sound speed ($\tilde{C}_{\rm{eff}}$) and normalized effective local vertical half-thickness ($\tilde{Z}_0$) as functions of the normalized external pressure $\tilde{P}_{\rm{ext}}$. The value of $\tilde{C}_{\rm{eff}}$ is maximum (=1.061) at $\tilde{P}_{\rm{ext}}=1/3$ (as shown by the dashed horizontal and vertical lines). }
\label{fig:ceff_pext}
\end{figure}

In \autoref{fig:ceff_pext}, we review properties of the normalized local effective sound speed ($\tilde{C}_{\rm{eff}}$) and initial vertical half-thickness ($\tilde{Z}_0$) as a function of dimensionless external pressure ($\tilde{P}_{\rm{ext}}$).
In the limit of low external pressure ($P_{\rm{ext}} \rightarrow 0$), the local effective sound speed reduces to the isothermal sound speed (see \autoref{eq:ceff}) i.e., $\tilde{C}_{\rm{eff}} = 1$ (see \autoref{eq:ceff_tilde}). 
We see that in this limit ($P_{\rm{ext}} \ll (\pi/2) G \sigma_{n,0}^2$), the half-thickness $Z_0 \propto 1/\sigma_{n,0}$ (see \autoref{eq:Z0}). 
In this case, the half-thickness increases in the direction of decreasing surface density, and the external pressure acts to contribute a force in the direction opposite that of the surface density increase. Hence, there is an increased restorative effect to density perturbations and therefore an increased effective sound speed.
Whereas in the regime of large external pressure ($P_{\rm{ext}} \gg (\pi/2) G \sigma_{n,0}^2$), $Z_0$ becomes proportional to $\sigma_{n,0}$, determined by the interplay between internal thermal pressure within the cloud and the external pressure. 
In this case, the half-thickness decreases in the direction of decreasing surface density, and the external pressure acts to contribute a force in the same direction as the surface density increase. Hence, there is a decreased restorative effect to density perturbations and therefore a decreased effective sound speed.
As a result, $\tilde{C}_{\rm{eff},0}$ attains a maximum (=1.061) at $\tilde{P}_{\rm{ext}} =1/3$ and thereafter gradually decreases, while $\tilde{Z}_0$ gradually decreases with increasing $\tilde{P}_{\rm{ext}}$, as can be seen from \autoref{eq:equn_Z0_Pext}.


\section{Notes on Generalized Toomre Criterion} \label{sec:gen_Toomre_criterion_app}
In the limit of flux-freezing the dispersion relation is
\begin{equation}
    \omega^2 = 4\Omega^2 + k^2 (C^2_{\rm{eff,0}} + V^2 _{A,0}) - 2\pi Gk \sigma_{n,0} (1- \mu_0 ^{-2}).
\end{equation}
To minimize $\omega^2$, the criteria are  $d (\omega^2)/dk = 0$ and $d^2 (\omega^2)/dk^2> 0$ 
at  ${k=k_{\rm{min}}}$, yielding
\begin{equation}
    k_{\rm{min}} = \frac{\pi G \sigma_{n,0} (1-\mu_0 ^{-2})}{(C^2_{\rm{eff,0}} + V^2 _{A,0})} \>.
\end{equation}
Now, to obtain the instability criterion, we set $\omega^2<0$ at ${k=k_{\rm{min}}}$, which gives
\begin{equation}
    4\Omega^2 + k^2 _{\rm{min}} (C^2_{\rm{eff,0}} + V^2 _{A,0}) - 2\pi G k_{\rm{min}} \sigma_{n,0} (1 - \mu_0 ^{-2})  < 0\, ,
\end{equation}
yielding
\begin{equation}
    \frac{\Omega (C^2_{\rm{eff,0}} + V^2 _{A,0})^{1/2}}{\pi G \sigma_{n,0} (1-\mu_0 ^{-2})} <\frac{1}{2}.
\end{equation}
In the dimensionless form it becomes
\begin{equation}
\begin{aligned}
    \frac{c_s \> \Omega}{\pi G \sigma_{n,0}} \frac{(\Tilde{C}^2 _{\rm{eff,0}} + \tilde{Z}_0 \mu_0 ^{-2} )^{1/2}}{(1-\mu_0 ^{-2})} & {} < \frac{1}{2} \> , \\
    {\rm or, \hspace{0.2cm}} Q \frac{(\Tilde{C}^2 _{\rm{eff,0}} + \tilde{Z}_0 \mu_0 ^{-2} )^{1/2}}{(1-\mu_0 ^{-2})} & {} < \frac{1}{2} \> ,\\
    {\rm or, \hspace{0.2cm}} Q_{\rm{eff}} & {} < \frac{1}{2} \>, \\
    {\rm or, \hspace{0.2cm}} Q & {} < \frac{1}{2} \frac{(1-\mu_0 ^{-2})}{(\Tilde{C}^2 _{\rm{eff,0}} + \tilde{Z}_0 \mu_0 ^{-2} )^{1/2}}= Q_{\rm crit,m} \>,
\end{aligned}    
\end{equation}
where 
\begin{equation}
Q_{\rm{eff}} = \frac{\Omega (C^2_{\rm{eff,0}} + V^2 _{A,0})^{1/2}}{\pi G \sigma_{n,0} (1-\mu_0 ^{-2})} = Q \frac{(\Tilde{C}^2 _{\rm{eff,0}} + \tilde{Z}_0 \mu_0 ^{-2} )^{1/2}}{(1-\mu_0 ^{-2})}.
\end{equation}
See \autoref{sec:toomreQeff} for further discussion.


\section{Notes on Stationary field limit} \label{sec:statBapp}
In the limit of stationary magnetic fields, $\omega \delta B'_{z, \rm{eq}} \rightarrow 0$, the resulting dispersion relation can be obtained using \Crefrange{eq:matrix}{eq:gamma}. It follows that
\begin{eqnarray}
&
\begin{vmatrix}
-\omega & k_x c_s & k_y c_s & 0\\
\frac{k_x}{k}A_1 & -\omega \> c_s & 0 & \frac{k_x}{k} A_2 \\
\frac{k_y}{k} A_1 & 0 & -\omega \> c_s & \frac{k_y}{k} A_2 \\
0 &  \frac{k_x}{\mu_0} c_s &  \frac{k_y}{\mu_0} c_s & -i(\theta+\gamma)
\end{vmatrix}
= 0 
& \\
& 
\Rightarrow \omega^2 + \omega \frac{i k A_2}{\mu_0 (\theta+\gamma)} - k A_1  =0 \, ,
\label{eq:statB_DR}
\end{eqnarray}
where $A_1$, $A_2$, $\theta$, and $\gamma$ are explicitly written in \autoref{sec:DR}. Recall that $\theta$ and $\gamma$ represent the case of AD and OD, respectively.
Simplifying each individual term of \autoref{eq:statB_DR} yields
\begin{equation}
    \omega^2 = \omega'^2 \> \frac{(2\pi G\sigma_{n,0})^2}{c_s^2},
\label{eq:omega2statB}    
\end{equation}
\begin{equation}
\begin{aligned}
    \omega \frac{i k A_2}{\mu_0 \theta} {} & = \omega \frac{i k}{\mu_0} \frac{\left(2 \pi G \sigma_{n,0} \> \mu_0 ^{-1} + k\>  V_{A,0} ^2 \> \mu_0 \right)}{\tau_{ni,0} \left(2\pi G \sigma_{n,0}  \mu_0 ^{-2} k + k^2 V_{A,0} ^2 \right)} \\
                                           & = i \frac{\omega'}{\taunit} \> \frac{(2\pi G\sigma_{n,0})^2}{c_s^2},
\end{aligned}    
\label{eq:omega1statB_AD} 
\end{equation}
\begin{equation}
\begin{aligned}
    \omega \frac{i k A_2}{\mu_0 \gamma}{} &  = i\omega \frac{\left(2 \pi G \sigma_{n,0} \> \mu_0 ^{-1} + k\>  V_{A,0} ^2 \> \mu_0 \right)}{\mu_0 \etaOD k} \\
                                        & = i \frac{\omega'}{\etaODt k' \mu_0^2}  \> (1 + k' \zt ) \> \frac{(2 \pi G \sigma_{n,0})^2}{c_s^2},
\label{eq:omega1statB_OD}   
\end{aligned}   
\end{equation}
\begin{equation}
    kA_1 = k (\ceff^2 k - 2 \pi G \sigma_{n,0}) = (\cefft^2 k'^2-k') \frac{(2\pi G \sigma_{n,0})^2}{c_s^2}.
\label{eq:omega0statB}    
\end{equation}
For the case of only OD (set $\theta=0$), combining \ref{eq:omega2statB}, \ref{eq:omega1statB_OD}, \ref{eq:omega0statB} we obtain the resulting dispersion relation and corresponding growth timescale as shown in \autoref{eq:OD_DR_statB} and \autoref{eq:OD_taug_statB}.
For the case of only AD (set $\gamma=0$), combining \ref{eq:omega2statB}, \ref{eq:omega1statB_AD}, \ref{eq:omega0statB} we obtain the resulting dispersion relation and corresponding growth timescale as shown in \autoref{eq:AD_DR_statB} and \autoref{eq:AD_taug_statB}.

\section{Additional figures of normalized growth timescale vs. length scale}
\label{sec:app_fig1ADODrot}

\begin{figure*}[ht!]
\gridline{\fig{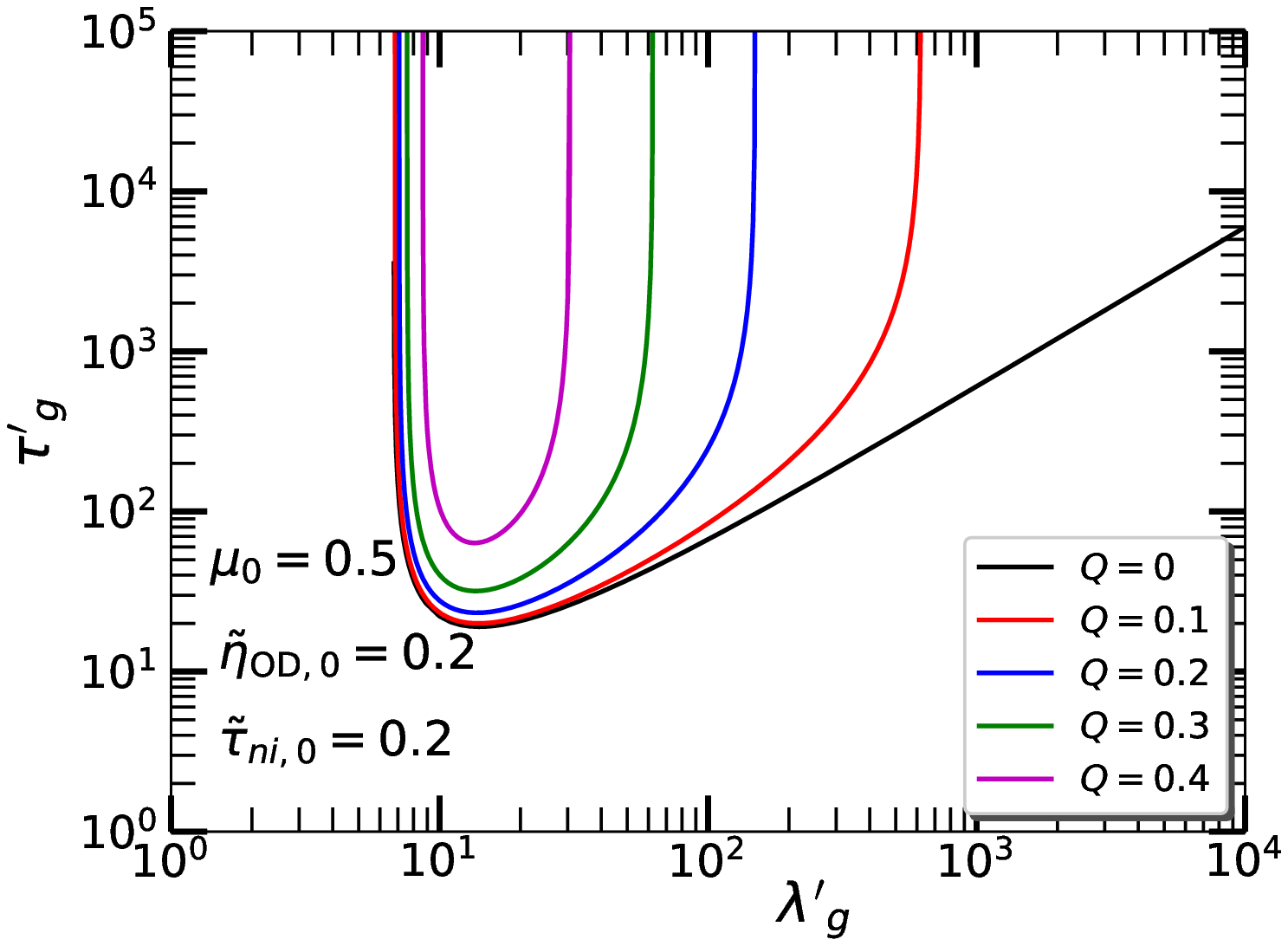}{0.33\textwidth}{(a)}
          \fig{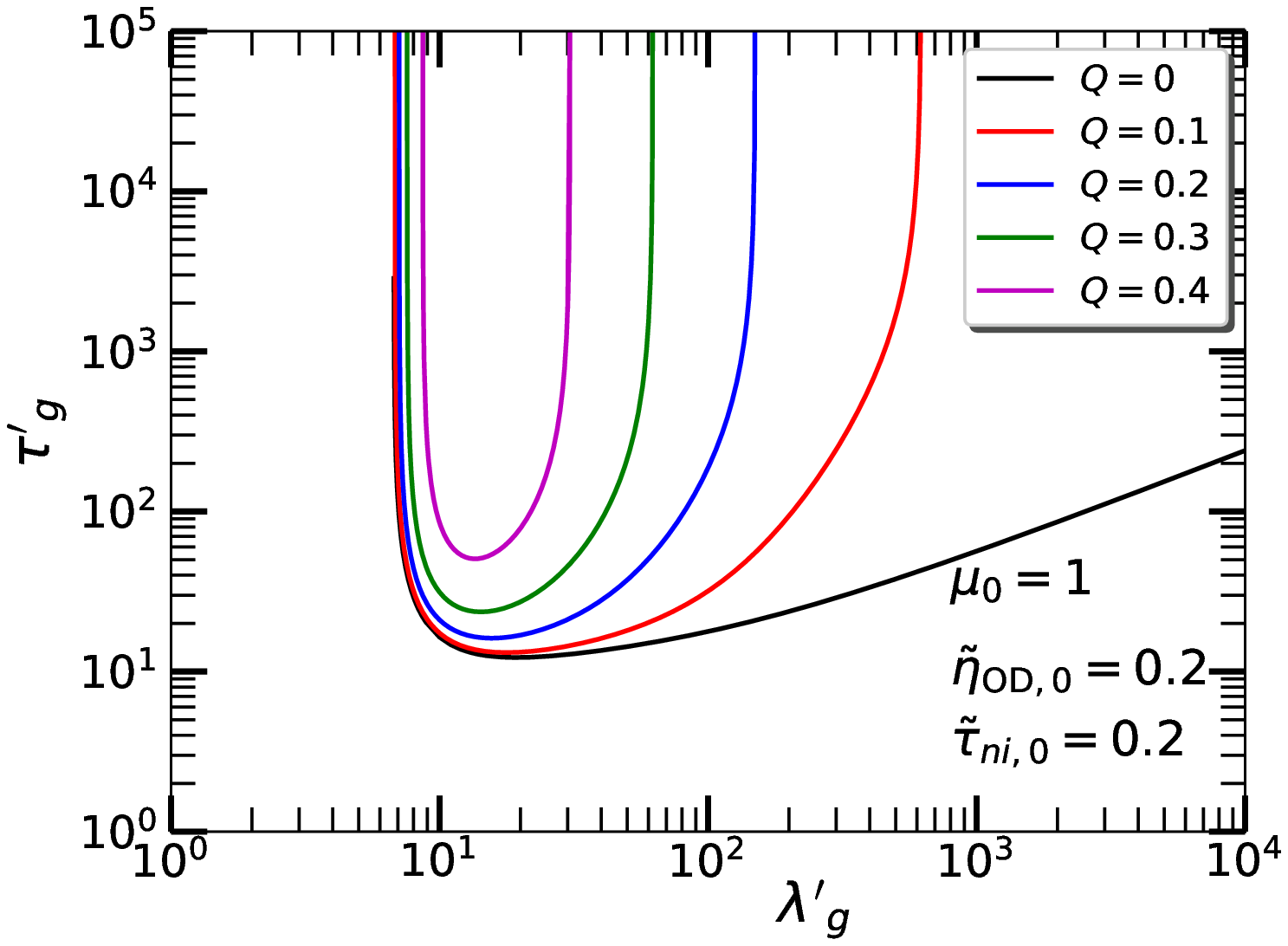}{0.33\textwidth}{(b)}
          \fig{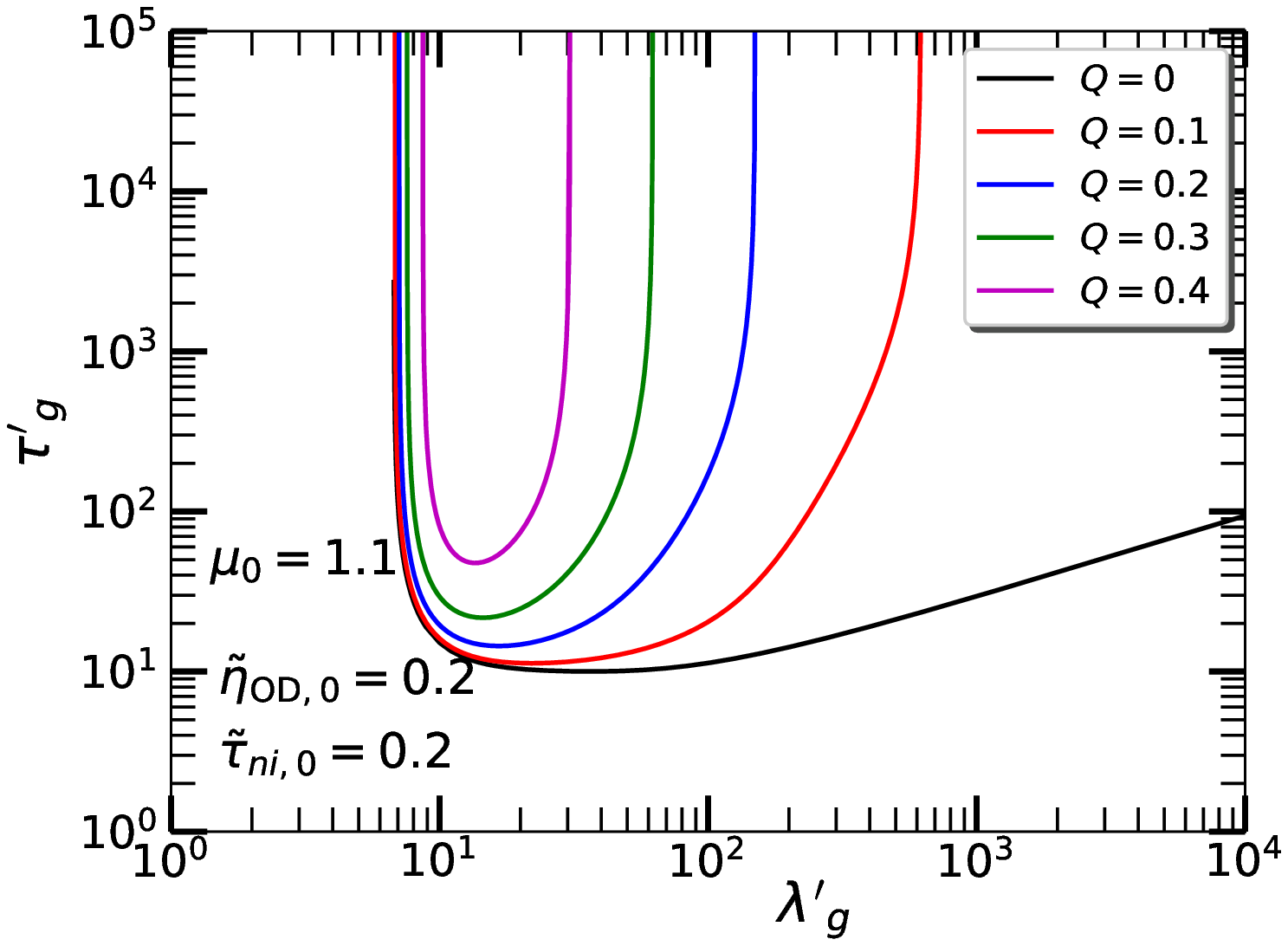}{0.33\textwidth}{(c)}
          } 
\gridline{\fig{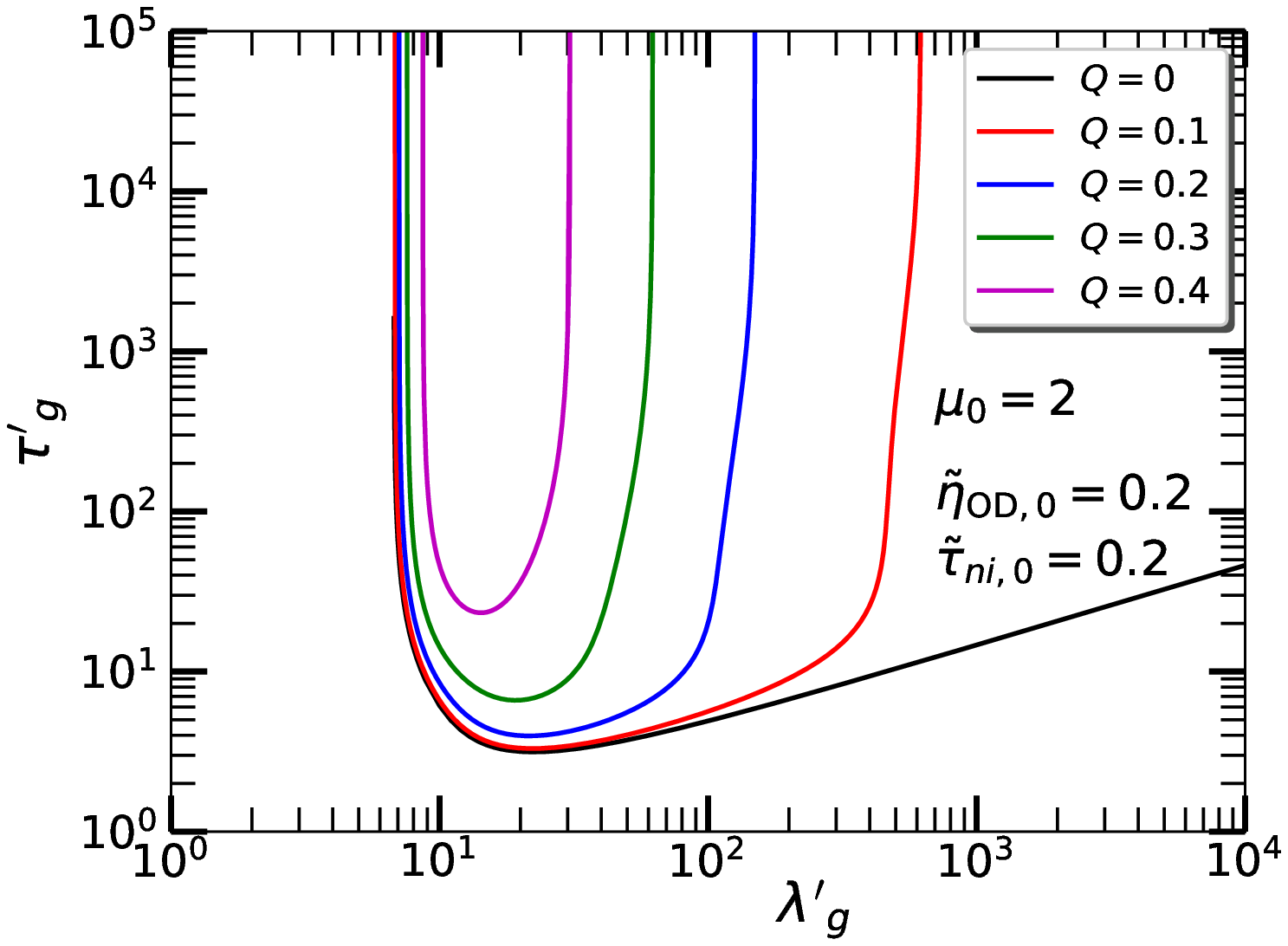}{0.33\textwidth}{(d)}
          \fig{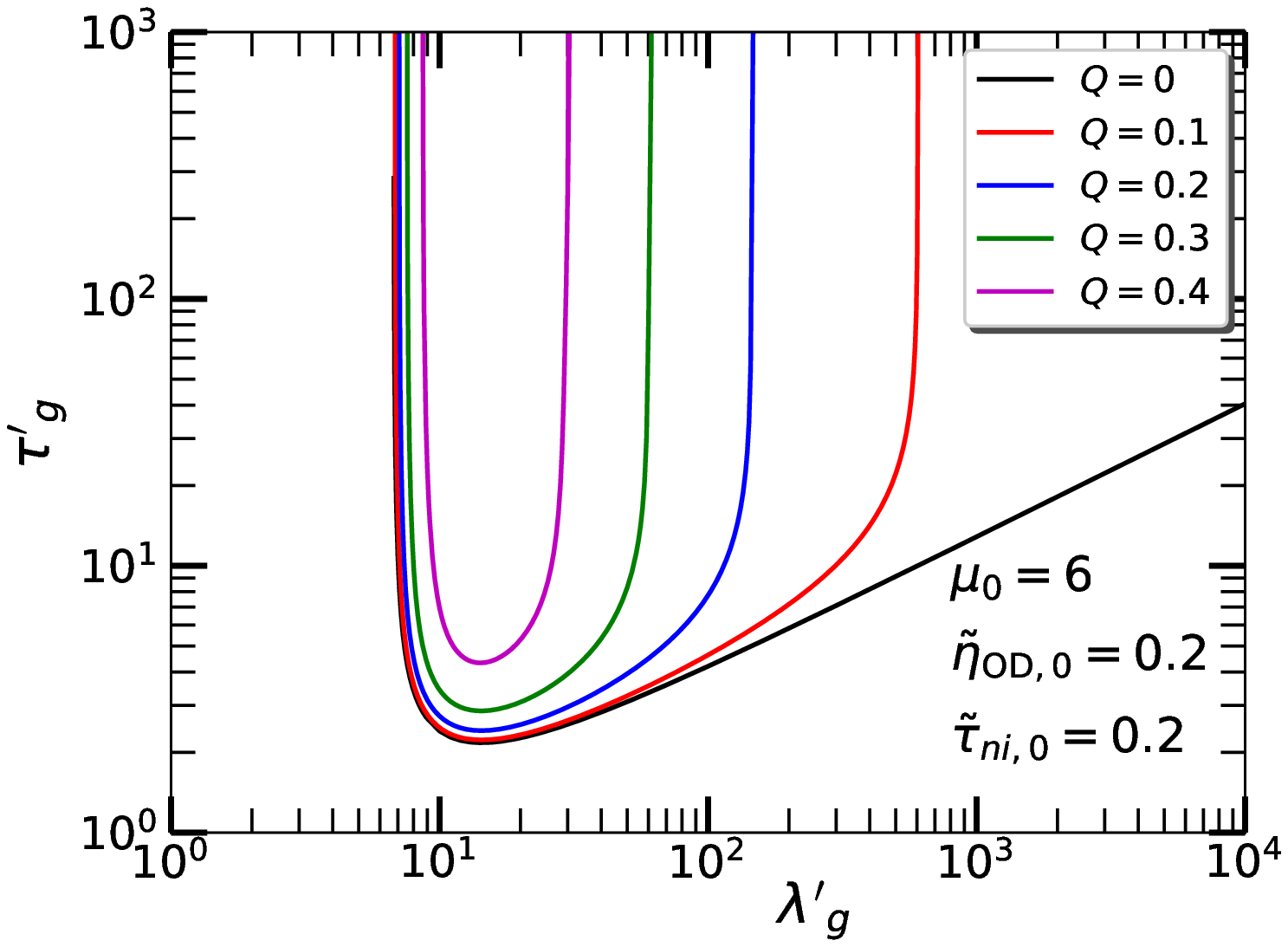}{0.33\textwidth}{(e)}
          }
\caption{Normalized growth time $\tau'_{g}=\tau_{g}/t_0$ of gravitationally unstable mode as a function of the normalized wavelength $\lambda'_{g}=\lambda_{g}/L_0$ for models with a fixed normalized Ohmic diffusivity $\etaODt = 0.2$ and neutral-ion collision time $\taunit = 0.2$, for different normalized mass-to-flux-ratio $\mu_0 = 0.5,1,1.1,2,6$. Each figure shows timescale curves for models with different normalized rotation $Q$ = 0 (black), 0.1 (red), 0.2 (blue), 0.3 (green), and 0.4 (magenta).}
\label{fig:fig1ADODrot}
\end{figure*}

\autoref{fig:fig1ADODrot} shows the curves of normalized growth timescale as a function of normalized length scale for different normalized rotation $Q=0, 0.1, 0.2, 0.3, 0.4$. This figure represents our model of the protostellar disk for distinct normalized mass-to-flux ratios $\mu_0 = 0.5, 1, 1.1, 2, 6$, with $\etaODt=0.2$ and $\taunit=0.2$, corresponding to $n_{n,0} = 10^{11} \ {\rm cm}^{-3}$. This is one of our fundamental results, which can be obtained by plotting the normalized form of the full dispersion relation as seen in \autoref{eq:normdr}. Because of the combination of both nonideal MHD effects, the timescale versus length scale curve attains a minimum at a smaller value of $\tau'_g$ even for the subcritical case $\mu_0=0.5$ as compared to the case when only one nonideal MHD effect is present. With the transition of $\mu_0$ from subcritical to supercritical, these curves gradually approach to that of the hydrodynamic case. See \autoref{sec:nonideal_theory_results}, \ref{sec:disk} for a detailed discussion.
\section{Monochromatic Perturbation}
\label{sec:perturbation}


\begin{figure*}[ht!]
\gridline{\fig{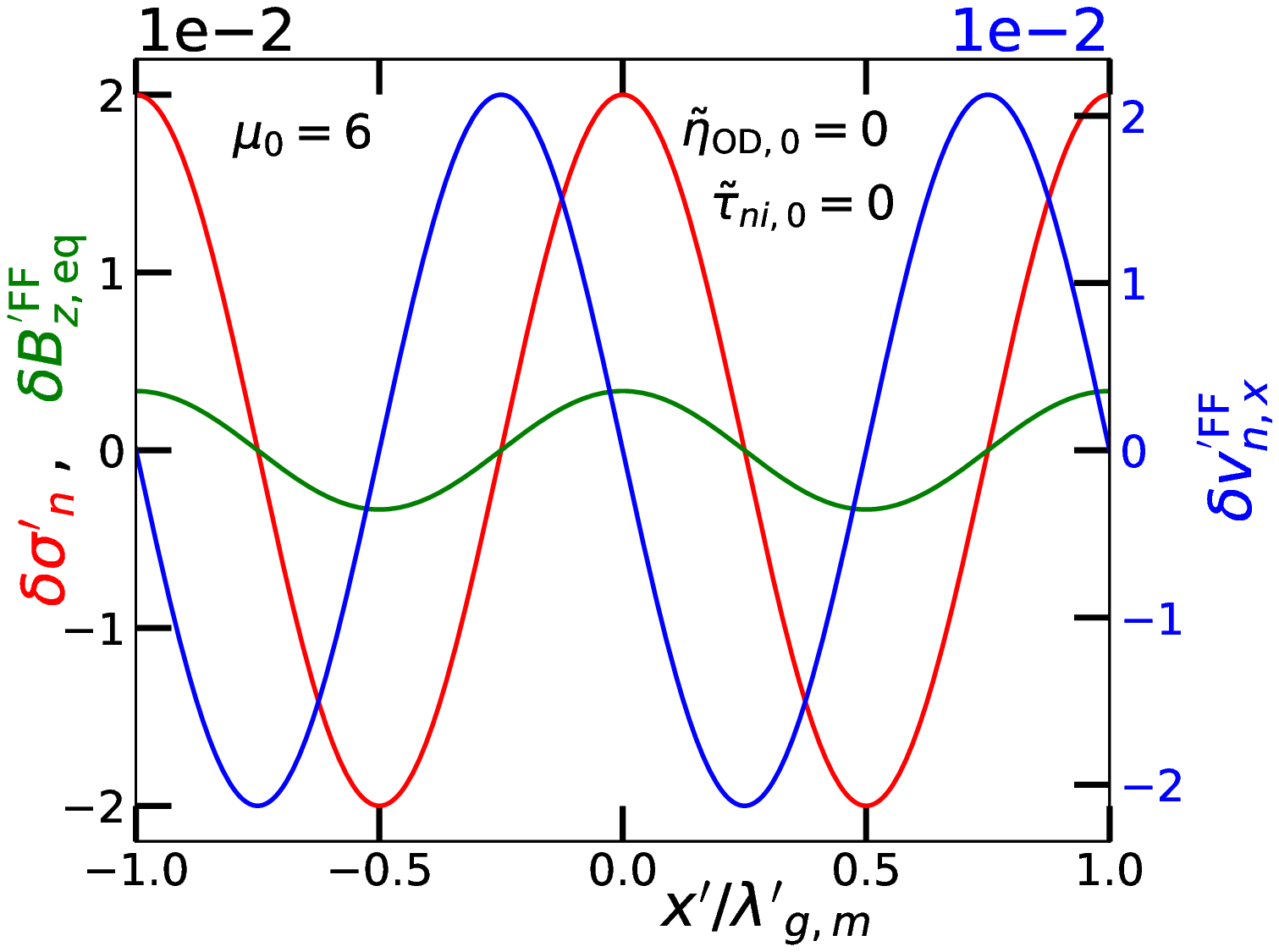}{0.33\textwidth}{(a)}
          \fig{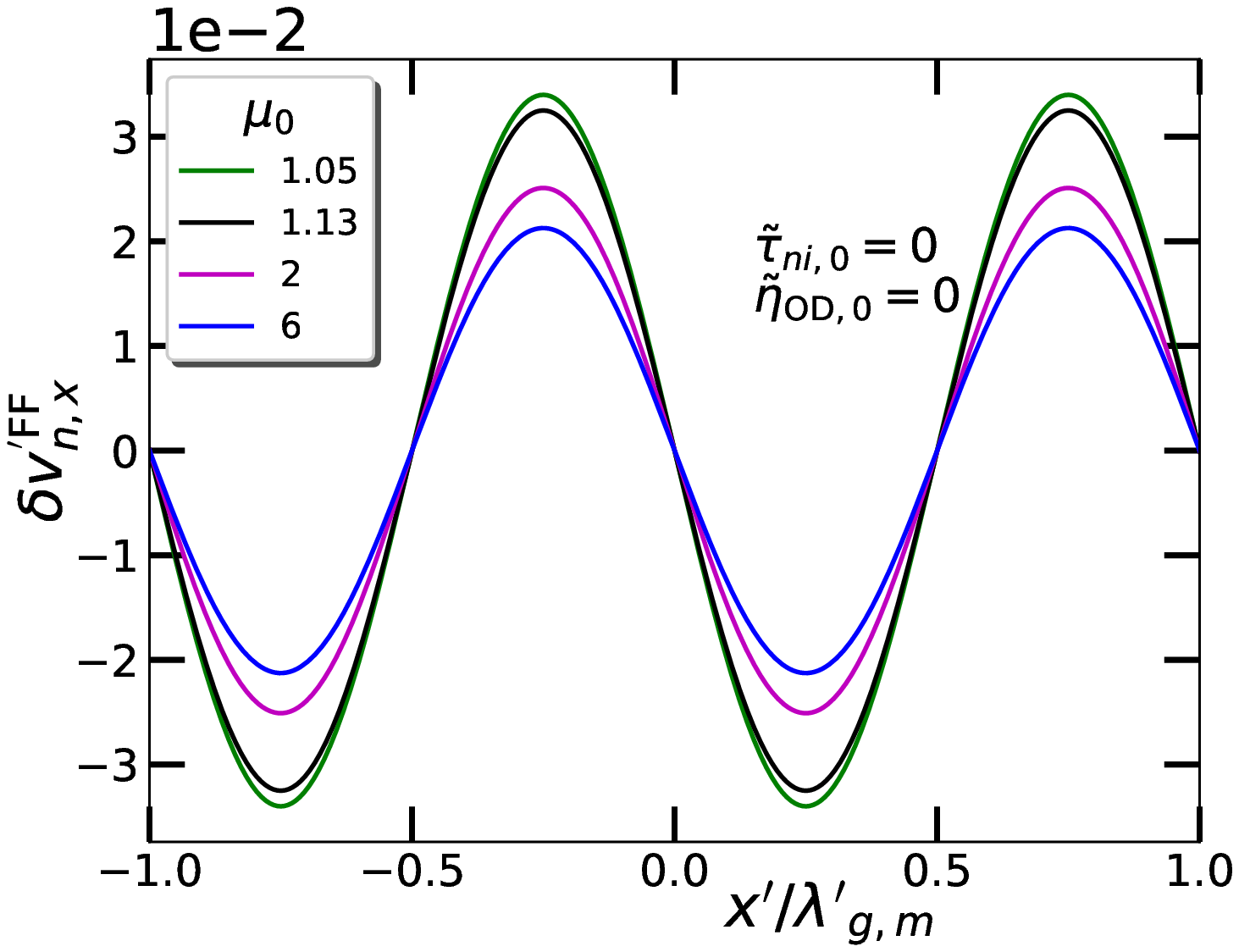}{0.33\textwidth}{(b)}
          \fig{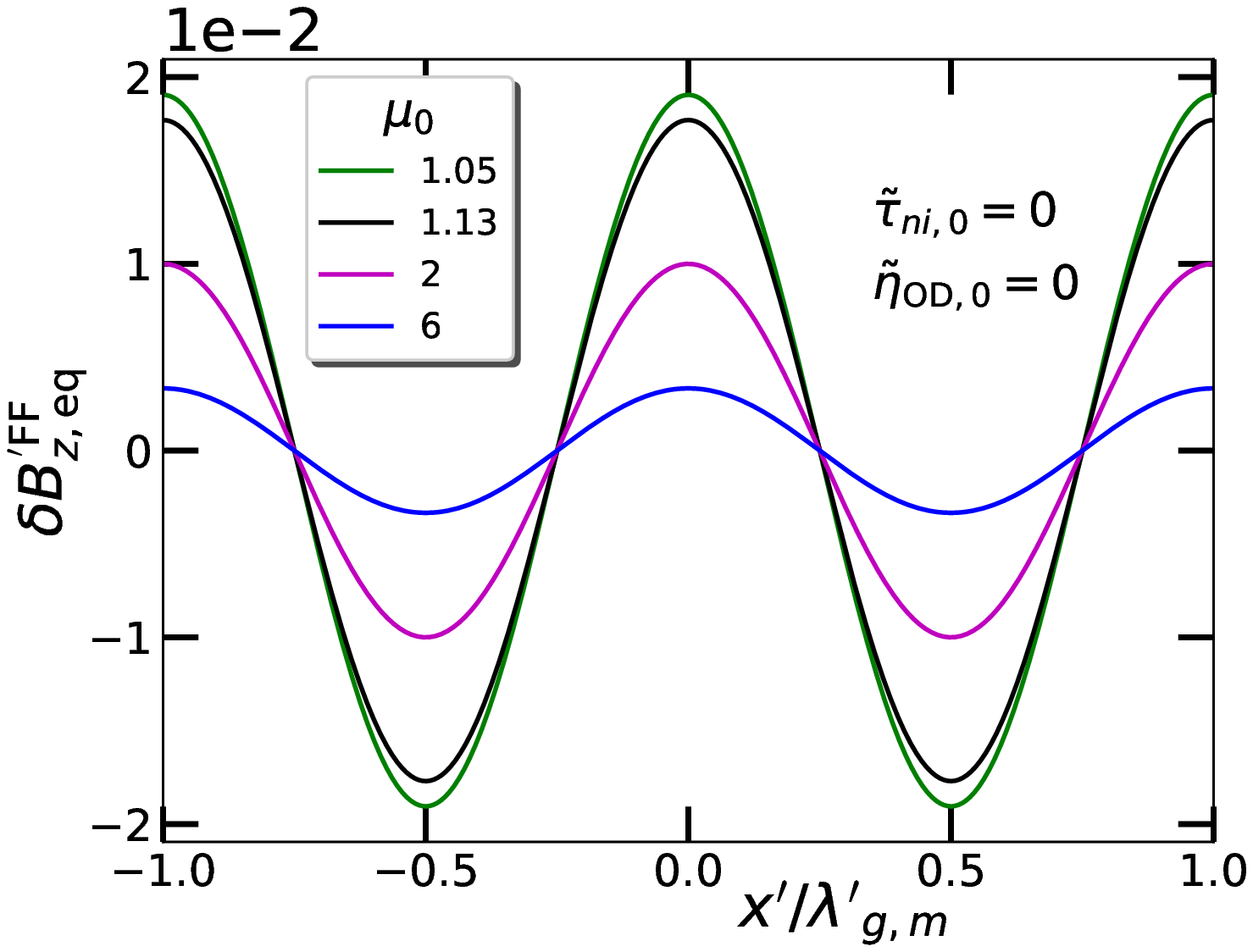}{0.33\textwidth}{(c)}
          }  
\gridline{\fig{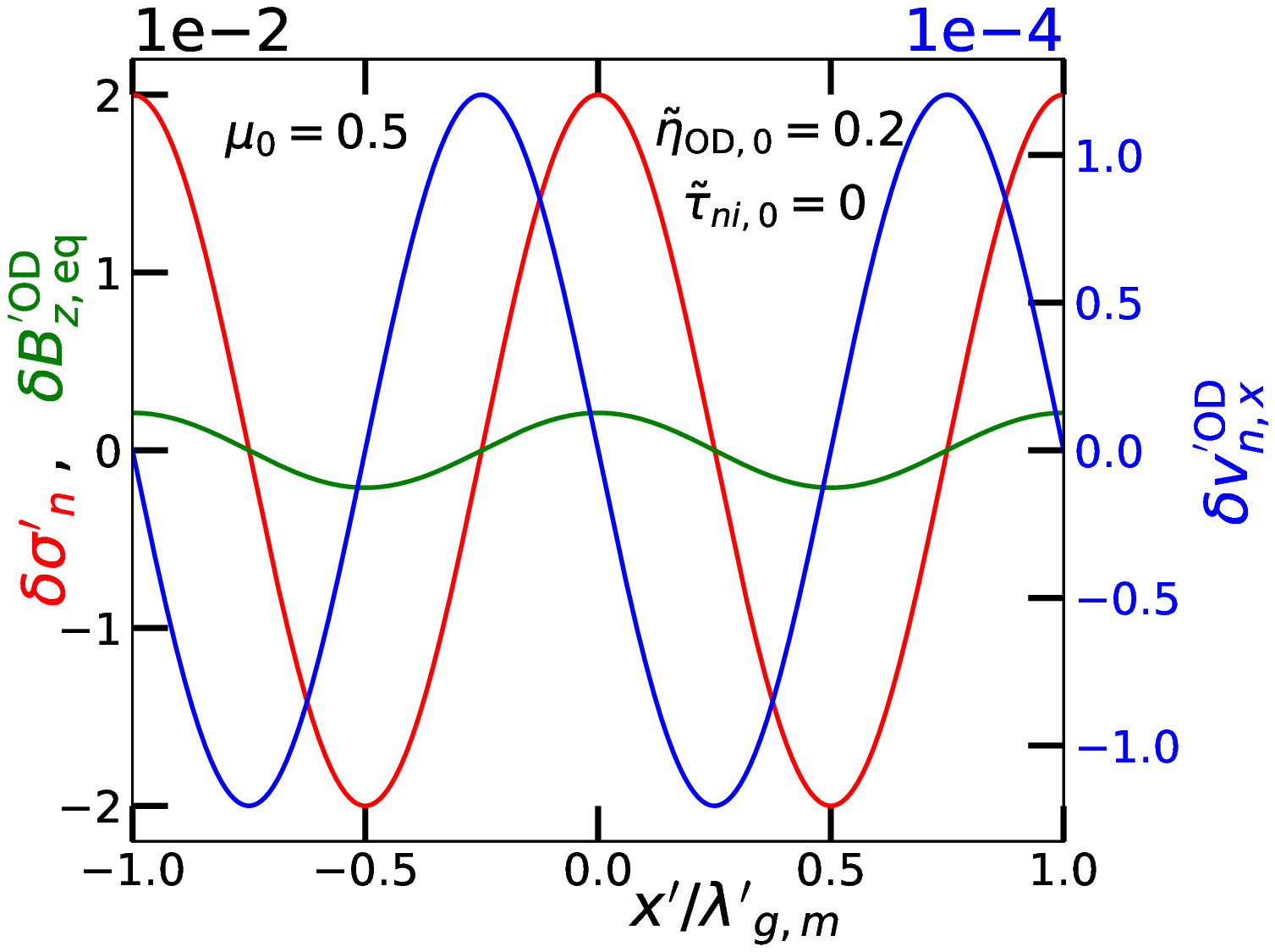}{0.33\textwidth}{(d)}
          \fig{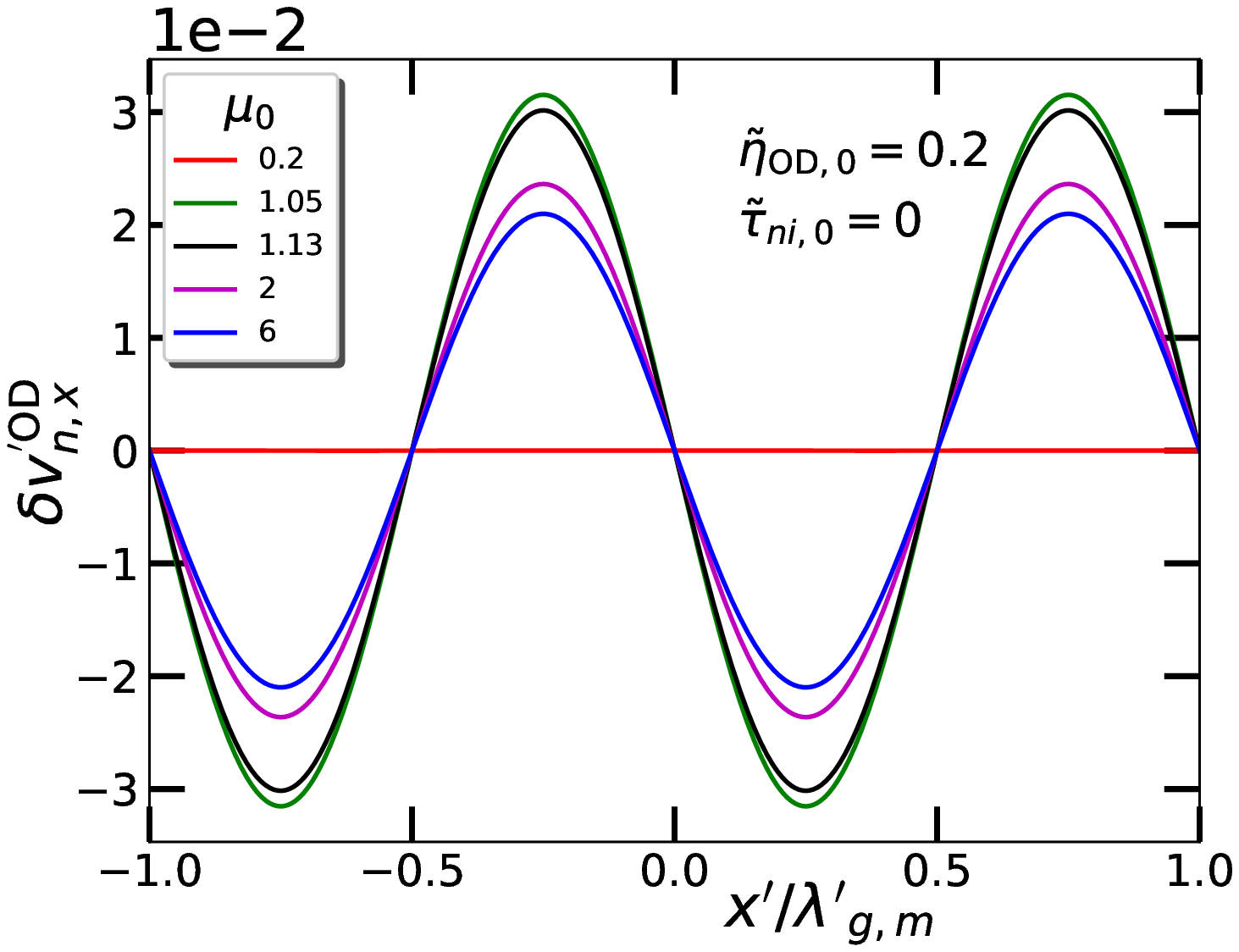}{0.33\textwidth}{(e)}
          \fig{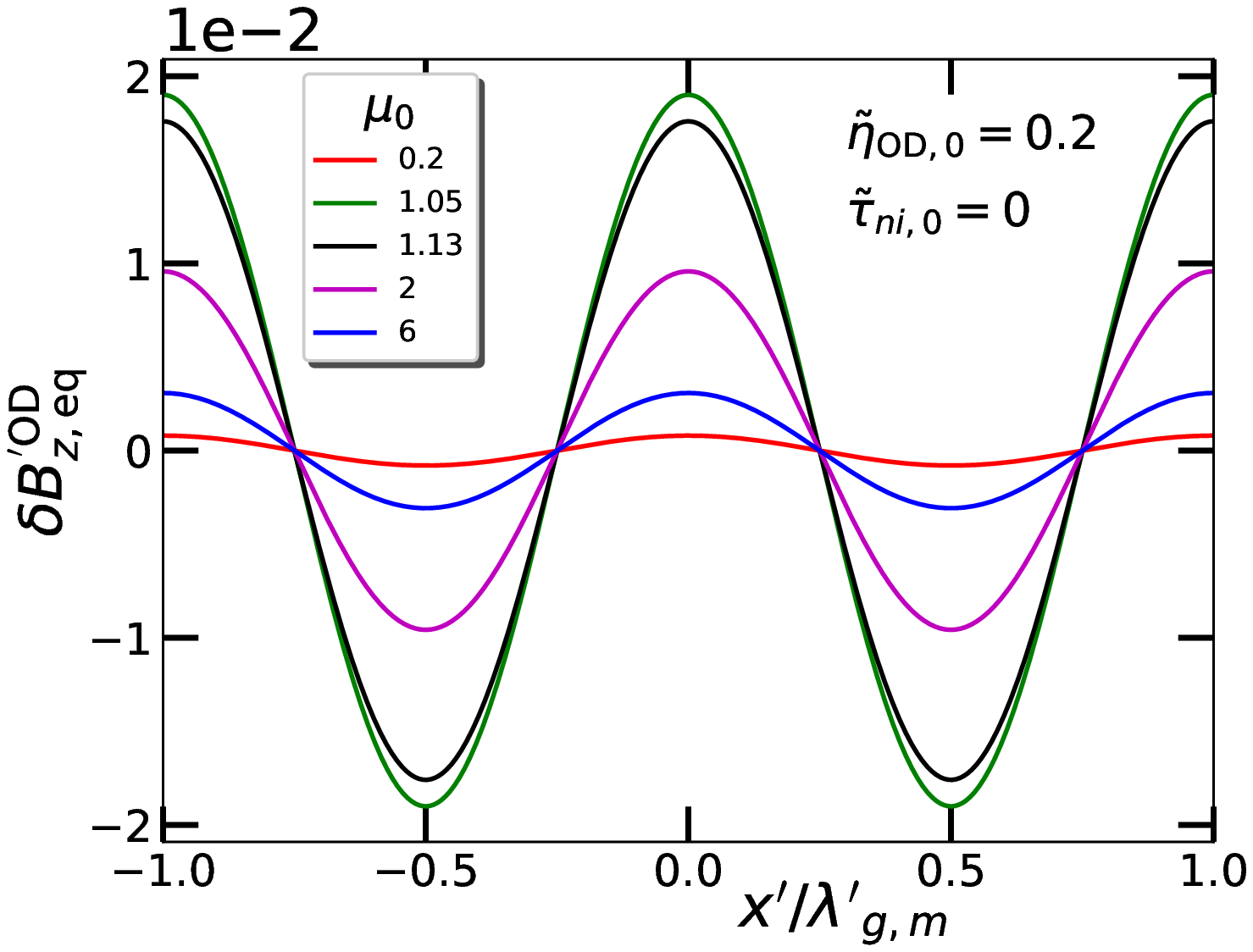}{0.33\textwidth}{(f)}
          }
\gridline{\fig{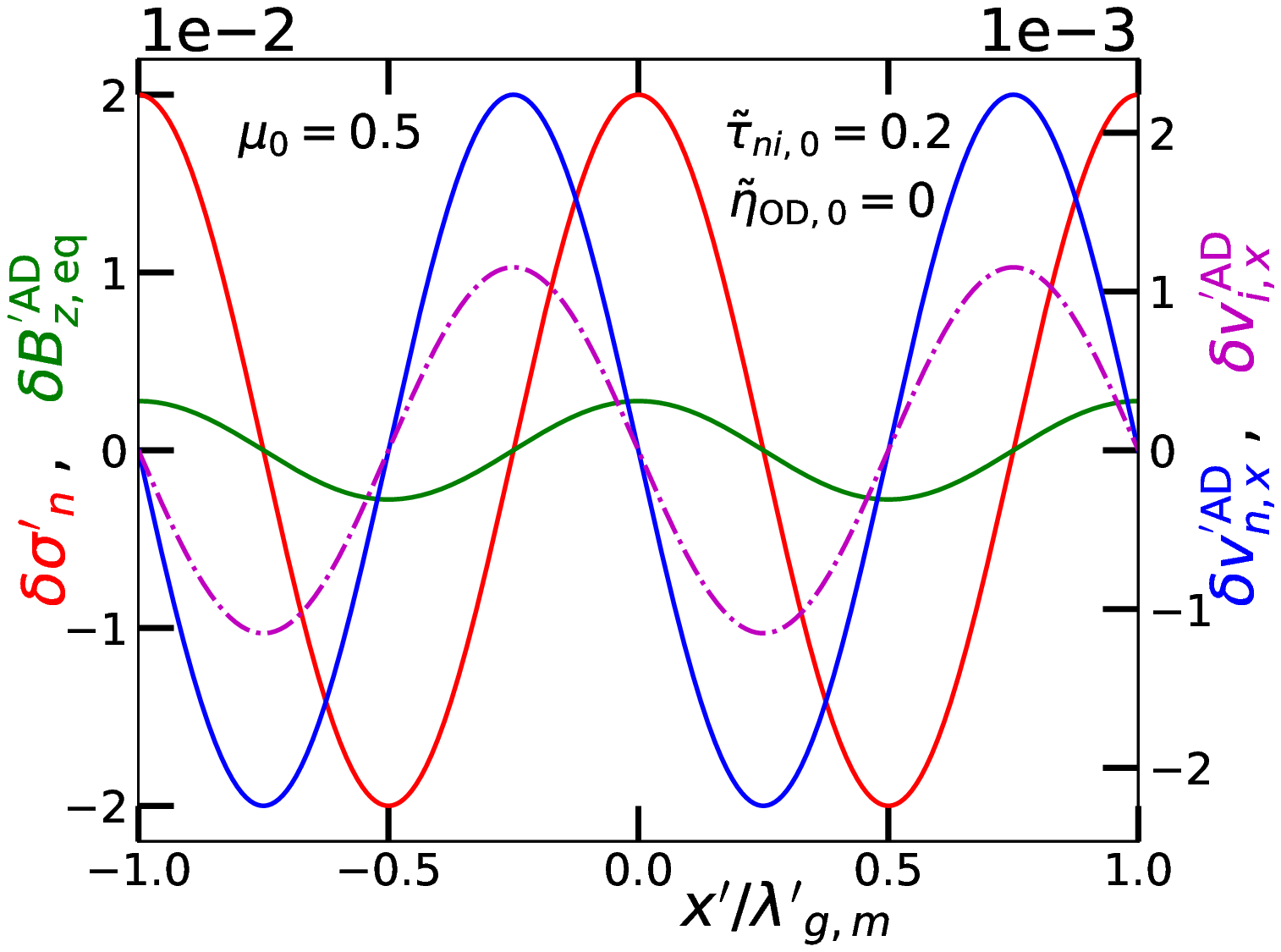}{0.33\textwidth}{(g)}
          \fig{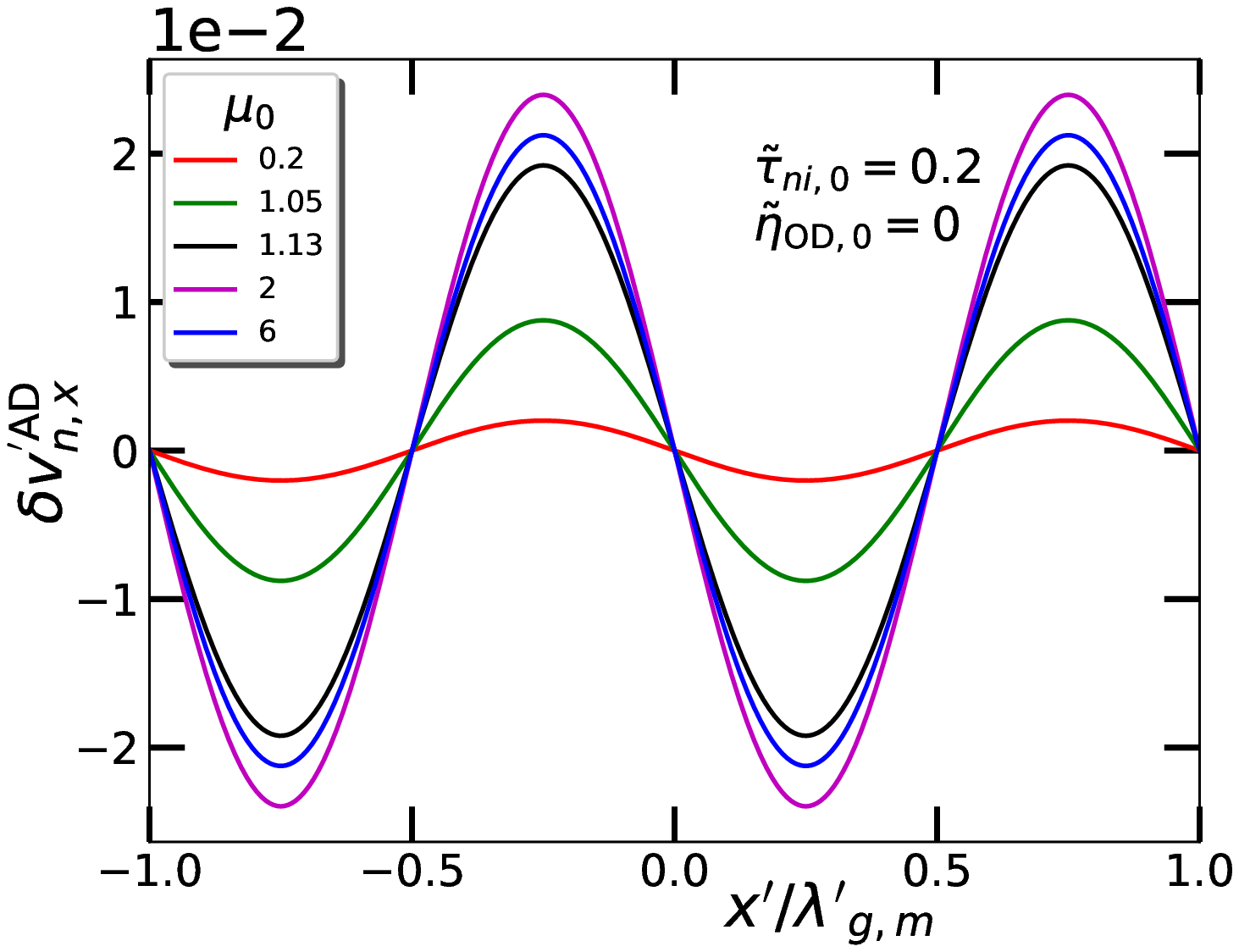}{0.33\textwidth}{(h)}
          \fig{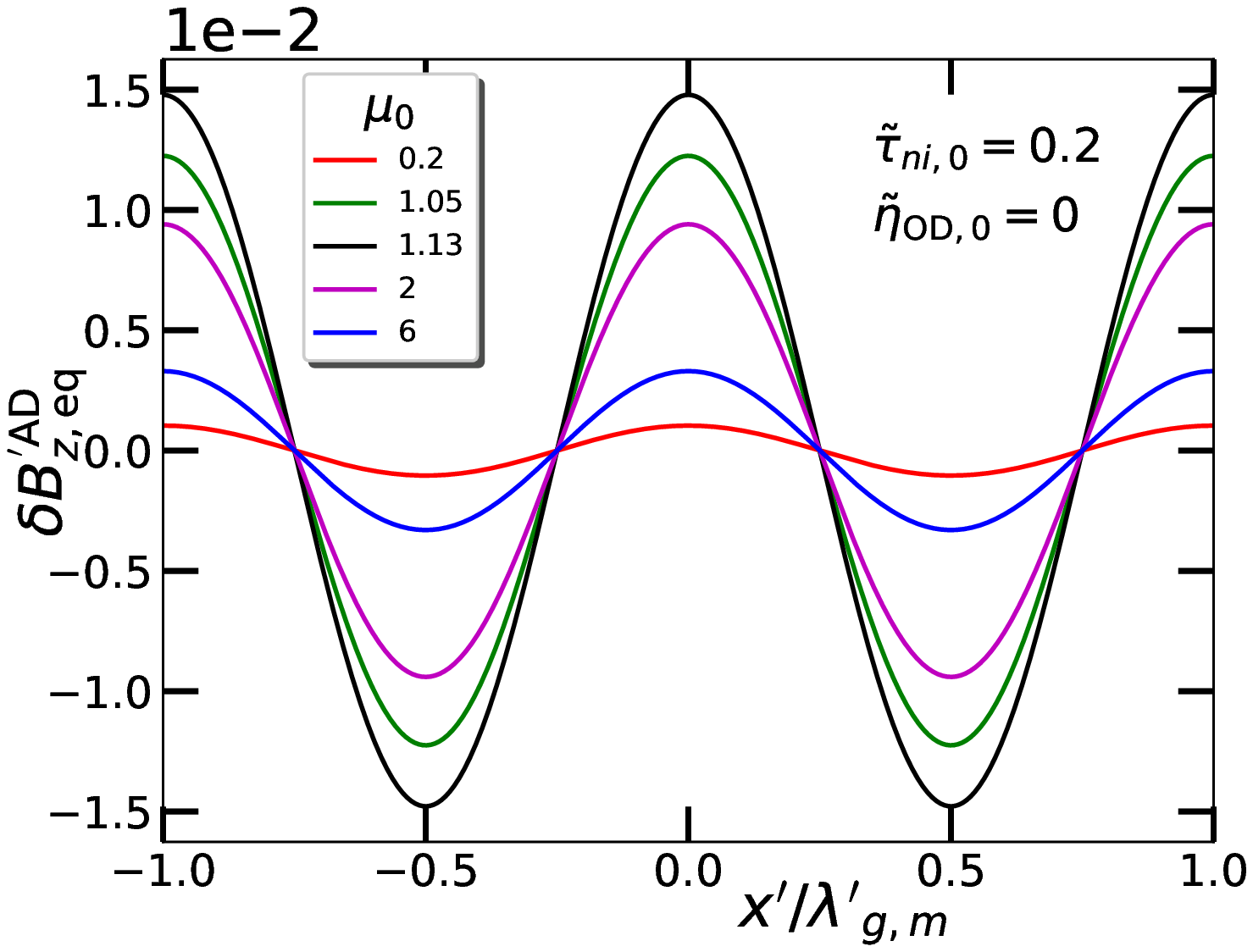}{0.33\textwidth}{(i)}
          }
\caption{Normalized amplitudes of perturbed eigenfunctions as a function of $x'/\lambda'_{g,m}$ ($x' = x/L_0$, $\lambda'_{g,m} = \lambda_{g,m}/L_0$). The upper, middle and lower panels show the cases of flux-freezing (FF), Ohmic dissipation (OD), ambipolar diffusion (AD), respectively. 
The first column (a, d, g) of each of three panels represents the spatial variation of normalized perturbed column density field ($\delta \sigma'_{n}$) (red), perturbed velocity field for neutrals ($\delta v'_{n,x}$), perturbed magnetic field ($\delta B'_{z,\rm{eq}}$) for a fixed $\mu_0$. 
For the case with AD the perturbed normalized velocity fields of ions ($\delta v'_{i,x}$) (magenta) is also shown in (g). The second (b, e, h) and third (c,f,i) column of these panels show the case of perturbed velocity field of neutrals ($\delta v^{'\rm{FF}}_{n,x}$, $\delta v^{'\rm{OD}}_{n,x}$, $\delta v^{'\rm{AD}}_{n,x}$) and magnetic field ($\delta B_{z,\rm{eq}}^{'\rm{FF}}$, $\delta B_{z,\rm{eq}}^{'\rm{OD}}$, $\delta B_{z,\rm{eq}}^{'\rm{AD}}$), respectively, for different values of $\mu_0 = $ 0.2 (red), 1.05 (green), 1.13 (black), 2 (magenta), 6 (blue). Note that for the flux-freezing case (a, b, c), only supercritical normalized mass-to-flux-ratios ($\mu_0> 1$) are considered.}
\label{fig:figpertODADFF}
\end{figure*}

Here we focus on the form of the eigenfunctions with a single wavenumber $k$. 
The column density perturbation is of the form
\begin{equation}
    \delta \sigma'_{n} (x,y,t) = \delta \sigma'_{n,a} \operatorname{Re} \left[ e^{i(kx-\omega t)} \right],
\end{equation}
where we take the uniform background state ($\sigma'_{n,0} = 1$) with a perturbed amplitude $\delta \sigma'_{n,a}$. Our dispersion analysis signifies that the linear disturbances are independent of the angle of their propagation ($\alpha$). Hence, the choice of direction of propagation becomes irrelevant to our context. The reference has been set up by making $\alpha = 0$ (parallel to the $x$- axis), which means that $k'_y = 0$, such that we can write $k'_x = k'_z \equiv k'$ (see \autoref{sec:LA}). By adding a column density perturbation in this way, we calculate the the initial velocity and magnetic field perturbations that are congruent with our system of
equations. We explicitly derive the equations for perturbed quantities from the dimensionless set of equations shown in \autoref{sec:norm}. Now, solving for the initial perturbations $\delta v'_{n,x}$, $\delta v'_{n,y}$, $\delta v'_{i,x}$, $\delta v'_{i,y}$,  $\delta B_{z,\rm{eq}}^{'\rm{FF}}$, $\delta B_{{z,\rm{eq}}}^{'\rm{OD}}$, and $\delta B_{{z,\rm{eq}}}^{'\rm{AD}}$ in terms of the given $\delta \sigma'_{n,0}$, $k'_x$, and $\tau'_g = i/\omega'$ (as a function of $\lambda' = 2\pi/k'$) yields

\begin{equation}
    \delta v'_{n,x} (x,y,t) = \frac{\lambda'}{2\pi \tau'_g} \delta \sigma'_{n,a} \> \operatorname{Re} \left[e^{i\left(\frac{\pi}{2}+ k x- \omega t \right)}\right], \>\> \delta v'_{n,y} (x,y,t) =0
\end{equation}

\begin{equation}
\begin{aligned}
    \delta v'_{i,x} (x,y,t) = \left[\frac{\lambda'}{2\pi \tau'_g} - K_1 \right] \delta \sigma'_{n,a} \> {} & \operatorname{Re} \left[e^{i \left(\frac{\pi}{2} + k x - \omega t \right)}\right], \\                                                                                    & \>\> \delta v'_{i,y} (x,y,t) =0  ,
\end{aligned}
\label{eq:mono_pert_delta_vix}
\end{equation}
where 
\begin{equation}
K_1 = \frac{\tilde{B}_{\rm{ref}} ^2  \lambda' \tilde{\tau}_{ni,0}}{\lambda' + 2\pi \tau'_g \tilde{\tau}_{ni,0} \tilde{B}_{\rm{ref}}^2 \left(1+ \frac{2\pi \tilde{Z}_0}{\lambda'}\right)}  \left(1+ \frac{2\pi \tilde{Z}_0}{\lambda'}\right).
\end{equation}
Linearization and normalization of $F_{\rm Mag,x}$ yields 
\begin{equation}
   F'_{\rm Mag,x,0} = \frac{F_{\rm Mag,x,0}}{(2\pi G \sigma^2 _{n,0})} = - \frac{ik'_x}{k'_z} \> \Tilde{B}_{\rm ref} \> (1+k'_z \> \tilde{Z}_0) \> \delta B^{' \rm AD }_{z, {\rm eq}} \> e^{i( k x -\omega t)} . \>\>
\end{equation}
Applying the condition $k'_x = k'_z \equiv k'$ to the above expression and plugging it back into the linearized and normalized form of \autoref{eq:velocityequn_in_xcomp} yields
\begin{equation}
\delta v'_{i,x}  = \delta v'_{n,x}  +  \taunit \left( -i \Tilde{B}_{\rm ref} \> (1+k' \> \tilde{Z}_0) \right) \> \delta B^{' \rm AD }_{z, {\rm eq}} \, .
\label{eq:deltavix_inter}   
\end{equation}
Further simplification of \autoref{eq:deltavix_inter} by substituting $\delta B^{' \rm AD }_{z, {\rm eq}}$ (see \autoref{eq:deltaB_AD_form}) gives  \autoref{eq:mono_pert_delta_vix}. 

Moving to the perturbed eigenfunctions for magnetic field,
in the limit of flux-freezing ($\etaODt \rightarrow 0$, $\taunit \rightarrow 0$),
\begin{equation}
    \delta B_{z,\rm{eq}}^{'\rm{FF}} (x,y,0) =  \tilde{B}_{\rm{ref}} \>\> \delta \sigma'_{n,a} \; \operatorname{Re} \left[e^{i(kx-\omega t)}\right].
\end{equation}
In the limit of only OD ($\tilde{\tau}_{ni,0} \rightarrow 0$), 
\begin{equation}
\begin{aligned}
    \delta B_{z,\rm{eq}}^{'\rm{OD}} (x,y,t) ={} & \left[\frac{\tilde{B}_{\rm{ref}}  \lambda'^{2}}{\lambda'^2 + 4\pi^2 \tau'_g \etaODt}\right]  \delta \sigma'_{n,a} \; \operatorname{Re} \left[e^{i(kx-\omega t)}\right].
\end{aligned}
\end{equation}
In the limit of only AD ($\etaODt \rightarrow 0$),
\begin{equation}
\begin{aligned}
    \delta B_{z,\rm{eq}}^{'\rm{AD}}(x,y,t) = \left[\frac{\tilde{B}_{\rm{ref}}  \lambda'}{\lambda' + 2\pi \tau'_g \tilde{\tau}_{ni,0} \tilde{B}_{\rm{ref}}^2 \left(1+ \frac{2\pi \tilde{Z}_0}{\lambda'}\right)} \right] \delta \sigma'_{n,a} \> \operatorname{Re} \left[e^{i(k x-\omega t)}\right].
\end{aligned}
\label{eq:deltaB_AD_form}   
\end{equation}
At $t = 0$, considering only the initial real amplitude, $\operatorname{Re}[e^{i(k x-\omega t)}]$ and $\operatorname{Re}[e^{i(\frac{\pi}{2}+ kx- \omega t)}]$ can be written as $\cos \left(2 \pi x / \lambda \right)$ and $-\sin  \left(2 \pi x / \lambda \right)$, respectively. By defining the correspondence between the perturbed physical variables in this way, we are selecting the eigenvector of the perturbation at a single wavelength ($\lambda'$). We call this a monochromatic perturbation that can excite a single eigenmode of our model cloud at $t=0$, corresponding to a particular $\lambda'$ for each different $\mu_0$ (recall $\tilde{B}_{\rm{ref}} = 1/\mu_0$). When one initiates the time evolution of a model cloud in this fashion, the subsequent evolution is the continuous growth of that specific excited eigenmode. At $t = 0$, the perturbed eigenmodes can be written as the following:

\begin{equation}
    \delta \sigma'_{n} (x,y,0) = \delta \sigma'_{n,a} \> \cos \left(\frac{2 \pi x }{\lambda} \right),
\end{equation}

\begin{equation}
    \delta v'_{n,x} (x,y,0) = - \frac{\lambda'}{2\pi \tau'_g} \> \delta \sigma'_{n,a} \> \sin \left(\frac{2 \pi x }{\lambda} \right),
\label{eq:timescale_equn}    
\end{equation}

\begin{equation}
    \delta v'_{i,x} (x,y,0) = - \left[\frac{\lambda'}{2\pi \tau'_g} - K_1 \right] \> \delta \sigma'_{n,a} \> \sin \left(\frac{2 \pi x }{\lambda} \right),
\end{equation}

\begin{equation}
    \delta B_{z,\rm{eq}}^{'\rm{FF}} (x,y,0) =  \tilde{B}_{\rm{ref}} \>\> \delta \sigma'_{n,a} \; \cos \left(\frac{2 \pi x }{\lambda} \right),
\label{eq:deltaBFF}
\end{equation}

\begin{equation}
    \delta B_{z,\rm{eq}}^{'\rm{OD}} (x,y,0) =  \left[\frac{\tilde{B}_{\rm{ref}}  \lambda'^{2}}{\lambda'^2 + 4\pi^2 \tau'_g \etaODt}\right] \delta \sigma'_{n,a} \; \cos \left(\frac{2 \pi x }{\lambda} \right),
\label{eq:deltaBOD}    
\end{equation}

\begin{equation}
\begin{aligned}
    \delta B_{z,\rm{eq}}^{'\rm{AD}} (x,y,0) = \left[\frac{\tilde{B}_{\rm{ref}}  \lambda'}{\lambda' + 2\pi \tau'_g \tilde{\tau}_{ni,0} \tilde{B}_{\rm{ref}}^2 \left(1+ \frac{2\pi \tilde{Z}_0}{\lambda'}\right)}\right]\> \delta \sigma'_{n,a} \> \cos\left(\frac{2 \pi x }{\lambda} \right)\, .
\end{aligned}
\label{eq:deltaBAD}
\end{equation}
Now, we are interested to study the spatial variation of these dimensionless perturbed real amplitudes, e.g., as a function of $x/\lambda \left( = x'/\lambda' =  (x/L_0)/(\lambda/L_0) \right)$. In this calculation, we take $\tau'_g$ as $\tau'_{g,m}$, the shortest growth time and $\lambda'$ as $\lambda'_{g,m}$, the preferred length scale corresponding to the shortest timescale. 

\begin{figure*}[htb!]
\gridline{\fig{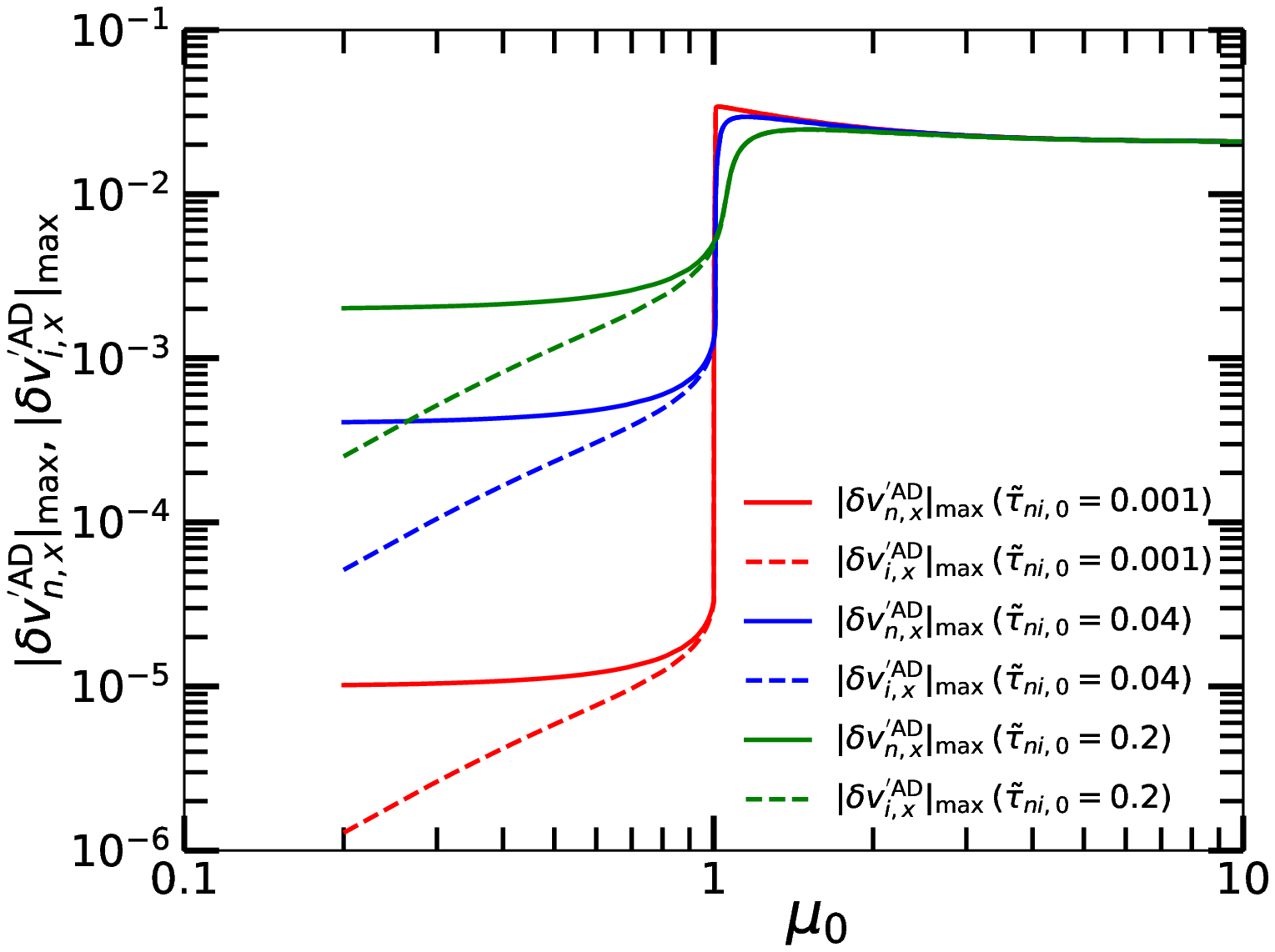}{0.45\textwidth}{(a)}
          \fig{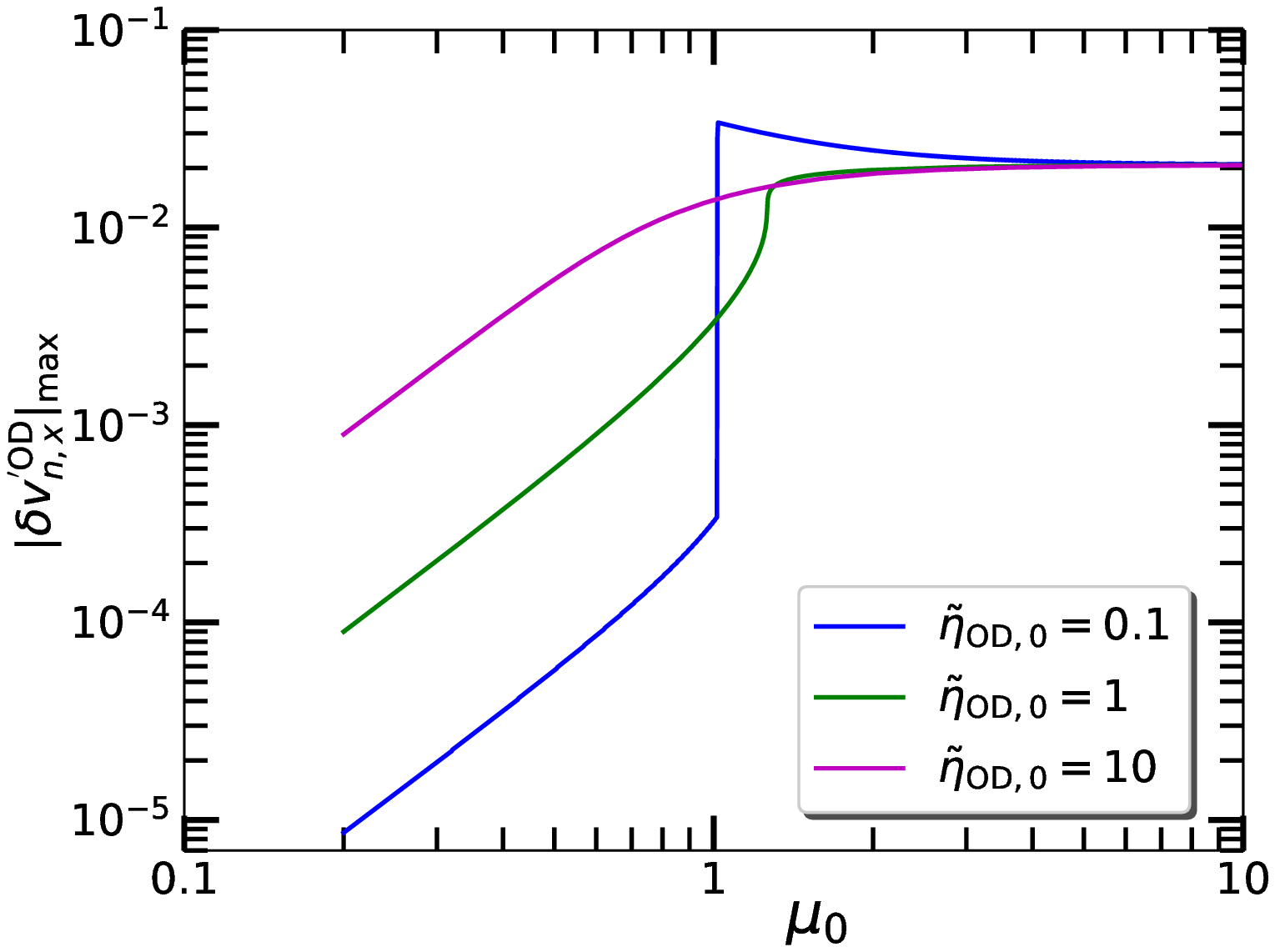}{0.45\textwidth}{(b)}
         }
\gridline{\fig{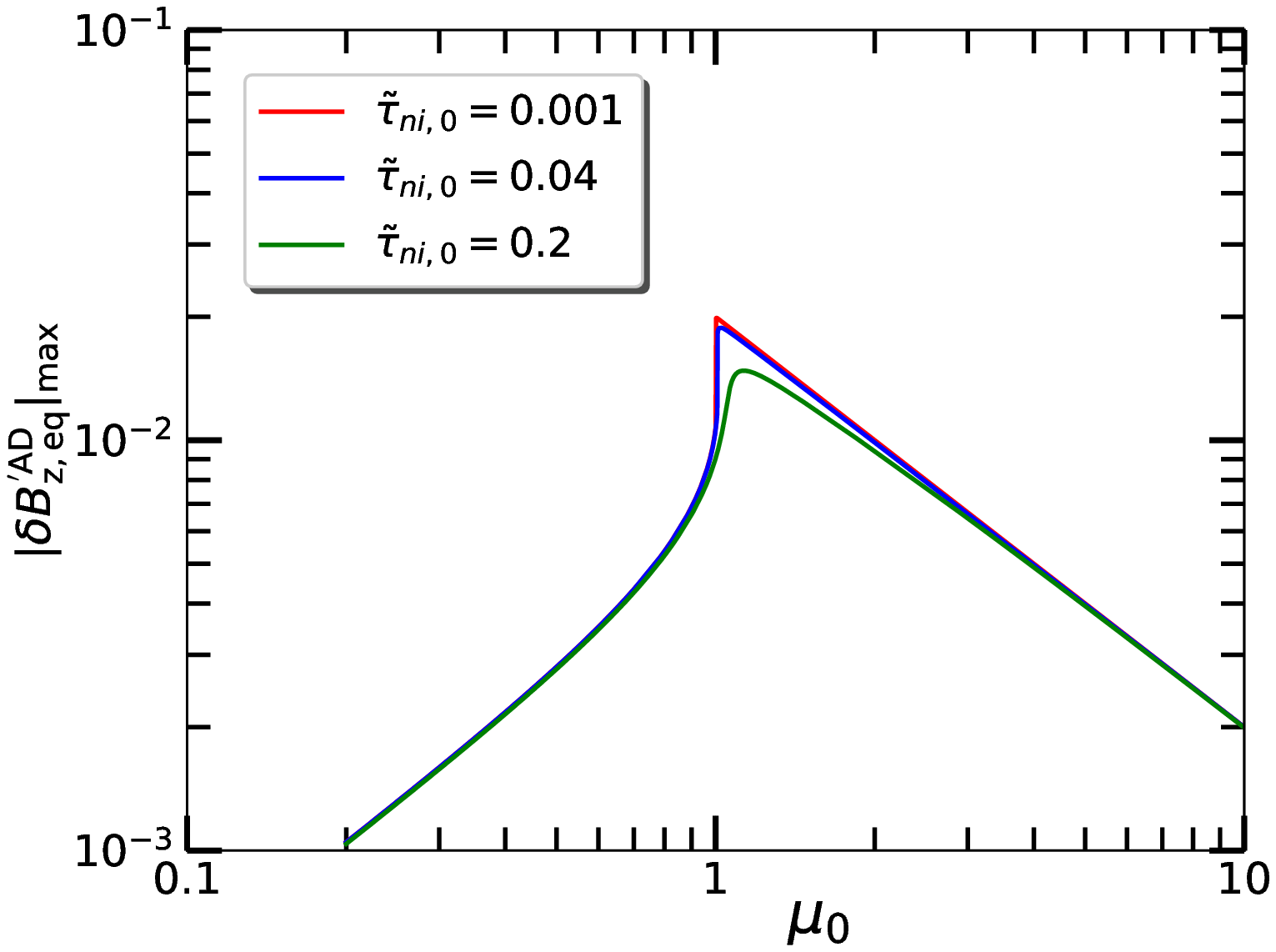}{0.45\textwidth}{(c)}
          \fig{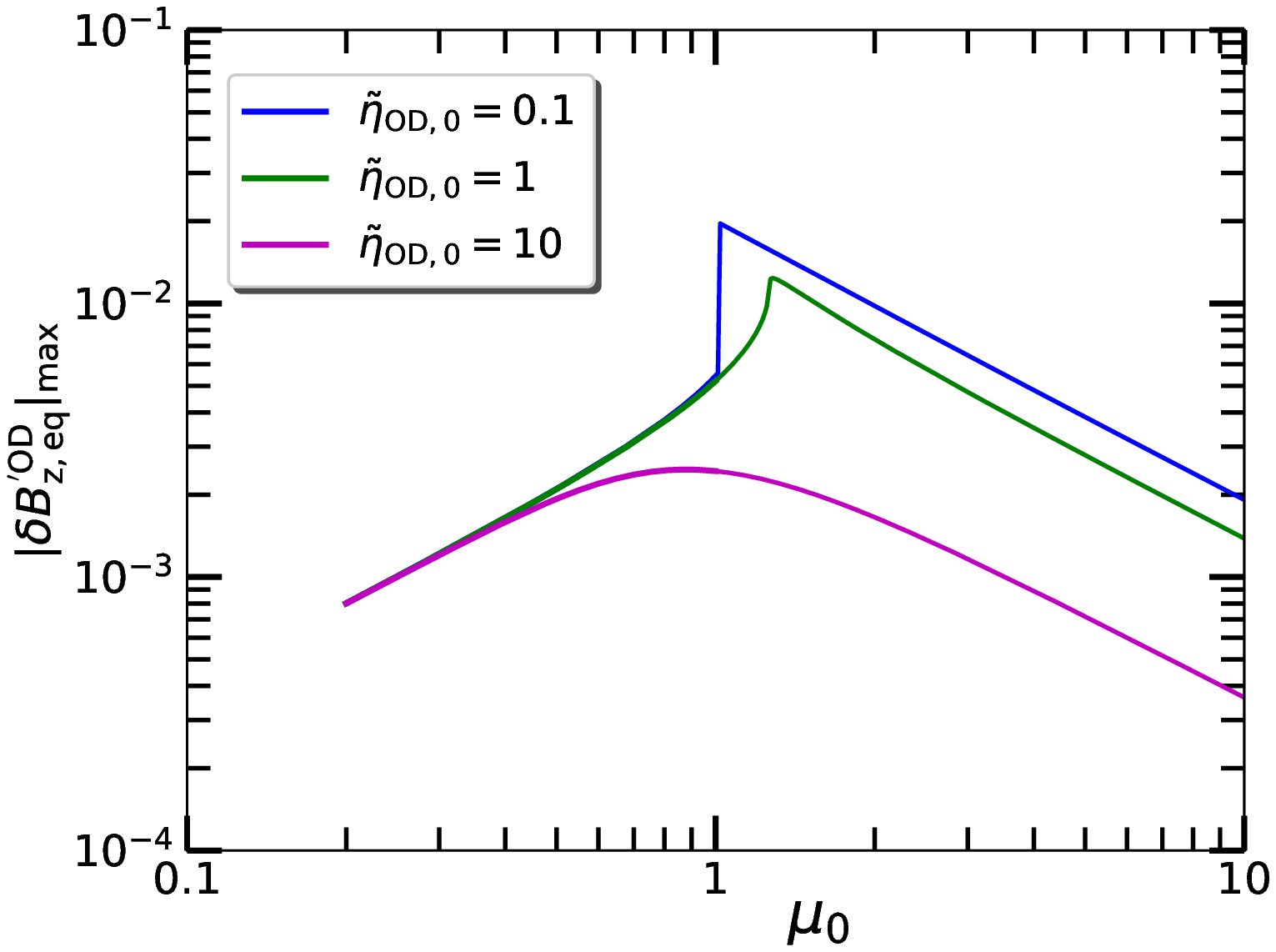}{0.45\textwidth}{(d)}
          }
\caption{Normalized maximum amplitude of perturbed velocity field and perturbed magnetic field as a function of normalized mass-to-flux-ratio ($\mu_0$). Upper left and lower left panel shows the cases for ambipolar diffusion with different normalized neutral-ion collision times $\tilde{\tau}_{ni,0}= 0.001, 0.04, 0.2$. Upper right and lower right panel shows the cases for Ohmic dissipation with different normalized Ohmic diffusivity $\etaODt= 0.1, 1, 10$.}
\label{fig:perturbedv_perturbedB}
\end{figure*}

\autoref{fig:figpertODADFF} shows the spatial variation of the perturbed column density function ($\delta \sigma'_n$),
the perturbed velocity field for neutrals ($\delta v'_{n,x}$) and ions ($\delta v'_{i,x}$), and
perturbed equatorial magnetic field ($\delta B'_{z, \rm{eq}}$) for three different MHD regimes: flux-frozen (FF), OD, and AD. 
We adopt $\delta \sigma'_{n,a} = 0.02$ to illustrate the regime of linear disturbances. 
Overall, we notice that adding a small amplitude perturbation to the initial column density ($\delta \sigma'_n$) gives rise to a perturbed magnetic field (as denoted by $\delta B^{'\rm{FF}}_{z, \rm{eq}}$, $\delta B^{'\rm{OD}}_{z, \rm{eq}}$, $\delta B^{'\rm{AD}}_{z, \rm{eq}}$) that follows the similar trend as $\delta \sigma'_n$ but has a relatively smaller amplitude as shown for each individual case (see the green line in \autoref{fig:figpertODADFF} a, d, g). 
This implies that the perturbation in the magnetic field will grow in the same way as the column density because the field lines are (at least partially) attached to matter. 
Whereas, the perturbed velocity field ($\delta v'_{n,x}$ and $\delta v'_{i,x}$) evolves keeping a phase-shift of $\pi/2$ with respect to the perturbed column density field for all three cases, denoting inward motion toward the density peak. 
For the case of OD, $\delta v'_{n,x}$ = $\delta v'_{i,x}$ and for the case of AD we need to study the velocities separately (see \autoref{fig:perturbedv_perturbedB} for the detailed discussion). 
For the case of flux-freezing, as shown by the upper panel of \autoref{fig:figpertODADFF} (see a, b, c), we study eigenfunctions for supercritical clouds. For the OD and AD cases we study the subcritical as well as supercritical clouds, as shown in the middle panel (see \autoref{fig:figpertODADFF} d, e, f) and lower panel (see \autoref{fig:figpertODADFF} g, h, i), respectively. 
In the the flux-frozen case, we find that the amplitude of $\delta v^{'\rm{FF}}_{n,x}$ gradually decreases as $\mu_0$ increases (see \autoref{fig:figpertODADFF}b). In contrast, for the case with OD and AD (see \autoref{fig:figpertODADFF}e and f, respectively), as $\mu_0$ goes from a subcritical value to a supercritical region, $\delta v^{'\rm{OD}}_{n,x}$ and $\delta v^{'\rm{AD}}_{n,x}$ attain a maximum at a nearly transcritical $\mu_0$.
Similarly, from \autoref{fig:figpertODADFF}(c) we notice that maximum amplitude of the perturbed magnetic field for the case of flux-frozen goes down rapidly as $\mu_0$ increases implying that the magnetic field contribution becomes less effective as it moves to a more supercritical regime.
Whereas, we see that maximum amplitude of the perturbed magnetic field for the case of OD and AD increase up to a certain $\mu_0$ and then drop off for greater $\mu_0$ as seen from \autoref{fig:figpertODADFF}(f) and (i) respectively. The value of $\mu_0$ with the peak perturbed magnetic field amplitude corresponds to the peak preferred length scale for the model with $\etaODt=0.2$ and $\taunit=0$, as well as for the model with $\taunit=0.2$ and $\etaODt=0$.


\autoref{fig:perturbedv_perturbedB}(a) and (b) show a compact depiction of the maximum amplitude of the perturbed velocity field as a function of $\mu_0$ for the cases with only OD and only AD, respectively. 
For the case with only OD, neutrals and all the charged particles move collectively as a single fluid. 
Hence, the perturbed velocity is the same and is identified only by $|\delta v^{'\rm{OD}}_{n,x}|_{\rm max}$ which is equal to $|\delta v^{'\rm{OD}}_{i,x}|_{\rm max}$, (recall \autoref{eq:velocityequn_in}). 
However, for the case of only AD, one can obtain the perturbed velocity field separately for ions and neutrals. We find that the maximum amplitude of perturbed velocity for ions ($|\delta v^{'\rm{AD}}_{i,x}|_{\rm max}$) is much less than that of neutrals ($|\delta v^{'\rm{AD}}_{n,x}|_{\rm max}$) over the entire subcritical region. 
With increasing $\etaODt$ and $\taunit$, the maximum amplitude of perturbed infall velocity gradually increases for $\mu_0<1$. 
Finally, in the highly supercritical regime ($\mu_0 \sim 10$), all the curves for nonideal MHD cases attain the limiting value obtained for the flux-frozen (FF) case. This is because in this limit, the motion is dominated by gravity and all the particles move together. 

\autoref{fig:perturbedv_perturbedB}(c) and (d) present the maximum amplitude of the perturbed magnetic field as a function of $\mu_0$ for OD and AD, respectively. 
For $|\delta B^{'\rm OD}_{z,{\rm eq}}|_{\rm max}$ and $|\delta B^{'\rm AD}_{z,{\rm eq}}|_{\rm max}$, the peaks occur at nearly transcritical values that correspond to the peak preferred length scale for each respective case as seen in \autoref{fig:ADonly_ODonly}(b) and (d).
This feature uncovers the fact that magnetic field provides the greatest support at a specific $\mu_0$ that corresponds to the peak preferred length scale of gravitational instability. 
In the hydrodynamic limit ($\mu_0 \rightarrow \infty$), all the curves for FF, OD, and AD (see \Crefrange{eq:deltaBFF}{eq:deltaBAD}) will diminish to zero.

Not surprisingly, the shortest growth timescale as obtained in \autoref{fig:ADonly_ODonly}(a) and (c) can also be deduced using the maximum amplitude of perturbed velocity field of neutrals for each respective case (i.e., FF, AD, OD), using
\begin{equation}
    \tau'_{g,m} = \frac{\lambda'_{g,m}}{\bigg|\delta v^{'}_{n,x}\bigg|_{\rm{max}}} \frac{\delta \sigma'_{n,a}}{2\pi}\, ,
\end{equation} 
which is derived from \autoref{eq:timescale_equn}.
Coming to the case of AD, we know that collisions between neutrals and ions give rise to a drift speed between the two fluids. For ion fluid there is a balance between the Lorentz force and the drag force due to friction with neutrals. While for the neutral fluid, the inward pull of gravity is opposed by collisions with ions and other forces. 
In the subcritical regime ($\mu_0 < 1$) the neutrals also come into an effective force-balance, between gravity and the collisions with ions. 
As a consequence, the infall motion of the neutrals ($|\delta v^{'\rm{AD}}_{n,x}|_{\rm max}$) gets plateaued at a terminal velocity and becomes independent of $\mu_0$. 
Therefore, the timescale of contraction reaches a saturation in the regime $\mu_0 < 1$. 
But the ions are still tied to the field lines and hence the infall motion of ions as denoted by $|\delta v^{'\rm{AD}}_{i,x}|_{\rm max}$ gradually increases toward the supercritical regime. 
For the case of only OD, all particles move together, 
but collisions cause a loss of induced current, which dissipates the magnetic flux. When this happens in the subcritical clouds, it is a cause for slow contraction of the perturbed column density field. 

\section{Calculation of $W_*$}   \label{sec:w_star_app}
Using the thin-disk formalism, we earlier calculated the thermal midplane pressure for the neutrals including the effects of the weight of the gas column, the external pressure, and the magnetic pressure. Now, we investigate the effect of a central star (once present) of mass $M_*$ by including it in our vertical pressure balance equation \citep{dapp12}, which becomes
\begin{equation}
    \rho_{n,0}c_s^2 = \frac{\pi}{2}G \sigma_{n,0}^2 + P_{\rm{ext}} + W_* ,
\label{eq:protostarW}
\end{equation}
where $W_*$ is the extra vertical squeezing due to the newly formed star's gravitational field, integrated up to the disk's local vertical finite half-thickness $Z_0$. Therefore, it is
\begin{equation}
    W_* = 2GM_* \rho_{n,0} \int_{0}^{Z_0} \frac{z\> dz}{(r^2 +z^2)^{3/2}} \, ,
\label{eq:Wstar_int}    
\end{equation}
where $z$ is the vertical coordinate and $r^2 = x^2 + y^2$. Using the one-zone approximation we integrate from $z=0$ to a fixed $Z_0$  ($=\sigma_{n,0}/(2\rho_{n,0})$). Then we do a negative binomial expansion of the integrated result $\left[1/r - 1/(r^2+Z_0^2)^{1/2}\right]$ under the approximation $Z_0/r \ll 1$, keeping leading order terms to yield 
\begin{equation}
    W_* = \frac{GM_* \rho_{n,0} Z_0 ^2}{r^3} = \frac{GM_* \sigma_{n,0} ^2}{4 \rho_{n,0}r^3} \, .
\label{eq:W_star_int}    
\end{equation}
Next, we calculate a surface density keeping $\rho_{n,0}$ fixed. Substituting $W_*$ into \autoref{eq:protostarW}, and using \autoref{eq:W_star_int}, one finds the modified expression
\begin{equation}
    \sigma_{n,0} = \left[\frac{\rho_{n,0} \; c_s^2}{\frac{\pi}{2}G(1+\tilde{P}_{\rm{ext}}) + \frac{GM_*}{4\rho_{n,0} r^3 }}\right]^\frac{1}{2} .
\label{eq:modifiedsigma}
\end{equation}
We choose $M_* = 0.01 \; M_{\odot}$ and $r = 50 \; \rm{AU}$, corresponding to a very early stage of star formation, yielding $M_*/(4\rho_{n,0} r^3) = 0.03$ and $\sigma_{n,0} = 59.36 \>  \rm{g} \> \rm{cm}^{-2}$ from \autoref{eq:modifiedsigma}, which is roughly the same as the protostellar disk surface density for $T=30 \, \rm{K}$ if we take $W_* =0$ in our model. Therefore, we see that \autoref{eq:modifiedsigma} can be simplified to the former \autoref{eq:pbal_lin} when $M_* =0$. Whereas, at a later stage of protostar formation, taking $M_* = 0.5 \>M_{\odot}$ and $r = 50 \;\rm{AU}$, we find $\sigma_{n,0} = 43.53\,  \rm{g} \, \rm{cm}^{-2}$ since $M_*/(4\rho_{n,0} r^3) = 1.54$. 
We can also write a generalized expression for $Z_0$ including the effects of $M_*$ as follows:
\begin{equation}
    Z_0 = \frac{1}{2 \rho_{n,0}} \left[\frac{\rho_{n,0} \; c_s^2}{\frac{\pi}{2}G(1+\tilde{P}_{\rm{ext}})+ \frac{GM_*}{4\rho_{n,0} r^3}} \right]^\frac{1}{2} \, .
\label{eq:general_Z0}    
\end{equation}
For the case of $M_* =0$, the above expression can be reduced to $Z_0 = \sigma_{n,0}/(2\rho_{n,0})$ using \autoref{eq:pbal_lin}. 
Using \autoref{eq:general_Z0}, the values of $Z_0$ are calculated to be $5.11 \, \rm{AU}$ and $3.74 \, \rm{AU}$ for $M_* =0.01\, M_{\odot}$ and $0.5\, M_{\odot}$, respectively.


\bibliography{myref}{}
\bibliographystyle{aasjournal}

\end{document}